\documentclass[11pt]{article}
\usepackage{microtype}
\usepackage{xspace}
\usepackage{bbm}
\usepackage{graphicx}
\usepackage{amsmath,amssymb,amsthm}

\usepackage{paralist}
\usepackage{bm}
\usepackage{xspace}
\usepackage{url}
\usepackage{fullpage, prettyref}
\usepackage{boxedminipage}
\usepackage{wrapfig}
\usepackage{ifthen}
\usepackage{color}
\usepackage{xcolor}
\usepackage{algpseudocode}

\usepackage{algorithm}
\usepackage{fullpage}
\usepackage{hyperref}
\usepackage[compact]{titlesec}
\usepackage{cleveref}
\usepackage{nicefrac}

\usepackage[bottom=1in,top=1in,left=1in,right=1in]{geometry}

\usepackage{framed}
\renewenvironment{leftbar}[1][\hsize]
{%
    \MakeFramed{\hsize#1\advance\hsize-\width\FrameRestore}%
}
{\endMakeFramed}

\usepackage{thmtools}
\usepackage{thm-restate}
\declaretheorem[numberwithin=section]{theorem}
\declaretheorem[sibling=theorem]{lemma}
\declaretheorem[sibling=theorem]{claim}

\crefname{claim}{claim}{claims}

\newcommand{\ex}{{\mathbb E}}
\newcommand{\pr}{{\mathbb P}}

\newcommand{\ignore}[1]{}

\newcommand{\I}{\ensuremath{{\cal I}}\xspace}

\newcommand{\cov}{\ensuremath{{\sf Cov}}\xspace}
\newcommand{\sse}{\ensuremath{\subseteq}\xspace}

\newcommand{\Z}{{\mathbb Z}}

\newcommand{\poly}{\mathrm{poly}}

\newcommand{\calL}{{\cal L}}
\newcommand{\calF}{{\cal F}}

\newcommand{\calS}{{\cal S}}

\newcommand{\calC}{{\cal C}}

\newcommand{\calA}{{\cal A}}

\newcommand{\bt}{\boldsymbol{t}}

\newcommand{\ceil}[1]{\lceil#1\rceil}

\newcommand{\initOneLiners}{%
    \setlength{\itemsep}{0pt}
    \setlength{\parsep }{0pt}
    \setlength{\topsep }{0pt}
}
\newenvironment{OneLiners}[1][\ensuremath{\bullet}]
    {\begin{list}
        {#1}
        {\initOneLiners}}
    {\end{list}}

\newcommand{\Opt}{\ensuremath{\mathsf{Opt}}\xspace}
\newcommand{\lp}{\mathsf{Lp}}

\newcommand{\seq}{\pmb{\sigma}}
\newcommand{\density}{\rho}
\newcommand{\univ}{A}

\renewcommand{\bt}{{t}}

\renewcommand{\cov}{{\tt cov}}

\newcommand{\St}{\calS_\bt}
\newcommand{\At}{\univ_\bt}
\newcommand{\Lt}[1]{\calL_\bt({#1})}
\newcommand{\Et}[1]{\At(#1)}
\newcommand{\dt}{\density_\bt}
\newcommand{\vt}{\mathsf{vol}}
\newcommand{\nt}{n_\bt}
\newcommand{\Nt}{N_\bt}
\newcommand{\ft}{f_\bt}
\newcommand{\lt}{{l_\bt}}
\newcommand{\covt}{\cov_\bt}
\newcommand{\ini}{\ensuremath{\mathsf{init}\xspace}}
\newcommand{\fin}{\ensuremath{\mathsf{fin}\xspace}}

\usepackage{caption}

\DeclareCaptionFormat{algor}{%
  \hrulefill\par\offinterlineskip\vskip1pt%
    \textbf{#1#2}#3\offinterlineskip\hrulefill}
\DeclareCaptionStyle{algori}{singlelinecheck=off,format=algor,labelsep=space}
\title{Online and Dynamic Algorithms for Set Cover}
\author{Anupam Gupta\thanks{Carnegie Mellon University. Email: {\tt anupamg@cs.cmu.edu}}
\and Ravishankar Krishnaswamy\thanks{Microsoft Research India. Email: {\tt rakri@microsoft.com}}
\and Amit Kumar\thanks{IIT Delhi. Email: {\tt amitk@cse.iitd.ac.in}}
\and Debmalya Panigrahi\thanks{Duke University. Email: {\tt debmalya@cs.duke.edu}}}
\date{}

\begin{document}

\maketitle

\begin{abstract}
	In this paper, we study the set cover problem in the fully dynamic model. 
    In this model, the set of active elements, i.e., those that must be covered
    at any given time, can change due to element arrivals and departures. The 
    goal is to maintain an algorithmic solution that is competitive with respect 
    to the current optimal solution. This model is popular in both the dynamic 
    algorithms and online algorithms communities. The difference is in the 
    restriction placed on the algorithm: in dynamic algorithms, the 
    running time of the algorithm making updates (called update time)
    is bounded, while in online algorithms, the number of updates made
    to the solution (called recourse) is limited.
    
	We give new results in both settings (all recourse and update time
    bounds are amortized):
    \begin{itemize}
	\item
	In the update time setting, we obtain $O(\log n)$-competitiveness with 
    $O(f\log n)$ update time, and $O(f^3)$-competitiveness with $O(f^2)$ 
    update time. The $O(\log n)$-competitive algorithm is the first one
    to achieve a competitive ratio independent of $f$ in this setting.
    The second result improves on previous work by removing an $O(\log n)$
    factor in the update time bound. This has an important consequence: we
	obtain the first deterministic constant-competitive, 
    constant update time algorithm for fully-dynamic vertex cover.
    \item 
    In the recourse setting,
    we show a competitive ratio of $O(\min\{\log n, f\})$
    with constant recourse.
    The most relevant previous result is the $O(\log m\log n)$ bound for
    online set cover in the insertion-only model with no recourse. Note
    that we can match the best offline bounds with $O(1)$ recourse, something
    that is impossible in the classical online model.
	\end{itemize}
    These results also yield, as corollaries, new results for the maximum $k$-coverage
    problem and the non-metric facility location problem in the 
    fully dynamic model.
    
    Our results are based on two algorithmic frameworks in the fully-dynamic 
    model that are inspired by the classic greedy and primal-dual algorithms
    for offline set cover. We show that both frameworks can be used for 
    obtaining both recourse and update time bounds, thereby demonstrating 
    algorithmic techniques common to these strands of research.
\end{abstract}

\thispagestyle{empty}
\setcounter{page}{0}
\clearpage

\section{Introduction}
\label{sec:introduction}

In the (offline) set cover problem, we are given a universe $U$ of $n$
elements and a family $\calF$ of $m$ sets with non-negative costs. The
goal is to find a subfamily of sets $\calS \sse \calF$ of minimum cost
that covers $U$. Several techniques achieve an approximation factor of
$\ln n$ for this problem, and we cannot achieve $(1 - \varepsilon)\ln n$
unless $P=NP$~\cite{SW10,DS14}. The set cover problem has been popular due to
its wide applicability. However, in many applications of this problem,
we want to cover some subset $A \sse U$ of the universe, and this set
changes over time. In general, each update to $A$ inserts or deletes an
element, and now we must allow our algorithm to change the solution to
restore the feasibility and approximation. We call this the
{\em (fully-)dynamic set cover} problem.

As in all dynamic algorithms, we want to avoid recomputing the solution
from scratch, and hence constrain our algorithm to make only ``limited''
changes. This means it cannot use the offline algorithm off-the-shelf.
Two different communities---in {\em online} algorithms and in {\em
  dynamic} algorithms---have approached such problems in slightly
different ways. Mainly, they differ in the restrictions they
place on the changes we can make after each update in the input. In online
algorithms, where decisions are traditionally irrevocable, the dynamic (or so-called recourse)
version of the model allows us to change a small number of past
decisions while maintaining a good competitive ratio. The number of
changes at each step is called its \emph{recourse}. In the context
of set cover, at each update, the algorithm is allowed to change a
limited number of sets in the solution.  In contrast, in dynamic
algorithms, the parameter of interest is the running time to implement
this change. This running time is usually called the {\em update time}.

Note the difference between the models: the online model ostensibly does
not care about run-times and places only an information-theoretic
restriction, whereas the dynamic model places a stronger, computational
restriction. Hence, a bound on the update time automatically implies the
same bound on recourse, but not the other way around. In most cases,
however, this observation cannot be used directly because the recourse
bounds one desires (and can achieve) are much smaller than the bounds on
update time. This is perhaps the reason that research in these two
domains has progressed largely independent of each other; one exception
is~\cite{LOPSZ15} for the Steiner tree problem.
Indeed, for set cover, online algorithms
researchers have focused on obtaining (poly)logarithmic approximations
in the insert-only model with no recourse~\cite{AAABN}. In dynamic
algorithms, this problem has been studied as an extension of dynamic
matching/vertex cover, and the current results give approximation
factors that depend on $f$, the maximum element frequency.\footnote{The
  {\em frequency} of an element is the number of sets it belongs to;
  hence $f=2$ for vertex cover instances, where the elements are edges
  and the sets are vertices, since an edge/element belongs to two
  vertices/sets.}  In this paper, we bring these two strands of research
closer together, both in terms of algorithmic techniques and achieved
results. We give new results, and improve existing ones, in both
domains, and develop a general framework that should prove useful
for other problems in both domains.


\subsection{Our Results} Following the literature, our
recourse/update time bounds, except where explicitly stated, are
\emph{amortized bounds}. In other words, the recourse/update time in a
single input step can be higher than the given bounds, but the average
over any prefix of input steps obeys these bounds.

Let $n_t$ denote the number of elements that need to be covered at time
$t$, and $n$ denote the maximum value of $n_t$, i.e., $n = \max_t n_t$.
Similarly, let $f_t$ be the maximum frequency of elements active at time
$t$, and $f = \max_t f_t$.
Our main results for the setting of online algorithms with recourse are:
\begin{theorem}[Recourse]
  \label{thm:main-recourse}
  There exist polynomial-time algorithms for the (fully-)dynamic set
  cover problem with $O(1)$ recourse per input step:
  \begin{OneLiners}
  \item[(a)] an $O(\log n_t)$-competitive deterministic
    algorithm, and
  \item[(b)] an $O(f_t)$-competitive randomized algorithm.
  \end{OneLiners}
  Moreover, these can be combined to give a single algorithm that
  achieves $O(\min(f_t,\log n_t))$-competitiveness with $O(1)$ recourse.
\end{theorem}

The only prior result known for online set cover is the seminal $O(\log
m \log n)$ competitive algorithm for the classical (insertion-only,
no recourse)
online model with $m$ sets and $n$ elements~\cite{AAABN}. Moreover,
there is a matching $\Omega(\log m \log n)$ lower bound, assuming
$P\neq NP$~\cite{Korman05}. Hence, our
result shows that this lower bound breaks down even if we allow only
(amortized) constant changes in the solution per input step; moreover, we
are able to remove the dependence on the number of sets $m$ altogether 
from the competitive ratio. 
This is particularly useful because many problems in combinatorial optimization
can be modeled as set cover using exponentially many sets. 
(See Appendix~\ref{sec:fl} for an example: the dynamic non-metric
facility location problem.) Observe that
our competitive ratio is asymptotically tight assuming $P \not= NP$.


Next, we give our results for dynamic set cover in the update-time setting:
\begin{theorem}[Update Time]
  \label{thm:main-update}
  There exist polynomial-time algorithms for the (fully-)dynamic set
  cover problem:
  \begin{OneLiners}
  \item[(a)] an $O(\log n_t)$-competitive deterministic algorithm with
    $O(f\log n)$ update time, and
  \item[(b)] an $O(f_t^3)$-competitive deterministic algorithm
    with $O(f^2)$ update time.
  \end{OneLiners}
  Moreover, we can combine these into a single algorithm that achieves
  $O(\min(f_t^3,\log n_t))$-competitive algorithm with $O(f(f + \log n))$ update time.
\end{theorem}
To the best of our knowledge, part~(a) above is the first result to
obtain a competitive ratio independent of the maximum element frequency
$f$, with guarantees on the update time. Indeed, the current-best
existing result for dynamic set cover obtains a competitive ratio of
$O(f^2)$ with $O(f\log n)$ update time~\cite{Bhatta-icalp}. Our result
of part~(b) is able to remove the dependence on $\log n$ in the update
time, although the competitive ratio worsens from $O(f^2)$ to $O(f^3)$.

For the case of vertex cover where $f=2$, we obtain an
$O(1)$-competitive deterministic algorithm
for the \emph{dynamic (weighted) vertex cover} problem, with $O(1)$
update time. The previous-best deterministic algorithm was
$(2+\varepsilon)$-competitive with $O(\varepsilon^{-2} \log n)$ update
time~\cite{Bhatta}. For randomized algorithms, Solomon~\cite{Solomon16}
recently gave an algorithm to maintain maximal matchings with $O(1)$
update time; this gives a $2$-competitive algorithm for (unweighted)
vertex cover. Our algorithm can be seen as giving (a)~a deterministic
counterpart to Solomon's result for vertex cover (see \cite{Bhatta}
for a discussion of the challenges in getting a deterministic
algorithm that matches the randomized results in this context),
and (b) extending it from unweighted
vertex cover to weighted set cover (a.k.a.\ hypergraph vertex
cover).

The algorithm in Theorem~\ref{thm:main-update}(a) can also be adapted
to obtain a constant-competitive, $O(\log n)$-recourse, and 
$O(f\log n)$-update time algorithm for the maximum $k$-coverage
problem in the fully-dynamic model. We give details of this application
in Appendix~\ref{sec:coverage}.

Finally, we remark that the competitive ratios in Theorem~\ref{thm:main-recourse} (a)
and Theorem~\ref{thm:main-update} (a) can be improved to $O(\log \Delta_t)$,
where $\Delta_t := \max |S\cap \At|$ is the maximum cardinality among 
all sets at time $t$. Although this a tighter bound since $\Delta_t \leq n_t$, 
for the sake of clarity, we will first prove the bound
of $O(\log n_t)$ in both cases, and then improve the analysis to 
obtain the tighter bound.

\textbf{Non-Amortized Results.} Given these amortized bounds on recourse
and update time, one may wonder about non-amortized bounds. We give a
partial answer to this question, for the case of recourse bounds. We
show that if the algorithm were given exponential time, the recourse
bound in Theorem~\ref{thm:main-recourse}(a) can be made
non-amortized. We leave the problem of obtaining polynomial-time
algorithms with non-amortized guarantees for these models an interesting open question.

\begin{theorem}
  \label{thm:scaling}
  There is a $O(\log n)$-competitive deterministic algorithm for the
  dynamic set cover problem, with $O(1)$ non-amortized recourse per
  input step. If all sets have the same cost (unweighted set cover),
  then the competitive ratio improves to $O(1)$. This algorithm runs in
  exponential time.
\end{theorem}

\subsection{Our Techniques}

We grouped our results based on recourse or update time, but the
techniques give a different natural grouping. Indeed, the $O(\log
n)$-competitive results are based on related greedy-based algorithms,
and the frequency-based results are based on primal-dual techniques.

\textbf{Greedy Algorithms.}  The results of
Theorems~\ref{thm:main-recourse}(a) and~\ref{thm:main-update}(a) are
based on a novel \emph{dynamic greedy} framework for the set cover problem.  A
barrier to making greedy algorithms dynamic is their sequential nature,
and element inserts/deletes can play havoc with this. However, we
abstract out the intrinsic properties that the analysis of greedy uses,
and show these properties can be maintained fast, and with small amounts
of recourse. Our algorithm does simple ``local'' moves, and the analysis
uses a delicate token scheme to ensure constant recourse and small
run-time.

In more detail, the greedy algorithm chooses sets one-by-one, minimizing
 the \emph{incremental cost-per-element} covered at each step. The analysis then shows that number of elements covered at incremental-costs $\approx 2^i ( \nicefrac{\Opt}{n} )$ is at most $n/2^i$, which will easily give us the desired $O(\log n)$ bound for approximation factor. So our abstraction in the dynamic setting is then the following: \emph{try to cover as many elements at low incremental costs!} Indeed, if at some time we can find a set $S$ along with some elements $X \subseteq S$ such that the \emph{new incremental-cost} $\nicefrac{c_S}{|X|}$ is at most half the current incremental-cost for each element in $X$, then we add such a set to the solution, and repeat.
Of course, changes beget other changes---if set $S$ now covers
element $e$ which was covered by $T$, the cost of $T$ is now shared
among fewer elements, causing their incremental-costs to increase. This can cause
cascades of changes. We nevertheless show that this works using a delicate token-based argument: each new element brings $L$
tokens with it, and any time a new set is bought, we expend one token. If we make sure that the total tokens remaining is always non-negative, we'd be done. Proving
that this can be done with $L = O(1)$ tokens per
element is the basis of Theorem~\ref{thm:main-recourse}(a).
The proof of
Theorem~\ref{thm:main-update}(a) is similar, albeit we need to now
argue about running times.




\textbf{Primal-Dual Algorithms.} The results of
Theorems~\ref{thm:main-recourse}(b) and~\ref{thm:main-update}(b) are
inspired by {\em primal-dual} algorithms. Offline, we raise some duals
until some set constraint becomes tight (so that the set is paid for):
we then pick this set in our solution. Elements pay for at most $f$
sets, hence the $f$-approximation. But if elements disappear and take
their dual with them, tight constraints can become slack, and we must
take recourse action.

For Theorem~\ref{thm:main-update}(b), let us use vertex cover to
illustrate our ideas. Inspired by previous
work~\cite{Baswana,Bhatta,Solomon16}, a natural idea is to place each
vertex $v$ at some \emph{integer level} $\ell(v)$, and set the dual $y_e$ for an edge $(u,v)$ to be $\nicefrac{1}{2^{\max(\ell(u),\ell(v))}}$. To define the solution, then we include all vertices whose dual constraints are approximately tight i.e., $\calS = \{ v \, : \, \nicefrac12 \leq \sum_{e \sim v} y_e \leq 2 \}$. The cost is automatically bounded because we only include approximately-tight vertices, and the dual solution is approximately feasible. Now, when edges arrive/depart, these bounds might be violated and then we move the vertices up or down accordingly. To bound the updates, we again use
a token scheme, where tokens are added by arrivals/deletions, and are spent
on updates. When vertices move, they also transfer some tokens to
their neighbors on incident edges, which then use up these tokens when they move, and so on.
A key idea in our analysis is to use asymmetry in token transfer---vertices
moving down transfer more tokens to their neighbors than vertices moving
up. Using this idea, we
obtain a \emph{deterministic} constant-competitive vertex cover algorithm with
constant update time. These ideas extend to the set cover setting,
giving $O(f^3)$-competitive $O(f^2)$ update time algorithms; this is the
first result for dynamic set cover with update time independent of $n$.

Finally, Theorem~\ref{thm:main-recourse}(b): getting $f$-competitiveness
and $O(f)$-recourse is easy, but making this $O(1)$-recourse is the
challenge. We show that an elegant randomized algorithm due to
Pitt~\cite{Pitt}---for an uncovered element $e$, pick a random set covering it from
the ``harmonic'' distribution, and
repeat---can be made dynamic with $O(1)$-recourse. We then use dual-fitting to bound the cost.

\textbf{Combiner Algorithm:} We then show how we can dynamically maintain the \emph{best-of-both} solutions with the same recourse/update-time bounds. In the offline setting, getting such an algorithm is trivial as we can simply output the solution with lower cost. However, this is not immediate in the online setting as the lower-cost solution could oscillate between the two solutions we maintain; so we exploit the values of the two competitive-ratio guarantees to design our overall combiner algorithm.

\subsection{Related Work}

Set cover has been studied both in the offline~\cite{SW10} and online
settings~\cite{AAABN}; under suitable complexity-theoretic assumptions,
both the $O(\log n)$ and $O(f)$-approximations are best possible.

Dynamic algorithms are a vibrant area of research; see~\cite{EGI-survey}
for many applications and pointers. Maintaining approximate vertex
covers and approximate matchings in a dynamic setting have seen much
interest in recent years, starting with~\cite{OnakR10};
see~\cite{Baswana,NeimanS16,GuptaP13,Bhatta,Bhatta-icalp,BernsteinS15,BernsteinS16,BHN16}.
For the exact version of the problem~\cite{Sank} gives polynomial update
times, and logarithmic times are ruled out by recent
works~\cite{abboud2014popular,HKNS15,KPP16}.

The study of online algorithms with recourse goes back at least to the
dynamic Steiner tree problem in~\cite{IW91}: motivated by lower bounds
for fully-dynamic inputs~\cite{AzarBK94,AKPPW93} for online scheduling
problems, \cite{PW98, Wes00, AGZ99, SSS, SV10, EL11} studied models with
job reassignments. Maintaining an exact matching with small recourse was
studied by~\cite{GKKV95,CDKL09, bosek2014online}; low-load assignments and
flows by~\cite{Wes00,GKS14-matching}, Steiner trees by~\cite{MSVW12,
  GuGK13, GK14-steiner, LOPSZ15}. As far as we know, \L{}acki et
al.~\cite{LOPSZ15} are the only previous work trying to bridge the
online recourse and dynamic algorithms communities.

\noindent
{\bf Bibliographic Note:}
Very recently, and independently of us, Bhattacharya et
al.~\cite{Bhatta2} get a deterministic $O(1)$-competitive, $O(1)$
update-time algorithm for dynamic vertex cover. Since the
manuscript became public very recently, we are yet to
investigate the similarities and differences.


\subsection{Notation}
\label{sec:notation}

The input is a set system $(U, \calF)$; $c_S$ is the cost of set $S$.
In (dynamic) set cover, the input sequence is $\pmb{\sigma} = \langle
\sigma_1, \sigma_2, \ldots \rangle$, where request $\sigma_t$ is either
$(e_t, +)$ or $(e_t, -)$. The initial \emph{active set} $\univ_0 =
\emptyset$. If $\sigma_t = (e_t, +)$, then $\univ_t \gets \univ_{t-1}
\cup \{e_t\}$; if $\sigma_t = (e_t, -)$ then $\univ_t \gets \univ_{t-1}
\setminus \{e_t\}$. 
We do not need to know either the universe $U$ or the entire family
$\calF$ up-front. Indeed, at time $t$, we know only the elements seen so
far, and which sets they belong to.

We need to maintain a feasible set cover $\calS_t \sse \calF$, i.e., the sets
in $\calS_t$ must cover the set of active elements $\univ_t$. Define $n_t :=
|\univ_t|$ and $n = \max_t n_t$. The \emph{frequency} of an element $e \in U$ is the number
of sets of $\calF$ it belongs to; let $f_t := \max_{e \in A_t}
\text{frequency}(e)$ be the maximum frequency of any active element at
time $t$.  Let $\Opt_t$ be the cost of the optimal set cover
for the set system $(\univ_t, \calF)$.



\textbf{Recourse and Update Times.} The \emph{recourse} is the number of sets we add
or drop from our set cover over the course of the algorithm, as a
function of the length of the input sequence. An online algorithm is
\emph{$\alpha$-competitive} with \emph{$r$ (amortized) recourse} if at
every time $t$, the solution $\calS_t$ has total cost at most $\alpha
\cdot \Opt_t$, and the total recourse in the first $t$ steps is at most
$r \cdot t$.  (Since every dropped set must have been added at some
point and we maintain at most $n_t$ sets at time $t$---at most one for
each active element---counting only the number of sets \emph{dropped}
until time $t$ changes the recourse only by a constant factor.) For
\emph{$r$ worst-case recourse}, the number of sets dropped at each
time-step must be at most $r$. In the \emph{update-time} model, we
measure the \emph{amount of time} taken to update the solution $\calS_{t-1}$ to
$\calS_t$. 
We use the terms \emph{amortized} or \emph{worst-case} \emph{update
  time} similarly.


\section{Dynamic Greedy Algorithms}
\label{sec:greedy}

In this section, we describe our greedy framework and sketch the main ideas for proving~\Cref{thm:main-update}(a) and~\Cref{thm:main-recourse}(a). Complete details appear in~\Cref{sec:logn1-full,sec:lognlogn-full} respectively.

\subsection{The Dynamic Greedy Framework}
\label{sec:framework}

We now describe a generic framework for greedy set cover algorithms in
the fully dynamic model. We will later instantiate this framework in two
different ways to obtain our results in the recourse and update time
settings. An algorithm for dynamic set cover maintains a solution
(denoted $\St$ at time $t$) with sets that cover the active elements
$A_t$.  In addition, our greedy framework also maintains an {\em
  assignment} $\varphi_t(e)$ of each active element $e$ to a unique set
in $\St$ covering it; define $\covt(S) := \{e \mid \varphi_t(e) = S\}$ to be the set of elements assigned to $S$, and $S$ is said to be \emph{responsible for
  covering} the elements in $\covt(S)$. Our algorithms also use the notions of volume and density.

\begin{itemize}
\item \emph{Volume.} At each time, our algorithm maintains for every element a notion of
  \emph{volume} $\vt(e) > 0$.
\item \emph{Density.} Define the {\em density} of a set $S$ in
  $\St$ as $\dt(S) := c(S)/\sum_{e\in\covt(S)} \vt(e)$, the ratio of its
  cost and the volume of elements it covers.
\end{itemize}
For concreteness, think of $\vt(e) = 1$ for all $e$, and hence the
density of a set $S$ is the standard notion of per-element-cost of covering elements in
$\covt(S)$. In fact this is what we will use for the update time algorithm; later, we
will use a different notion of volume in our constant recourse algorithm.
\emph{The notion of element volumes is all that we need to characterize our algorithms.}

\begin{itemize}
\item \emph{Density Levels.}  We also place each set in $\St$ in some
  {\em density level}. Each density level $\ell$ has an associated range
  $R_\ell := [2^\ell, 2^{\ell+10}]$ of densities.
  Any set $S$ at level $\ell$ must have density
  $\dt(S)$ in the interval
  $R_\ell$. 
  We say that element $e$ is at level $\ell$ if its covering set
  $\varphi_t(e)$ is at level $\ell$.
\end{itemize}
Note that adjacent density ranges overlap across multiple levels; this range gives the necessary friction which prevents too many changes in our algorithm which helps in bounding both recourse and update time. The algorithm will dynamically make sure that each set $S \in \St$ covering a set $\covt(S)$ elements at some time $t$ will be placed in one of its allowed levels.
 We are now armed to define the crucial ``greedylike'' concept in the dynamic setting --- the notion of {\em
  stability}. 
\begin{itemize}
\item \emph{Stable Solutions.} A solution $\St$ is \emph{stable} if
  for every density level $\ell$, there is no subset $X$ of elements
  currently at level $\ell$ (perhaps covered by different sets) that can
  all be covered by some set $S$, such that the density of the resulting set $c(S)/\sum_{e\in
  X} \vt(e) < 2^\ell$, i.e., the set $S$ (if added to $\St$) and these elements $X$ would belong in a \emph{strictly lower density level}.
\end{itemize}

Loosely, stability means there is no collection of elements that can
jointly ``defect'' to be covered by a set $S$ which charges them less
cost-per-element. Now the dynamic algorithm just tries to maintain a
stable solution.  Suppose the current solution $\calS_{t-1}$ is
stable.

\begin{leftbar}
\noindent {\bf Arrival of $e$.} Add the cheapest set
covering $e$ to $\St$, and run {\sf Stabilize.}

\noindent {\bf Departure of $e$.} Remove $e$ from its covering set $S = \varphi_{t-1}(e)$, and update $S$'s density and $\covt(S) = \cov_{t-1}(S) \setminus \{e\}$. If $\covt(S) = \phi$, delete $S$ from $\St$. Else, if its density falls outside the range of its level, move it to the \emph{highest level} which can accommodate it. Run {\sf Stabilize.}

\noindent {\bf Stabilize.}  Repeatedly perform the following steps until the
solution is stable: if there exists level $\ell$ and
elements $X$ currently at level $\ell$, such that $X \sse S$ for some $S \in
\calF$, and the density $c(S)/\sum_{e \in X} \vt(e) < 2^\ell$: add $S$ to $\St$,  and reassign the elements in $X$ to $S$ by updating $\varphi_t(\cdot)$ for elements in $X$; place $S$ at the \emph{highest density level} where it is allowed. Also update $\covt(\cdot)$ for the sets previously covering elements in $X$. As a result, if the updated density of such a set $S'$ previously covering some elements in $X$ increases beyond $2^{\ell+10}$, we move it to the \emph{highest
level} that can accommodate it.
\end{leftbar}

This completes the description of the algorithm framework, and also completes the description of our dynamic algorithm for update time (since the notion of volume is $\vt(e) = 1$ always). The bulk of
our analyses is in showing that such algorithms terminate, and moreover,
that they make a small number of updates. However, we can already show
that {\em if} we find a stable solution, the cost is small.


\begin{lemma}
  \label{lma:cost}
  The sum of costs of sets in a level $\ell$ in any stable
  solution at time $t$ is at most $2^{10}\,\Opt_t$.
\end{lemma}
\begin{proof}
  Suppose not, and some level $\ell$ contains sets of total cost $c \cdot \Opt_t$ where
  $c > 2^{10}$. Let the total volume of elements at level $\ell$ be
  $\vt_{\ell}$. Then there \emph{exists a set} in $\St$ at this level with density at least $c \cdot (\Opt_t/\vt_\ell)$, and therefore the upper density limit of this level is at least as large. In turn, this implies that the smallest density allowed at this level is at
  least $(c/2^{10}) \cdot (\Opt_t / \vt_{\ell}) > (\Opt_t / \vt_{\ell})$.
  On the other hand, the optimal solution covers all elements at level $\ell$ and has an average density of $\Opt_t / \vt_{\ell}$; in particular, there is some set with density \emph{at most}
  $\Opt_t/ \vt_\ell$. But the density of this set is too low for level $\ell$,
  contradicting the stability condition.
\end{proof}
To complete the cost analysis, we will then argue that our algorithms
maintain only $O(\log n_t)$ non-trivial density levels, so
Lemma~\ref{lma:cost} implies $O(\log n_t)$-competitiveness.

\subsection{An $O(\log n_t)$-Competitive $O(f \log n)$-Update-Time Algorithm}
\label{sec:lognlogn}

We now present our $O(\log n_t)$-competitive algorithm with amortized
update time of $O(f \log n)$, where $f$ is the maximum element
frequency.
To completely describe the algorithm, we simply define the volume of
elements. We then bound the competitive ratio, and finally the update
time.

\begin{leftbar}
   The volume of every element $\vt(e) = 1$ at all times.
\end{leftbar}

{\bf Competitive Ratio.}
The total cost of all sets $S$ with density $\dt(S) \leq \Opt_t/\nt$ is at
most $\Opt_t$, since there are $\nt$ active elements at time
$t$. Moreover, by Lemma~\ref{lma:cost}, the highest cost set in any
stable solution has cost at most $2^{10}\cdot \Opt_t$. Since the element
volumes are all $1$, the maximum density of any set is at most
$2^{10}\cdot \Opt_t$. Consequently we only need to consider the $O(\log
\nt)$ levels with density between $\Opt_t/\nt$ and $2^{10}\cdot
\Opt_t$. Using Lemma~\ref{lma:cost} again, the algorithm is $O(\log
\nt)$-competitive.

\medskip\noindent {\bf Update Time.} We bound the update time in two
steps. We first bound the number of level changes made by elements
to $O(\log n)$ per element
arrival. Then, we bound the total update time by $O(f)$ times
the number of level changes by elements. This gives the amortized update
time bound of $O(f\log n)$. For this second step we use that the data
structures we maintain change only when an element changes level, and
that these data structures have an update time of $O(f)$ per
element level change. These details are presented in
\Cref{sec:data-struc}; here we focus on the first step.

\newcommand{\lo}{\mathtt{lo}\xspace}

We use a {\em token-based} scheme for this amortization. When an element
arrives, it brings with it $O(\log n)$ tokens. Whenever an element
changes level, it expends $1$ token.  In addition, tokens are transferred
between elements during the algorithm. We always ensure that
each element has a non-negative number of tokens while it remains
active; this implies the amortized bound of
is $O(\log n)$ for element level changes.
To this end, we maintain a strong invariant on
the number of tokens elements will have with them: define the
\emph{base level} of an element $e$ (denoted $b(e)$) as the largest
integer $i$ such that $2^i \leq c_{S_e}$, where $S_e$ is the minimum-cost set
in $\calF$ that contains $e$. The base level of $e$ decides the range of
density levels it can reside in, as we observe now.
\begin{lemma}
  \label{lma:min-level}
  At all times, an element $e$ resides at a level $\lo(e) := b(e) - \log
  n - 2$ or higher.
\end{lemma}
\begin{proof}
  Any set $S$ below level $b(e) - \log n - 2$ has cost $< c_{S_e}$,
  so $S$ cannot contain $e$.
\end{proof}

\begin{leftbar}
 {\bf Token Invariant:} Any element $e \in \At$
  covered at level $i$ 
  has at least $2(i -
  \lo(e)) \geq 0$ tokens.
\end{leftbar}

{\bf Token Re-distribution:} Our basic idea of
token re-distribution is simple. There are two reasons for an element
$e$ to change its density level.  The first situation is that the
element $e$ being covered by set $S'$ in our solution is now covered by a
new set $S$ that enters the solution at a lower level. In this case, $e$
can spare tokens, since the token invariant at the lower level requires
fewer tokens from $e$. So $e$ spends one token to pay for the move, and
contributes one token to $S'$'s \emph{emergency fund} (to ``atone'' for
deserting its siblings in $S'$).
This emergency fund will come in handy when $S'$ and its remaining elements move to a higher
density level if it violates the density range for its current level.

Indeed, the second situation for an element to change its
density level is when $\covt(S)$ for a set $S$ decreases to the extent
that $S$ no longer satisfies the upper bound on the density range in its current
level. In that case $S$ ``floats up'' to a higher level. Now the
elements in $\covt(S)$ moving up have to spend one token for the move to
the higher level, but also have to satisfy a more demanding token
invariant. But they can satisfy this by sharing the tokens in the
emergency fund, as we show next.
\begin{lemma}
  \label{lem:token-invt}
  The {\em Token Invariant} is always satisfied.
\end{lemma}
\begin{proof}
  We prove this by induction over the sequence of moves. Case I: if an
  element moves to a lower level, it requires at least $2$ fewer tokens,
  and hence the invariant is satisfied even after $e$ spends $1$ token
  for the move and gives the other to the emergency fund of the set it
  is leaving.

  In Case II: suppose set $S$ floats up from level $i$ to level
  $i+\ell$. We need to show that each element covered by $S$ has an
  excess of at least $2 \ell + 1$ tokens, which will be sufficient for
  it to satisfy the new token invariant at level $i + \ell$, and also to
  expend one token for the move. Let $t_\ini$ be the time when set $S$
  was added to level $i$, and let $\cov_{\ini}(S)$ be the initial set of
  elements $S$ covers at that stage. Also, let $t_\fin$ be the current
  time, when set $S$ moves to level $i + \ell$. Let the set of elements
  covered by $S$ at this time be $\cov_{\fin} (S)$. Since a set is always moved to the \emph{highest level} that can
  accommodate it, we know that level $i+1$ could not accommodate $S$ at time $t_\ini$, and
  so $c_S/|\cov_{\ini}| < 2^{i+1}$. On the other hand, since level
  $i+\ell$ can accommodate $S$ at time $t_\fin$, we have
  $c_S/|\cov_{\fin}| \geq 2^{i+\ell}$.  It follows that
  \begin{equation*}
    |\cov_{\ini}|~/~|\cov_{\fin}| \geq 2^{\ell-1}.
  \end{equation*}
  Moreover, the emergency fund contains $|\cov_{\ini}| - |\cov_{\fin}|$
  tokens, since each element that left $S$ gave one token into this fund.
  Sharing this among the remaining $|\cov_{\fin}|$ elements
  gives each element in $\cov_{\fin}$ at least $(|\cov_{\ini}| -
  |\cov_{\fin}|)/|\cov_{\fin}| \geq 2^{\ell - 1} - 1$ new tokens. So the
  token invariant holds for set $S$ at level $i + \ell$ if
  \begin{equation}
    \label{eq:req}
    2^{\ell - 1} - 1 \geq 2 \ell + 1.
  \end{equation}	
  Next, note that set $S$ moves out of level $i$ at time $t_\fin$
  because its density exceeds $2^{i+10}$. Since it settles at the
  highest level that can accommodate it, it could not move to level
  $i+\ell+1$, and so its density is less than $2^{i+\ell+1}$. Taking
  logs, $i+10 < i+\ell+1$, i.e., $\ell > 9$, and so~\eqref{eq:req} easily holds.
\end{proof}

In Appendix~\ref{sec:lognlogn-full} we show a more nuanced recourse bound of
$O(\sum_t \log n_t)$, and also give details of the data structures for
the updated update time bound of $O(\sum_t f \log n)$.

\subsection{An $O(\log n_t)$-Competitive $O(1)$-Recourse
  Algorithm}
\label{sec:logn1}

The above algorithm has $O(\log n)$ amortized recourse, since the total
number of \emph{set changes} (which is the quantity of interest) in
$\St$ is bounded by the number of times that \emph{elements} change
levels.  We now improve this recourse bound to $O(1)$. Intuitively, one
way of reducing recourse is to slacken the stability condition. But we
must slacken it carefully, since it affects the competitive ratio.  Our
idea is to carefully use the flexibility we have in defining the element
volume. As earlier, the base level $b(e)$ for element $e$ is the highest
level $i$ such that $2^{i} \leq c_{S_e}$, where $S_e$ is the min-cost
set containing $e$. We now use the following definition of element
volume:

\begin{leftbar}
   The volume of an element $e$ \underline{at density level $i$} is given by
    $\vt(e, i) = 2^{i-b(e)}$.
\end{leftbar}
Recall that the density $\dt(S)$ of a set $S$ is the ratio of its cost
and the total volume of elements in $\covt(S)$. Now note that the
density depends on the level of the set. To show our algorithm is valid,
we need that each set $S$ can be accommodated by at least one density
level. (Proof in Lemma~\ref{lma:valid-real}.)
\begin{lemma}
  \label{lma:valid}
  Every set covering any set of elements can be placed at some
  density level.
\end{lemma}

\smallskip\noindent {\bf Competitive Ratio.}  Again, we need to show
there are $O(\log \nt)$ ``interesting levels''. We first show that an
element cannot lie above its base level in a stable solution.
\begin{restatable}{claim}{BaseLevel}
  \label{cl:baselevel}
  In a stable solution, the density level of any element $e$ is $b(e)$
  or lower. So the volume of $e$ is at most $1$.
\end{restatable}

\begin{proof}
  Suppose element $e$ is at level $i \geq b(e) +1$.  Since
  $\vt(e,b(e)) = 1$, we could possibly add the set $S_e$ at level
  $b(e)$ and set $\varphi_t(e) = S_e$. This contradicts stability of
  $\St$.
\end{proof}

The total cost of all sets $S$ with density $\dt(S) \leq \Opt_t/\nt$
is at most $\Opt_t$ since there are $\nt$ active elements, each with
volume at most $1$ by Claim~\ref{cl:baselevel}.
Moreover, for any element $e$, the cost of the min-cost set containing
$e$ is at most $\Opt_t$, so $b(e) \leq \log (\Opt_t)$. Hence, by
Claim~\ref{cl:baselevel}, all sets are in levels $\log (\Opt_t)$ or
lower. So, we only need to consider sets in levels $\log (\Opt_t)$ down
to $\log (\Opt_t/\nt) - 10$. Using Lemma~\ref{lma:cost}, we get the
competitive ratio to be $O(\log \nt)$.

\smallskip\noindent {\bf Recourse.}  Our token scheme is now more
involved: a new element now brings just $1$ token. Whenever a new set
is added to the solution $\St$, the system expends $\Omega(1)$ tokens.
Note that because we are just measuring the recourse, i.e., the number
of sets that change in our solution, we do not need to spend tokens when
a set floats up. In addition, we show how to transfer tokens between
elements to maintain the following invariant, which ensures that each
element has non-zero tokens.

\begin{leftbar} {\bf Token Invariant:} An element $e \in \At$ covered at
  level $i$ 
  has at least $\vt(e, i) = 2^{i-b(e)}$ tokens.
\end{leftbar}
A new element $e$ is initially covered by the minimum cost set
containing it at level $b(e)$. Since the element comes in with $1$
token, the invariant is initially satisfied.


\smallskip \noindent {\bf Token Re-distribution:} Our token
re-distribution scheme is similar to that in the update time
setting. When a new set $S$ is added to the solution $\St$ at some level
$i$, and some element $e \in \covt(S)$ is reassigned from its current
level $i'$ to be covered by $S$, the token requirement of $e$
decreases, and hence it has excess tokens. Element $e$ expends half of
these excess tokens towards the addition of
$S$ to the solution (we show in Claim~\ref{cl:rec-set-form} that a total
of $\Omega(1)$ token can be expended per new set created); next $e$ contributes
the remaining half to the emergency fund for the set $S'$ at level $i'$
it used to belong to. When such a set $S'$ floats up, it now takes the tokens in
the emergency fund and distributes it among the remaining elements in
$S$ {\em in proportion to their respective volumes}, since their token
requirement is larger at the new level. In Lemma~\ref{lem:token-invt-2}
we show this re-distribution suffices to maintain the token
invariant.

\begin{restatable}{claim}{RecSetForm}
  \label{cl:rec-set-form}
  Whenever a new set $S$ is added to $\St$, the elements in
  $\covt(S)$ expend $\Omega(1)$ tokens.
\end{restatable}
\begin{proof}
  Suppose $S$ is added at level $i$. Note that the volume, and therefore
  the token requirement, of every element covered by $S$ at level $i$ is
  at most half of the token requirement at its previous level $i' \geq
  i+1$.  So it suffices to show that $\sum_{e \in \covt(S)} \vt(e, i)
  \geq \Omega(1)$.  Fix an element $e\in \covt(S)$. If $e$ is above its
  base level, i.e., $i \geq b(e)$, then $\vt(e, i)\geq 1$. So, assume $i
  < b(e)$. Since the density of every set at level $i$ is at most
  $2^{i+10}$, $\dt(S, i) \leq 2^{i+10}$. On the other hand, by
  definition of base level, $c_S \geq 2^{b(e)}$. Therefore, the total
  volume of elements in $\covt(S)$ is equal to $c_S / \dt(S, i) \geq
  2^{b(e)-i-10} > 2^{-10}$ since $i < b(e)$.
\end{proof}

Now, we show that the token invariant holds at all times.
\begin{lemma} \label{lem:token-invt-2}
  The {\em Token Invariant} is always satisfied.
\end{lemma}
\begin{proof}
  The proof is by induction on the sequence of moves.  When elements
  move down, their token requirement decreases, and the token invariant
  trivially holds. So, we focus on the case where a set $S$ moves up
  from level $i$ to level $i+\ell$.  Let $t_\ini$ be the time when set
  $S$ was added to level $i$, and let $\cov_{\ini}(S)$ denote the
  initial set of elements it covers at that stage. Also, let $t_\fin$ be
  the current time, when set $S$ moves to level $i + \ell$. Let the set
  of elements covered by $S$ at this time be $\cov_{\fin}
  (S)$. Correspondingly, for some density level $j$, let $\vt(\cov_{\ini},j)$ and
  $\vt(\cov_{\fin}, j)$ denote the sum of level-$j$ volumes of elements in
  $\cov_{\ini}(S)$ and $\cov_{\fin}(S)$ respectively.

  Recall that a set is always moved to the highest level that can
  accommodate it. This implies that at time $t_\ini$, level $i+1$ could
  not accommodate $S$, i.e., $c_S/\vt(\cov_{\ini}, i+1) < 2^{i+1}$. On
  the other hand, since level $i+\ell$ can accommodate $S$ at time
  $t_\fin$, we have $c_S/\vt(\cov_{\fin}, i+\ell) \geq 2^{i+\ell}$.  It
  follows that
  \begin{equation*}
    \frac{\vt(\cov_{\ini}, i+1)}{\vt(\cov_{\fin}, i+\ell)} \geq 2^{\ell-1}.
  \end{equation*}
  We now normalize this comparison of the two volumes at the same level:
  \begin{equation}
    \frac{\vt(\cov_{\ini}, i)}{\vt(\cov_{\fin}, i)}
    = \frac{\vt(\cov_{\ini}, i+1) / 2}{\vt(\cov_{\fin}, i+\ell) / 2^{\ell}}
    = \frac{\vt(\cov_{\ini}, i+1)}{\vt(\cov_{\fin}, i+\ell)} \cdot 2^{\ell-1}
    \geq 2^{2(\ell-1)}. \label{eq:5}
  \end{equation}

  Remember that the emergency fund is distributed among the elements in
  $\cov_{\fin}$ in proportion to their volumes. Since each departing element in
  $(\cov_{\ini}\setminus \cov_{\fin})$ contributed half its excess tokens, i.e., at least $1/4$ of its
  level-$i$ volume, the number of tokens element
  $e$ in $\cov_{\fin}$ receives is at least
  \begin{gather*}
    (1/4)\cdot (\vt(\cov_{\ini},i) - \vt(\cov_{\fin},i)) \cdot
    \frac{\vt(e, i) }{\vt(\cov_{\fin},i)} \geq \vt(e, i)\cdot
    \left(2^{2(\ell - 2)} - (1/4)\right).
  \end{gather*}
  The inequality follows from~(\ref{eq:5}).
  Adding this to the $\vt(e, i)$ tokens that element $e$ had at time
  $t_\ini$ (using the token invariant inductively), element $e$ has at
  least $2^{2(\ell-2)} \cdot \vt(e, i)$ tokens at $t_\fin$. Therefore,
  the token invariant holds at level $i + \ell$ if this is at least
  $\vt(e, i + \ell) = 2^{\ell} \cdot \vt(e, i)$. In other words, we want
  \begin{equation}
    \label{eq:req2}
    2^{2(\ell - 2)} \geq 2^{\ell}.
  \end{equation}	
  The rest of the proof is simple given our definition of volume. Set $S$ is moving out of
  level $i$ at time $t_\fin$ is because its density exceeds
  $2^{i+10}$. It settles at the highest level that can accommodate it,
  so it could not move to level $2^{i+\ell+1}$ and its density at level
  $i+\ell+1$ was less than $2^{i+\ell+1}$. When a set moves up $\ell+1$
  levels, its density decreases by a factor of $2^{\ell+1}$ because of a
  corresponding increase in the volume of elements it covers. It follows
  that $i+10 < i+\ell+1 + (\ell+1)$, i.e., $\ell \geq 4$. To complete
  the proof, note that \eqref{eq:req2} holds for $\ell \geq 4$.
\end{proof}

A detailed description of the algorithm, along with more detailed
proofs, appear in~\Cref{sec:logn1-full}.


\newcommand{\yt}{\ensuremath{y}\xspace}
\newcommand{\levelt}{\ensuremath{\mathsf{level}}\xspace}
\newcommand{\out}{{\sf out}\xspace}
\newcommand{\ins}{{\tt in}\xspace}
\newcommand{\new}{\ensuremath{\mathsf{new}\xspace}}
\newcommand{\old}{\ensuremath{\mathsf{old}\xspace}}

\newcommand{\pdapx}{\beta}
\newcommand{\ytil}{\tilde{y}}

\newcommand{\Stabi}{\mathsf{Stabilize}}

\section{Dynamic Primal Dual Algorithms}
\label{sec:pd-mainbody}

In \S\ref{sec:greedy} we saw dynamic algorithms inspired by the greedy
analysis of set cover. However, the natural greedy algorithms do not
yield competitive ratio better than $O(\log n)$ even for special cases
like the vertex cover problem. So we turn our attention to dynamic
algorithms based on the \emph{primal-dual} framework, which typically have approximation ratios depending on the parameter $f$, the maximum number of sets containing any element. Our first result is a deterministic fully-dynamic algorithm in the update time model
with $O(f^3)$-competitiveness and an update time of $O(f^2)$
(establishing~\Cref{thm:main-update}(b)). Note that for vertex cover
this gives constant-competitiveness with \emph{deterministic} constant
update time. (The randomized version of this result for the special case
of \emph{unweighted} vertex cover follows from the recent dynamic
maximal-matching algorithm of~\cite{Solomon16}.)  Our algorithm follows
a similar framework as~\cite{Bhatta-icalp}, but we
use a more nuanced and apparently novel {\em asymmetric} token transfer scheme in the
analysis to remove the dependence on $n$ in the update time.
We then outline our
algorithm with improved bounds in the recourse model: we get
$f$-competitiveness with $O(1)$ recourse
(establishing~\Cref{thm:main-recourse}(b))\footnote{In fact, we can get stronger guarantees for both the algorithms to have competitive ratio depend only on $\ft$, the \emph{maximum frequency at time $t$} and not $f = \max_t \ft$. We show how in~\Cref{sec:unknown-f}.}; complete details of these
algorithms appear in~\Cref{sec:pd-dyn,sec:f1} respectively.


\subsection{An $O(f^3)$-Competitive $O(f^2)$-Update-Time
  Algorithm} \label{sec:short-pd-update}

Given a set system $(U,\calF)$, and an element $e$, let $\calF_e$ denote
the sets containing $e$, and let $f_e := |\calF_e|$. For now we assume
we know $f$ such that $f_e \leq f$ for all $e \in A_t$; we discharge
this assumption in \Cref{sec:unknown-f}. Like the algorithms
in~\Cref{sec:greedy}, our algorithm assigns each set to a \emph{level},
but the intuition now is different. Firstly, only the sets in the
solution $\St$ were assigned levels in~\Cref{sec:greedy}, whereas
\emph{all} sets will be assigned levels in the primal-dual
framework. Secondly, the intuition of a level in the greedy framework
corresponds to the density (or incremental cost-per-element covered) of
the sets, whereas the intuition of a level here loosely corresponds to
how quickly a set becomes tight if we run the standard primal-dual
algorithm. Sets that become tight sooner are
in higher levels. We also define \emph{base levels}, but now these are
defined not for elements but for sets.

\begin{itemize}
\item \emph{Set and Element ``Levels''.}
   At all times, each set $S \in \calF$ resides at an integer level,
   denoted by $\levelt(S)$. For an element $e \in \At$, define
   $\levelt(e) := \max_{S \, : e \in S} \levelt(S)$ to be the largest
   level of any set covering it.
 \item \emph{Set and Element ``Dual Values''.}  Given levels for sets
   and elements, the \emph{dual value} of an element $e$ is defined to
   be $\yt(e) := 2^{-\levelt(e)}$. The dual value of
   a set $S \in \calF$ is the sum of dual values of its elements,
   $\yt(S) := \sum_{e \in S \cap \At} \yt(e)$.
\end{itemize}

Recall the dual of the set cover LP:
\begin{gather}
  \textstyle \max\{ \sum_{e \in A_t} \ytil_e \mid \sum_{e \in A_t \cap
    S} \ytil_e \leq c_S ~\forall S \in \calF, ~~~\ytil_e \geq 0 \} .\label{eq:6a}
\end{gather}
Now our solution at time $t$ is simply all sets whose dual constraints
are \emph{approximately tight}, i.e., $\St = \{S \, : \, \yt(S) \geq
\nicefrac{c_S}{\pdapx}\}$ for $\pdapx := 32
f$. In addition, we will try to ensure that the duals  $\yt(e)$ we maintain will be approximately feasible for~\eqref{eq:6a}, i.e., for every set $\yt(S) \leq \beta c_S$. Then, bounding the cost becomes a simple application of LP duality.

The main challenge is in maintaining such a dual solution dynamically.
We achieve this by defining a \emph{base level}
for every set to indicate non-tight dual constraints, and ensuring that all sets
strictly \emph{above their base levels} always have approximately tight dual
constraints. Formally, the \emph{base level} for set $S$ is defined to
be $b(S) := - \ceil{ \log ( \pdapx {c_S}) } - 1$, and our solution
$\St$ consists of \emph{all sets} that
are located strictly above their respective base levels, i.e., $\St
= \{S \in \calF \, : \, \levelt(S) > b(S)\}$. (To initialize, each
set $S$ is placed at level $b(S)$.)

We now give the intuition behind this definition of base levels. Suppose a new element $e$
arrives and it is uncovered, i.e., all its covering sets are at their base levels. Then,
element $e$ has a dual value of $\yt(e) = \nicefrac{1}{2^{b(S_e)}} > 2 \pdapx c_{S_e}$,
where $S_e$ is the minimum-cost set containing $e$. This implies that
$\yt(S_e) > 2 \pdapx c_{S_e}$, and so our algorithm moves $S_e$ up to a higher level
and includes it in the solution.

We ensure the approximate tightness and approximate feasibility of dual constraints using
the \emph{stability property} below.

\begin{itemize}
\item \emph{Stable Solutions.} A solution $\St$ is \emph{stable} if: for
  all sets $S$ with $\levelt(S) \geq b(S)$ we have $y(S) \in
  [\nicefrac{c_S}{\pdapx}, \pdapx c_S]$, and for all $S$ with $\levelt(S) <
  b(S)$ we have $y(S) < \pdapx c_S$.
\end{itemize}

Loosely, the algorithm follows the principle of least effort toward ensuring stability of all sets: if at some point $\yt(S)$ is too large, $S$ moves up the least number of levels so that the resulting $\yt(S)$ falls within the admissible range -- observe that as $S$ moves up one level, every element $e$ in $S$ for which $\levelt(e) = \levelt(S)$ halves its current dual value. Similarly if $\yt(S)$ is too small, it moves down until $\yt(S) \geq \nicefrac{c_S}{\pdapx}$. Indeed, since the competitive ratio is defined purely by the two barriers $\nicefrac{1}{\pdapx}$ and $\pdapx$, this lazy approach is very natural if the goal is to minimize the number of updates.

\begin{leftbar}
  \noindent {\bf Arrival:} When $e$ arrives, the current levels for sets
  define $\yt(e) := 1/\max_{S: e \in S} 2^{\levelt(S)}$. Update
  $\yt(S)$ for all sets. Run $\Stabi$.

  \medskip \noindent {\bf Departure:} Delete $e$ from $\At$. Update
  $\yt(S)$ for all sets. Run $\Stabi$.

  \medskip \noindent {\bf Stabilize:} Repeatedly perform the following
  steps until the solution is stable: If for some set $S$ at level
  $\levelt(S)$ we have $\yt(S) > \pdapx c_S$, find the \emph{lowest
    level} $\ell' > \levelt(S)$ such that placing $S$ at level $\ell'$
  results in $\yt(S) \leq \pdapx c_S$.
  Analogously, if $\yt(S) < \nicefrac{c_S}{\pdapx}$, find the \emph{highest}
  level $\ell' < \levelt(S)$ such that placing $S$ at level $\ell'$
  results in $\yt(S) \geq \nicefrac{c_S}{\pdapx}$. If such an $\ell' < b(S)
  -1$, we place $S$ at level $b(S) -1$ and drop $S$ from $\St$.
\end{leftbar}

\begin{lemma}[Stability $\Rightarrow$ Approximation]
  \label{lma:pd-cost}
  Any stable solution $\St$ has cost $O(f^3)\, \Opt_t$.
\end{lemma}
\begin{proof}
  The cost of $\St$ is
 $$\sum_{S \in \St} c_S \leq \pdapx \sum_{S \in \St} \yt(S)
  = \pdapx \sum_{S \in \St} \sum_{e \in S \cap \At} \yt(e) \leq f \pdapx
  \sum_{e \in \At} \yt(e).$$
	Now since $\yt(e)/\pdapx$ is a feasible dual solution (see the set
    cover dual in (\ref{eq:6a})), the lemma follows from LP duality.
\end{proof}

\subsubsection{Bounding the Update Time}

As in~\Cref{sec:greedy}, most updates happen when the solution
stabilizes itself at various points. Indeed, even termination of $\Stabi$ is a priori not clear.
In any call to $\Stabi$ when a set $S$ moves, let $\out^{\old}(S)$
and $\out^{\new}(S)$ respectively denote the \emph{out-elements}
of $S$ at the beginning and end of the move. These are the elements
whose level is equal to the level of $S$, i.e., elements whose dual
value changes when $S$ moves up/down. Following
Solomon~\cite{Solomon16}, we first show that the total
update time for maintaining our data structures in an upward move
is $O(f\cdot |\out^{\new}(S)|)$, and that in a downward move is
$O(f\cdot |\out^{\old}(S)|)$. The details of the data structures
establishing these bounds are given in Appendix~\ref{sec:pd-dyn}.

Given these observations, it suffices to control the sizes of
$\out^\new(S)$ and $\out^\old(S)$, and charge them to
distinct arrivals. Again, we use a token scheme for the amortized analysis of these costs.

\smallskip \noindent {\bf Token Distribution:} When an element arrives,
it gives $20f$ tokens to each of the $f$ sets containing it, totaling
$O(f^2)$ tokens. When an element departs, it gives $1$ token to each
of the $f$ sets covering it. When a set moves up from some level $\ell$
to level $\ell^*$ in $\Stabi$, it expends $f\cdot |\out^\new(S)|$ tokens
to pay for the update-time, and for all $e \in \out^\new(S)$ it
transfers $1$ token to each set $S' \in \calF_e$. Similarly when a set
moves down in $\Stabi$, it expends $f\cdot |\out^\old(S)|$ tokens, and for
all $e \in \out^\old(S)$ it transfers $20f$ tokens to each $S' \in
\calF_e$. \emph{Note the asymmetry between the number of tokens
  transferred in the two cases!}

Note that at most $20f^2$ tokens are injected per element arrival/departure.
Moreover, from the discussion above, the total
update time for each up\-/down\-ward move is $O(1)$ times the number
of tokens expended from the system. In what follows, we show that we never expend more tokens
than we inject -- thus obtaining an amortized $O(f^2)$ bound on update times.
To this end,
we divide the movement of a set $S$ into
\emph{epochs}, where each \emph{up-epoch} is a maximal set of contiguous
upward moves of $S$ and a \emph{down-epoch} is a maximal set of
contiguous downward moves of~$S$. (Epochs may span
different calls to $\Stabi$.)



\begin{lemma}[Up-Epochs]
  \label{lem:epochup}
  Consider an up-epoch of a set $S$ ending at level $\ell$. The total
  number of tokens expended or transferred by $S$ during this epoch is
  at most $2^{\ell+8} f^2 c_S$. Moreover, the total number of tokens
  that $S$ gained during this epoch is at least $2^{\ell+8} f^2
  c_S$.
\end{lemma}

\begin{proof}
  Consider an up-epoch where $S$ moves from $\ell_0 \rightarrow \ldots
  \rightarrow \ell_k = \ell$ via a sequence of up-moves.
   The tokens expended and transferred during the upward move
   $\ell_{i-1} \rightarrow \ell_i$ is
   $(f+f)\cdot |\out^\new(S)| \leq (2f) \pdapx\, 2^{\ell^*} c_S$.
   The inequality holds because $S$ satisfies the stability condition
   when it settles at level $\ell_i$, and each out-element has
   dual value $1/2^{\ell_i}$. Therefore, the \emph{total tokens}
   transferred/expended in this \emph{entire up-epoch} is at most $2f
  \sum_{i=1}^{k} \pdapx 2^{\ell_i} c_S \leq 4\pdapx f 2^\ell c_S$. Using
  $\pdapx = 32f$ proves the first claim.

  For the second claim, consider the moment when this epoch started. We
  first observe that when $S$ had moved down to level $\ell_0$ at the
  end of the previous epoch say at time $t_0$, then $y_{t_0}(S) <
  \nicefrac{2c_S}{\pdapx}$. Indeed, if this epoch is the first ever
  epoch for set $S$ then $y_S$ was $0$. Else it was preceded by a
  down-epoch, in which case the final down-move in the previous epoch
  happened because $y_S < \nicefrac{c_S}{\pdapx}$ and the down-move can
  at most double the $y_{t_0}(S)$ value at level $\ell_0$. Let $S(t_0)$ be the elements of $S$
  which were active at that time. Now suppose at time $t_0$, we
  hypothetically placed $S$ in level $\ell-1$ without changing the
  levels of other sets. Let $y'_{t_0}(e)$ be the corresponding dual
  values for elements given by the sets being in these levels, and let
  $y'_{t_0}(S) := \sum_{e \in S(t_0)} y'_{t_0}(e)$. Clearly,
  $y'_{t_0}(S) < \nicefrac{2c_S}{\pdapx}$ as well, since we are
  hypothetically placing $S$ at a higher level $\ell-1$ instead of $\ell_0 \leq
  \ell -1$. Now consider the ending time $t_1$ of the up-epoch, just
  before we move $S$ from $\ell-1$ to $\ell$. Let $S(t_1)$ be the
  elements of $S$ active at this time, and let $y_{t_1}(e)$ be the dual
  values for all elements in $S(t_1)$. Then, $y_{t_1}(S) := \sum_{e \in
    S(t_1)} y_{t_1}(e)$. Clearly $y_{t_1}(S) > \pdapx c_S$, else we will
  not move $S$ from $\ell-1$ to $\ell$.

  Note that we placed $S$ at level $\ell-1$ in both settings, so the
  increase from $y'_{t_0}(S)$ to $y_{t_1}(S)$ of more than $(\pdapx-2)
  c_S$ can only happen because~(a) there are elements in $S(t_1)$ that
  are not present in $S(t_0)$, or (b)~there are elements in $S(t_1) \cap
  S(t_0)$ that have moved down since time $t_0$. Each such element can
  contribute at most $2^{-(\ell-1)}$ to $y_{t_1}(S) -
  y'_{t_0}(S)$, so there must be at least $2^{\ell-1} \cdot (\pdapx-1)
  c_S \geq 15 f c_S 2^{\ell}$ events in total, and our token distribution scheme now says that
  $S$ would have collected at least $300f^2 c_S 2^{\ell} \geq 2^{\ell+8} f^2 c_S$
  tokens in this epoch, completing the proof.
\end{proof}

\noindent {\em Remark:} Note that in the above lemma, the update time is bounded
in terms of tokens expended, and this seemingly depends on set cost $c_S$. At first
glance, this might appear strange because update times should be invariant to cost scaling.
A closer scrutiny, however, reveals that the update time is indeed scale-free since
higher set costs imply lower levels in the algorithm, and vice-versa. In other words,
on scaling, the change in $c_S$ is compensated by an opposite change in $2^{\ell}$.

A similar lemma holds for down-epochs (see Appendix~\ref{sec:pd-dyn}
for the proof).

\begin{restatable}[Down-Epochs]{lemma}{EpochDown}
  \label{lem:epochdown}
  Consider a down-epoch for set $S$ starting at level $\ell$. The number
  of tokens expended/transferred by $S$ during this epoch is at most
  $2^{\ell+1} f c_S$. Moreover, the total number of tokens $S$ gained in
  the beginning of this epoch is at
  least $2^{\ell+1} f c_S$.
\end{restatable}
The above two lemmas show that for each set $S$ and each epoch, the
total number of tokens $S$ expends or transfers is no more than what it
receives via transfers or arrivals/departures. This shows that the
total tokens never becomes negative,
and hence proves the amortized bound.

\subsection{An $O(f)$-Competitive $O(1)$-Recourse Algorithm}
\label{sec:f1-short}

In this section, we consider the recourse model and give an algorithm
with stronger guarantees than the one in the previous section. Our
algorithm is inspired by the following offline algorithm for set cover
-- arrange the elements in some arbitrary order and probe them in this
order. We maintain a tentative solution $\calS$ which is initially
empty. When we probe an element $e$, two cases arise: (i) the element
$e$ is already covered by an element $e \in \calS$:~in this case, we do
nothing, or (ii) there is no such set in $\calS$:~in this case, we pick a
random set from $\calF_e$, where a set $S \in \calF_e$ is chosen with
probability $\frac{1/c_S}{\sum_{S' \in \calF_e} 1/c_{S'}}$. This
algorithm is $O(f)$-competitive in expectation~\cite{Pitt}.

We now describe our \emph{dynamic implementation} of this algorithm.
Recall that $A_t$ denotes the set of
active elements at time $t$.  At all times $t$, we maintain a partition
of $A_t$ into two sets $P_t$ (called the \emph{probed set}) and $Q_t$
(the \emph{unprobed set}). Elements on which we have performed the
random experiment outlined above are the ones in the probed set.
We also maintain a bijection $\varphi$ from $P_t$ to $\calS_t$,
the set cover solution at time $t$, i.e., for every probed element,
there is a unique set in $\calS_t$ and vice-versa.
Elements can move from $Q_t$ to $P_t$ at a latter point of time (i.e.,
an element
currently in $Q_t$ can be in $P_{t'}$ for some $t' > t$), but once
an element is in set $P_t$ it stays in $P_{t'}$ for all $t' \geq
t$ (till the element departs).

We now describe the procedures which will be used our algorithm. At certain times, our
algorithm may choose to \emph{probe} an unprobed element. This will happen when there is
no set in the current solution covering this element.

\smallskip\noindent {\bf Probing an element.} When an unprobed element
$e \in Q_{t-1}$ is probed by the algorithm at time $t$, it selects a
single set that it belongs to, where set any $S$ containing $e$
is chosen with probability
$\frac{1/c_S}{\sum_{S'\in \calF_e} 1/c_{S'}}$.  This chosen set $S$ is
added to the current solution of the algorithm: $\calS_t := \calS_{t-1}
\cup \{S\}$. Element $e$ moves from the unprobed set to the probed set:
$P_t = P_{t-1} \cup \{e\}$ and $Q_t = Q_{t-1} \setminus \{e\}$. As long
as $e$ remains in the instance, i.e., is not deleted, the set $S$ will
also remain in the solution. We say that $e$ is responsible for
$S$ and denote $\varphi(e) := S$.

\noindent
Having defined the process of probing elements, we explain how elements
are probed. These probes are triggered by element insertions and
deletions as described below.

\smallskip
\noindent
{\bf Element Arrivals.}
Suppose element $e$ arrives at time $t$.  If $e$ is already covered in
the current solution $\calS_{t-1}$, it is added to the unprobed set
(i.e., $Q_t = Q_{t-1}\cup \{e\}$), and the solution remains unchanged
($\calS_t = \calS_{t-1}$). Else, if $e$ is not covered in the current
solution, then it is probed (which adds a set $\varphi(e)$ to $\calS_t$
as described above), and we set $P_t = P_{t-1} \cup \{e\}$.

\smallskip
\noindent
{\bf Element Departures.}
Suppose element $e$ departs from the instance at time $t$. If $e$ is
currently unprobed, then we set $Q_t = Q_t\setminus \{e\}$, but the
solution remains unchanged: $\calS_t = \calS_{t-1}$. Else if $e$ is a
probed element, then we set $P_t = P_{t-1}\setminus \{e\}$. In addition,
the set $\varphi(e)$ is also removed from the previous solution
$\calS_{t-1}$. This might lead to some elements in $Q_t$ becoming
uncovered in the current solution. We pick the first uncovered
element\footnote{We assume an arbitrary but fixed ordering on the
  elements.} and probe it, which adds a set covering it to $\calS$.
  This set might cover some previously uncovered elements. This
process continues, with the first uncovered element in the
unprobed set being probed in each iteration. The probing ends
once all elements in the unprobed set (and therefore, all elements
overall) are covered by the chosen sets. We then define this solution
as $\calS_t$.

It is easy to see the recourse bound: no set is deleted when elements arrive,
and at most $1$ set is deleted when an element departs. (Recall that we piggyback
set additions on deletions.) The proof of the
competitive ratio proceeds via a randomized dual fitting argument, and is described
in~\Cref{sec:f1}.

\section{Combining Two Dynamic Algorithms}

As our results show in~\Cref{thm:main-recourse}, we have obtained two
different algorithms; one has a competitive ratio of $O(\log \nt)$, and
another works for instances with maximum frequency $f$ and gives
$O(f)$-competitiveness, both with constant recourse. In fact, we can get stronger guarantees for the second algorithm: a competitive ratio of $\ft$, the \emph{maximum frequency at time $t$}. Note that $\ft$ over time can change drastically if very high-frequency elements arrive/depart (we give details of this in~\Cref{sec:unknown-f}). Now, in this section we
show how we can dynamically maintain the \emph{best-of-both} solutions
with constant recourse. Note that this is not as simple as
maintaining the solution with lower cost---the identity of the
lower-cost solution could oscillate between the two (depending on $\ft$ and $\nt$, both of which can change over time), and there is no
clear way to bound the recourse (or update-time) this way. 

We take a more problem-specific approach to overcome this difficulty:
indeed, we maintain different instances of the set cover problem, one
corresponding to each $f$ value (upto a power of two), and send each
element with frequency (i.e., number of sets covering it) roughly $2^i$
to the $i^{th}$ instance $\I^i_{{\rm pd}}$; for elements with frequency
more than $(\log \nt)$, we send them to a different instance $\I_{\rm
  g}$. Then, we run the primal-dual algorithm
(\Cref{thm:main-recourse}(ii)) for instances $\I^i_{{\rm pd}}$ which has
competitive ratio $O(2^i)$ and recourse $O(1)$, and run greedy
(\Cref{thm:main-recourse}(i)) for $\I_{\rm g}$ which has competitive
ratio $O(\log \nt)$ with recourse $O(1)$. Finally, if $\nt$ itself
doubles/halves thereby changing the competitive ratio guarantee of the
greedy algorithm, we reassign all elements in $\I_{\rm g}$ to the
appropriate instance according to its frequency. We thus get the
following theorem for the case of recourse.

\begin{theorem} \label{thm:best-of-two-main} There is an efficient
  $O(\min (f_t, \log \nt))$-competitive algorithm with recourse
  $O(1)$.
\end{theorem}

We can similarly build a $O(\min (f_t^3, \log \nt))$-competitive
combiner for the update-time model, whose update-time is $O(f(f + \log
n))$. The full details appear in~\Cref{sec:combiner}.

\subsection*{Acknowledgments.} We thank Yuting Ge and Archit Kulkarni
for enlightening discussions. Theorem~\ref{thm:scaling} was obtained in
conversations with Yuting.


\newpage

\section*{Appendix}

\appendix

\section{Dynamic Greedy Algorithm (Update-Time): Full Details}
\label{sec:lognlogn-full}

We now furnish full details of the proof of~\Cref{thm:main-update}(i).
While we gave most details in Section~\ref{sec:lognlogn}, here the
proofs are more formal and give nuanced results, namely bounds in terms
of $n_t$ (the number of active elements at time $t$) instead of $n$ (the
total number of elements seen until time $t$). Moreover, we deferred the
details of the implementation and data structures, which we present here.

{\bf Notation.} Recall that our algorithm maintains a solution (denoted $\St$ at time $t$) with sets that cover the active elements
$A_t$.  In addition, it also maintains an {\em assignment} $\varphi_t(e)$ of each active element $e$ to a unique set in $\St$ covering it; define $\covt(S) := \{e \mid \varphi_t(e) = S\}$ to be the set of elements assigned to $S$. Clearly, the sets $\{ \covt(S)
\mid S \in \calS_t \}$ are mutually disjoint and their union is
$\At$. For any set $S \in \St$  in the current solution which covers the elements $\covt(S)$,  we define its  {\em current density} to be
$\dt(S) := \nicefrac{c_S}{|\covt(S)|}$.
At each time $t$, we also maintain a partition of the sets in $\St$ into \emph{levels} $\{\Lt{i}\}_{i \in \Z}$, with
each set in $\St$ belonging to exactly one of these levels --- these levels correspond to the \emph{current densities} of the sets, rounded to the nearest power-of-two. Consequently, each element $e \in \At$ is also present in a unique level, corresponding to level of the set $\varphi_t(e)$ which currently covers it.
For a level $i$, we let $\Et{i}$ denote the set of active elements which are assigned to sets at level $i$; i.e., $\Et{i} := \{e \in \At
\mid \varphi(e) \in \Lt{i} \}$. Define $\Lt{> i}$ and $\Et{> i}$
similarly to denote the collection of sets (and elements) in levels $i+1$ and greater.
Finally, we use $\nt= |\At|$ to denote the current number of active elements. 
Our competitive ratio and recourse bounds at time $t$ will be in terms of $\nt$. 

\begin{algorithm}
\caption{Dynamic$(e_\bt, \pm)$}
\label{alg:dynamic}
\begin{algorithmic}[1]
\If{the operation $\sigma_\bt$ is $(e_\bt, +)$}
\State  let $S$ be the cheapest set containing $e_t$. (so, $\covt(S)$ is just $\{e_\bt\}$)
\State let $i$ be the highest level such that $2^i \leq c_S = \dt(S)$. Move $S$ to $\Lt{i}$
\ElsIf{the operation $\sigma_\bt$ is $(e_\bt, -)$}
\State  remove $e_t$ from the set $S$ which covers it; let $S \in \Lt{i}$
\If {$S$ becomes empty}
\State remove $S$
\ElsIf{the current density of $S$ exceeds $2^{i+10}$}
\State move $S$ to the the highest level $\ell$ such that $2^\ell \leq \dt(S)$
\EndIf
\EndIf
\State call {\sf Stabilize($t$)}
\end{algorithmic}
\end{algorithm}

\begin{algorithm}
  \caption{{\sf Stabilize($t$)}}
  \label{alg:fixdyn}
  \begin{algorithmic}[1]
    \While{there exists set $S \in \calF$ and level $i$ such that $\nicefrac{c_S}{|S \cap \Et{i}|} < 2^{i}$}
    \State add a copy of set $S$ to $\St$ and set $\covt(S)$ to $S \cap \Et{i}$
    \State assign $S$ to the highest level $i^\star$ for which $2^{i^\star} \leq \dt(S)$
    \While{there exists set $X \in \Lt{i}$ such that $\covt(X) \cap \covt(S) \neq \emptyset$}
    \State set $\covt(X) \leftarrow \covt(X) \setminus
    \covt(S)$, and update $\dt(X)$ accordingly
    \If{$\covt(X) = \emptyset$}
    \State remove $X$ from the solution
   \ElsIf{$\dt(X) > 2^{j+10}$}
   \State assign $X$ to the highest level $\ell$ such that
    $2^\ell \leq \dt(X)$
    \EndIf
    \EndWhile
    \EndWhile
\end{algorithmic}
\end{algorithm}

\subsection{Analysis Preliminaries}

We now analyze the algorithm. In~\Cref{sec:cost-analysis}, we show that the cost of the solutions $\St$ are always at most a logarithmic factor off the optimal solution at time step $t$, and in~\Cref{sec:recourse-analysis}, we bound the total amortized recourse. But first, we show that {\sf Stabilize} terminates in finite steps.
\begin{claim}
\label{cl:terminate}
  Algorithm~{\sf Stabilize} terminates.
\end{claim}
\begin{proof}
Suppose some change in the algorithm (either an arrival or departure) triggered a call of the algorithm {\sf Stabilize}. Now, we show that the while loop terminates after finitely many steps. To this end, consider the vector $\mathbf{v}_\bt=(\nt^\ell,
  \nt^{\ell+1}, \nt^{\ell+2}, \ldots)$, where $\ell$ is the lowest level with non-zero elements, and $\nt^i$ is the number of elements at level $i$ in the current solution, i.e., $\nt^i = | \Et{i} |$. We claim that this vector
  always increases lexicographically when we perform an iteration of the
  outermost while loop in Algorithm~{\sf Stabilize}. Indeed, when pick a set $S$ (and a
  corresponding index $i$), the new level for
  $S$ is $i^\star \leq i$ by definition, and hence all elements in $\covt(S)$ move
  down from a level  $\geq i$ to a level $< i$. Some elements at
  level above $i$ (covered by sets corresponding to $X$ in the
  description of the algorithm) may move to levels higher than $i$, but
  none of the levels at or below $i$ lose any elements.
  Hence the vector $\mathbf{v}_\bt$ increases lexicographically. Since
  none of the coordinates of this vector can be larger than $\nt$, this
  process must terminate finitely.
\end{proof}

The curious reader might wonder whether we need the above proof given that we would in any case need to bound the total update time of the algorithm. However, our update time analysis uses the finite termination of {\sf Stabilize} and so we provided the above proof. We next show some crucial invariants the algorithm satisfies:

\begin{leftbar}
\begin{itemize}
\item[(i)] A set $S \in \Lt{i}$ has current density $2^{i} \leq \dt(S) \leq  2^{i+10}$.
\item[(ii)] For each level $i \in \Z$, there exists no set $S \in \calF$ such that $\nicefrac{c_S}{|S \cap \Et{i}|} < 2^i$, i.e., if we include a new copy of $S$ into the solution and cover the elements in $S \cap \Et{i}$, then all the elements in $S \cap \Et{i}$ will strictly improve their density level (from $\geq i$ to $<i$)
\end{itemize}
\end{leftbar}

It is now easy to check that the algorithm maintains both the invariants.
\begin{claim}
\label{cl:inv}
  The solution $\St$ satisfies both the invariants at all times.
\end{claim}
\begin{proof}
Suppose the invariants are satisfied by
  $\St$. Then during the next operation, each set that is added (in both procedures {\sf Dynamic} and {\sf Stabilize}) is
  placed at a level that satisfies invariant~(i). Moreover, the
  algorithm~{\sf Stabilize} iterates till invariant~(ii) is satisfied (and
  always moves sets to the right levels to satisfy invariant~(i)). So provided this procedure terminates, we know that it satisfies both invariants at the end of time $t+1$ as well. The proof for termination will in fact follow from the proofs which bound the amortized update time.
\end{proof}

\subsection{Cost Analysis} \label{sec:cost-analysis}

Next we bound the cost of our solution. Recall that $\Opt_\bt$ denotes
the cost of the optimal solution at time $t$. Let $\dt$ denote
$\Opt_\bt/\nt$, and $i_\bt$ be the index such that $2^{i_\bt -1 } < \dt \leq 2^{i_\bt}$.
\begin{lemma}
  \label{lem:cost}
  The solution $\St$ has cost at most $O(\log \nt) \, \Opt_\bt$.\footnote{In fact we show (\Cref{lem:cost-better}) using a more nuanced argument that the cost is at most $O(\log \Delta_t) \, \Opt_\bt$ where $\Delta_t$ is the maximum set size $\max_{S \in \calF} |S \cap \At|$ at time $t$.}
\end{lemma}
\begin{proof}
  By Invariant~(i), the density of any set at levels $i_\bt$ or lower is at most $
  2^{i_\bt +10} < 2^{11} \cdot \dt$. So the total cost of sets in $\Lt{\leq
    i_\bt}$ is at most $\nt \cdot 2^{11} \cdot \dt \leq 2^{11} \cdot \Opt_\bt$.

In order to bound the cost of sets at higher levels, we first claim that for all non-negative integers $i \geq 0$, the number of
  elements covered by sets at levels $i_\bt + i$ is at most
  $\nt/2^{i}$; i.e., $|\Et{i_\bt+i}| \leq \nt/2^{i}$.  Suppose
  not.  Consider the optimal solution for $\At$ restricted to the
  elements in $\Et{i_\bt + i}$. By averaging, there exists a set
  $S$ in this solution for which $\nicefrac{c_S}{|S \cap \Et{ i_\bt +
      i}|} \leq \nicefrac{\Opt_\bt}{|\Et{i_\bt + i}|} < 2^{i+i_\bt}$.
  But then this set $S$ and level $i+i_\bt$ would violate
  invariant~(ii), a contradiction. This proves the claim;

  Now, for each such level $i_\bt + i$, where $1 \leq i \leq \log_2 \nt + 2$, the
  density of sets in the solution at this level is at most $2^{i_\bt + i + 10}$ (due to invariant~(i)). therefore, the total cost of sets in this level (volume times density) is at most $2^{i_\bt + i + 10} \cdot \nicefrac{\nt}{2^{i}}
  \leq 2^{11} \cdot \Opt_\bt$.

  Finally, the above claim also
  implies that all elements are covered by level $i_\bt + \log \nt + 2$. So summing the above cost-per-level bound over levels $i_\bt, \ldots, i_\bt +\log \nt + 2$ completes the proof.
\end{proof}

\subsection{Update Time I: Bounding Element-Level Changes} \label{sec:recourse-analysis}

We bound the update time in two parts: in the first part, we assign \emph{tokens} to elements, so that
whenever a new set $S$ is created in \textsf{Stabilize}, we will \emph{expend} $\Omega(1)$ tokens from the system and redistribute the remaining tokens to satisfy some token invariants. In addition, we also expend $\Omega(1)$ tokens for \emph{each element which changes its level}. Then, in~\Cref{sec:data-struc} we show how to maintain data structures so that the amortized time to update them can be charged to $O(f \log \Nt)$ times the number of tokens expended. So if the total number of tokens injected into the system is at most $O(\log \Nt)$ per element insertion or deletion, we would get our amortized update time to be $O(f  \log^2 \Nt)$ per element insertion or deletion. We now proceed with the details of the first part.

In order to clearly describe our token invariant, we again recall the base level of $e$ defined as $b(e)$, which is the largest integer $i$ such that $2^i \leq c_S$ where $S$ is the cheapest set covering $e$.

Before presenting the token invariant, we show a simple which lower bounds the levels above which an element can be covered.
\begin{claim}
  \label{cl:lowerbd}
  Each element $e$ is covered by a
  set at level at least $\lo_t(e) = b(e) - \ceil{\log \nt} - 10$ or above.
\end{claim}

\begin{proof}
Suppose $e$ is covered by a set $S$ in level $i$ at the beginning of update operation at time $t$. At this time, density of $S$ is at most $2^{i+10}$. Since $c_S \geq 2^{b_e}$, and density of $S$ is
at least $\frac{c_S}{n_t}$, it follows that $i \geq b_e  - \ceil{\log n_t} - 10$. During the operation at time $t$, $e$ could change levels. If we add a new set $S$ containing $e$ during Algorithm~{\sf Stabilize}, then $\rho_t(S) \geq c_S/n_t$, and
$c_S \geq 2^{b_e}$. So if we add $S$ at level $i$, we
know that $\rho_t(S) \geq 2^i$, and so, $i \leq b_e -
\ceil{\log \Nt}$. Any other operation will only move $e$ to a higher level. This proves the claim.
\end{proof}

\begin{leftbar}
 {\bf Token Invariant:} Any element $e \in \At$
  covered at level $i$ 
  has at least $2(i -
  \lo_t(e)) \geq 0$ tokens.
\end{leftbar}


\medskip \noindent {\bf Token Distribution Scheme:}
We inject tokens when elements arrive, and redistribute them when elements move sets or depart. Crucially, our redistribution would ensure that every time a new set is added to $\St$, a constant $\lambda$ units of tokens are expended from the system. More formally:

\begin{enumerate}
\item[(a)] {\bf Element arrival $\sigma_t = (e_\bt, +)$:} We give $e_\bt$ a total of $2 \lceil \log \nt \rceil + 7$ tokens, and $e_\bt$ immediately \emph{expends} one unit of token for the formation of the singleton set covering it.
\item[(b)] {\bf Element departure $\sigma_t = (e_\bt, -)$:} Suppose $e$ was covered by some set $S$ at level $i$. Then $e_\bt$ has at least $2$ tokens by Claim~\ref{cl:lowerbd} and so it expends one token from the system, and equally distributes at least one token to the remaining elements covered by $S$.
\item[(c)] {\bf {\sf Stabilize} operation:} Suppose we add a set $S$ to the solution $\St$ at some level $i^\star$ during an iteration of the {\bf While} loop in~{\sf Stabilize}. Consider any element $e$ now covered by $S$ but earlier covered by some set $U_j$ at level $i$. Suppose $e$ has $\tau_e \geq 2(i- \lo_t(e))$ tokens (since it currently satisfies the token invariant). To satisfy $e$'s new token requirement, $e$  retains $\tau'_e = 2(i^\star- \lo_t(e))$ tokens. Then, $e$ \emph{expends $\nicefrac12 (\tau_e - \tau'_e)$ tokens} from the system toward the creation of this set, and equally distributes the remaining $\nicefrac12 (\tau_e - \tau'_e)$ tokens among the remaining elements still covered by $U_j$, i.e., the set of elements $\cov^\bt(U_j) \setminus
\cov^\bt(S)$. Now if a set violates invariant~(i) and moves up, each remaining element \emph{expends $1$ token from the system} (which we show can be done while satisfying the new token requirement for these elements).
\item[(d)] {\bf Phase transition:} When $\nt$ becomes a power of two, say $2^{k}$, and suppose the previous (different) power-of-two value of $\nt$ was $2^{k-1}$, then we give each element in $\At$ an additional $2$ tokens. This is because each element's token invariants has changed now (because $\lo_t(e)$ has increased by $1$) and needs two extra tokens.\footnote{If the previous (different) power-of-two value of $\nt$ was $2^{k+1}$, then we have seen departures and the token requirement is weaker, so we don't need to give extra tokens in this case.}
\end{enumerate}

We begin with the following easy claims.

\begin{claim} \label{cl:elt-arr}
The total number of tokens introduced into the system is $O(\sum_t (\log \nt))$.
\end{claim}

\begin{proof}
New tokens are introduced only when elements arrive, and when $\nt$ becomes a power of two. Every element arrival introduces $O(\log \nt)$ tokens with it. Moreover, when $\nt$ becomes a power-of-two, say, $2^k$ and the previous power-of-two value of $\nt$ was $2^{k-1}$, then we add $2 \cdot 2^k$ tokens into the system in step (d) in the token distribution scheme. But we can \emph{charge} these extra tokens to the $2^{k-1}$ arrivals which must have happened in the period when $\nt$ increased from $2^{k-1}$ to $2^k$. So each arrival is associated with introducing $O(\log \nt)$ tokens, which completes the proof.
\end{proof}

\begin{claim} \label{cl:set-form}
Whenever an element moves its level up or down, one unit of token is expended from the system.
\end{claim}

\begin{proof}
This is easy to see, as whenever we create a new set $S$, we explicitly make each element in $S$ expend one token in step (c) above. Likewise, when a set moves up levels for violating invariant~(i), we make each element expend one token from the system.
\end{proof}

\begin{lemma} \label{lem:token-invt-app}
The {\em Token Invariant} is always satisfied.
\end{lemma}
\begin{proof}
We prove this by induction: indeed, suppose the invariant holds for all times up to $t-1$ and consider a time-step $t$. Firstly, we show that when $\nt$ becomes a power-of-two, the token invariant is satisfied. Indeed, suppose we have $\nt = 2^{k}$, and the previous (different) power-of-two value of $\nt$ was $2^{k-1}$. Then the token invariant for each element increases by $2$, which we make sure in step (d) in the token distribution scheme. On the other hand, suppose the previous power-of-two was $\nt = 2^{k+1}$, then the token invariant only becomes weaker.

Next, we show that the invariant holds if the operation $\sigma^\bt$ is $(e, +)$. Indeed, we add a new set at level $b(e)$ to cover $e$ and also give it $2 \ceil{\log \nt} + 7$ tokens in step (a) of the token scheme. It expends one token for its data structure update, and the rest clearly satisfy its token invariant at level $b(e)$. Similarly, if the operation is $(e^\bt,-)$, we simply delete $e^\bt$. Since it had at least $2$ tokens by Claim~\ref{cl:lowerbd}, we are able to make it give $1$ token to the remaining elements covered by the set which was covering $e^\bt$ and also expend $1$ token from the system.

Now we show that the operation~{\sf Stabilize} maintains the invariant. To this end, suppose we find a set $S$ during
an iteration of the {\bf While} loop of procedure~{\sf Stabilize}. Firstly note that the elements which are covered by the new set have requisite number of tokens since they each drop their level by at least one and we ensure in the token scheme property (c) that they retain sufficiently many tokens to satisfy their token invariant at the new level.

Now we turn our attention to the more challenging scenario when sets move up in level having lost many elements and their current density exceeds the threshold for their current level. So consider the setting when a set $S$ moves up from level $i$ to level $i+\ell$: we now show that each remaining element covered by $S$ can be given $2 \ell + 1$ tokens, which will be sufficient for it to satisfy the new token invariant at level $i + \ell$, and also to expend one token which we do in step (c) of the token distribution scheme. To this end, consider the first time $t_\ini$ when a set $S$ was added to level-$i$, and let $\cov_{t_\ini}(S)$ denote the initial set of elements it covers. Also consider the first time $t_\fin$ when it violates the invariant~(i) for level $i$, and so it moves to some higher level, say $i + \ell$. Let the set of remaining elements at this time be $\cov_{t_\fin} (S)$. We now make some observations regarding these cardinalities.

Firstly, because this set moves up from level $i$, its current density $\nicefrac{c_S}{|\cov_{t_\fin} (S)|}$ must be greater than $2^{i+10}$. That is,
     \begin{equation} \label{eq:dcurr1}
     {|\cov_{t_\fin} (S)|} < \nicefrac{c_S}{2^{i+10}} \, .
     \end{equation}

     Likewise, since it does not move up higher than $i+\ell$, its density $\nicefrac{c_S}{|\cov_{t_\fin} (S)|}$ must be strictly less than $2^{i+\ell+1}$ (recall that when a set is relocated, it
	is placed in the highest level that can accommodate it). Therefore we have,

\begin{equation}\label{eq:dcurr2}
     {|\cov_{t_\fin} (S)|} > \nicefrac{c_S}{2^{i+\ell+1}} \, .
     \end{equation}

Thus we get that $2^{i+\ell+1}  > 2^{i+10}$, i.e., $\ell > 9$.

Next, we note that when the set first entered
	level $i$ at time $t_\ini$, its density $\nicefrac{c_S}{|\cov_{t_\ini}(S)|}$ must have been less than $2^{i+1}$
	(otherwise we would have placed it at level $i+1$ or above); therefore, we have

     \begin{equation} \label{eq:dcurr3}
     {|\cov_{t_\ini} (S)|} > \nicefrac{c_S}{2^{i+1}} \, .
     \end{equation}

Similarly, when the set (and the remaining elements it covers) moves from level $i$ to level $i+\ell$ at time $t_\fin$,
its density is at least $2^{i+\ell}$, i.e.,
     \begin{equation} \label{eq:dcurr5}
     {|\cov_{t_\fin} (S)|} \leq \nicefrac{c_S}{2^{i+\ell}} \, .
     \end{equation}

So from~\cref{eq:dcurr3,eq:dcurr5}, we have that

	\begin{equation*}
		\nicefrac{|\cov_{t_\ini} (S)|}{|\cov_{t_\fin} (S)|} = 2^{\ell-1}.
	\end{equation*}

	To complete the proof, note that each element in $\cov_{t_\ini}(S) \setminus \cov_{t_\fin}(S)$ left $1$ token with the remaining elements $\cov_{t_\fin}(S)$ (since it either departed from the instance or moved down at least one level due to some new sets being formed). So the tokens which the remaining elements gained is at least
\begin{eqnarray}
& & |\cov_{t_\ini}(S)| - |\cov_{t_\fin}(S)| \\
&\geq& 2^{\ell-1} |\cov_{t_\fin}(S)| - |\cov_{t_\fin}(S)|  \\
&=& (2^{\ell-1}-1) \cdot |\cov_{t_\fin}(S)|  \\
&\geq& (2 \ell+1)\cdot |\cov_{t_\fin}(S)|
\end{eqnarray}
The last inequality uses $\ell \geq 9$. Hence, when the set (along with the remaining elements) moves up to level $i+\ell$,
	each of these elements get $2\ell$ extra tokens to meet their new token requirement.
\end{proof}

Thus, we have shown that the token invariant holds at all times. Therefore, by Claims~\ref{cl:elt-arr},~\ref{cl:set-form} and~\Cref{lem:cost,lem:token-invt} we get the following theorem.

\begin{theorem} \label{thm:wtd-tokens}
The above algorithm satisfies the following properties:
\begin{enumerate}
\item At all times $t$, the solution $\St$ is feasible for $\At$ and has cost $O(\log \nt)$ times the optimal cost.
\item Every time a new set is formed, each element expends one token from the system.
\item Every time a set changes level, each covered element expends one token from the system.
\item The total number of tokens expended till time $T$ is $O(\sum_{t=1}^{T} \log \nt)$.
\end{enumerate}
\end{theorem}

\newcommand{\densitylevel}{{\tt densityLevel}}
\newcommand{\covernode}{{\tt coverNode}}
\newcommand{\levelset}{{\tt levelSet}}
\newcommand{\unstablelist}{{\tt unstableList}}
\newcommand{\levelof}{{\tt level}}
\newcommand{\templist}{{\tt tempList}}

\subsection{Update Time II: Details of Data Structures}
\label{sec:data-struc}

We now give details of the data structures needed to complete the
description of our fully dynamic algorithm. The crucial step to
implement fast is in the ${\sf Stabilize}$ procedure where we find the
best density set using elements currently covered at level
  $i$ for some $i$. The main observation is that we will look for such
sets only when we \emph{move an element to a new level, say $i$}.  Therefore, it should
suffice to look at the $f$ sets containing an element whenever we change
its level, and see if their current densities $c_S/|A_t(i)| < 2^i$. Since each element expends one token when it changes its level, the total update time would then be within a factor $f$ of the
total number of tokens spent by the elements, if we can maintain these data structures consistently. Indeed, implementing this idea requires us to be careful with the data-structure, although we
use only simple data-structures such as doubly-linked lists
and arrays.

\paragraph{Data Structures.}
We maintain the following data structures.

\begin{itemize}
\item For each $S \in \St$, $\cov(S)$ is a doubly-linked list\footnote{All lists will be doubly-linked to allow
  constant-time inserts and deletes of elements (as long as we have a
  pointer to the element we want to delete). Moreover, each list element
  also maintains a pointer to the head of the list, where we maintain
  the current length of the list.}  whose cells contain the different active elements $S$ is responsible for covering in the current solution; we also store the level of $S$ using $\levelof(S)$.
\item For each set $S \in \calF$ and each level $i$\footnote{While there are many density levels overall, each set $S$ will only concern with a range of $O(\log \nt)$ density levels between $c_S/n_t$ and $c_S$. So even though we index this array using a global density level of $i$, we only store  $\log \nt$ entries here and simply use an appropriate offset on the index. For clarity of presentation, we omit this detail.}, $\densitylevel_S[i]$ is an linked list of the set of elements in $S \cap A_t(i)$, i.e., the set of elements which are currently in level $i$ which can be covered by $S$. This is the crucial data structure which helps us quickly see if there is an unstable set. The first cell in $\densitylevel_S[i]$ also has the size of this list, which we denote by $n(S,i)$.

\item For each level $i$, there is one list $\unstablelist(i)$ which has the candidate list of sets which may be unstable. Over time, as our algorithm makes changes by moving elements up and down, a set which entered this list may need to be removed since it may no longer be unstable. Moreover, if a set $S$ is unstable at level $i$, then its cell in $\unstablelist(i)$ will contain a pointer to $\densitylevel_S[i]$ and vice versa.
\item We also have a global $\unstablelist$ which is a linked list of pointers to the heads of all non-empty $\unstablelist(i)$ for different $i$'s. If an individual $\unstablelist(i)$ becomes empty, then it deletes its corresponding node from this list.
\item A global list $\templist$ which contains the list of sets which need to float up after losing some elements.
\item Finally, every element $e \in \At$ stores the following: (i) $\covernode(e)$ is a pointer to $e$'s cell in the list $\cov(S)$ where $S$ is the set currently covering $e$; (ii) a linked list $\levelset(e)$ which contains a pointer to each of the cells corresponding to $e$ in the $f$ different lists  $\densitylevel_S[i]$ where $S$ covers $e$, and $i$ is the current level of $e$.
\end{itemize}

\paragraph{Update Procedures.}

Now we give details of the procedures which will form building blocks of
the update operations.

\begin{itemize}
\item Algorithm~\ref{alg:remove-element} (\textsf{RemoveElement}) takes
  as input an element $e$ which is currently at level $i$, and {\em
    removes} $e$ from this level, either because we want to move it to a
  different level or delete it. The procedure, by traversing $\levelset(e)$, removes $e$ from the
  corresponding lists $\densitylevel_S[i]$ for every set $S$ containing
  $e$. Since we are only removing an element from level $i$, we will not
  create any new unstable sets at level $i$, but it is possible that $S$
  was unstable at level $i$ and now becomes stable after removal of $e$
  from it --- this is why we have a pointer from $\densitylevel_S[i]$ to the node corresponding to $S$ in $\unstablelist(i)$, if $S$ is present in the latter list. In this case, we remove $S$ from the list of unstable sets at level $i$. Observe that $\levelset(e)$ becomes empty after this step since we are yet to place $e$ in its new level. 
  $\{\densitylevel_S\}_{S \in \calF_e}$ as long as $e \in A_t$).
\item 
  Algorithm~\ref{alg:insertset} ({\sf InsertSet$(S,X,i)$}) takes a set
  $S$ along with a subset $X \sse S$ and index $i$, adds $S$ to our
  solution $\calS_t$, places it at level $i$ and sets $\cov(S)$ to
  $X$. For every element $e \in X$, and every set $S' \in \calF_e$, we
  add $e$ to $\densitylevel_{S'}[i]$. This may make some of these $S'$ unstable, in
  which case we add them to the list of unstable sets at level $i$.
\item Next, Algorithm~\ref{alg:stabilizeimplement} describes the
  implementation of {\sf Stabilize}, really just filling in the data-structure details in
  Algorithm~\ref{alg:fixdyn}. We first find an unstable set $S$ at some
  level $i$.  Let $i^\star$ be the level where it should be
  placed. After inserting $S$ at level $i^\star$, we need to remove
  elements getting covered by it -- these elements have come down from
  level $i$ to level $i^\star$.  Now, it is possible that the sets which
  were covering these elements (at level $i$) will float up to higher
  levels. So, whenever we move such an element down, we also add the set
  previously covering it to a temporary list. Finally, we check all the sets in
  this list, and move them up if needed. Moving such a set up involves a
  call to {\sf InsertSet} algorithm.

\item Having described {\sf Stabilize}, the actual insert or delete
  operations are easy; Algorithm~\ref{alg:dynamicimplement}
  (\textsf{Dynamic}) gives the implementation details for
  Algorithm~\ref{alg:dynamic}. To add an element $e$, we check for the
  cheapest set containing it, and call~${\sf InsertSet()}$ with suitable
  parameters. To delete an element $e$ currently covered by a set $S$,
  we update $\cov(S)$ and check if it needs to move up (or if $\cov(S)$
  becomes empty, we will just delete it). These steps are formalized in
  Algorithm~{\sf Update()}.
\end{itemize}

\paragraph{Running Time Analysis.}
Note that ${\sf RemoveElement}$ takes $O(f)$ time; ${\sf RemoveSet}$ and
${\sf InsertSet}$ take $O(f \cdot |X|)$ time. It follows that the outer
\emph{while} loop in {\sf Stabilize} takes time $O(f \cdot |X|) + O(f
\cdot \sum_{S'} |\cov(S')|)$, where the summation is over those sets $S'
\in \templist$ which move up. But notice that $|X|$ elements from $S$
have moved down at least one level, and elements in $S'$ move up. So the
total update time can be bounded in terms of $f$ times the total number
of times elements change levels. The same argument holds for the~{\sf
  Dynamic} algorithm. Hence we get the main theorem:

\begin{theorem}
  There is a fully-dynamic $O(\log \nt)$-factor approximation algorithm
  for set cover with total update complexity of $O(\sum_{t=1}^{T} (f
  \log ( \nt))$ over $T$ element operations.
\end{theorem}

\begin{algorithm}
\caption{{\sf RemoveElement$(e,i)$}}
\label{alg:remove-element}
\begin{algorithmic}[1]
\For{every set $S$ containing $e$ (use the list $\levelset(e)$)}
\State remove $e$ from $\densitylevel_S[i]$ and decrement $n(S,i)$.
\State \textbf{if} ($S \in \unstablelist(i)$ and $c_S/n(S,i)
\geq 2^i$) \textbf{then} remove $S$ from this list.
\EndFor
\end{algorithmic}
\end{algorithm}


\begin{algorithm}
\caption{{\sf InsertSet$(S,X,i)$}}
\label{alg:insertset}
\begin{algorithmic}[1]
\State add $S$ to $\calS_t$, and set $\cov(S) \leftarrow X, \levelof(S) \leftarrow i.$
 \For{every $e \in X$}
 \State initialize $\levelset(e) \gets \emptyset$.
 \For{each set $S'$ containing $e$}
 \State add $e$ to $\densitylevel_{S'}[i]$, and add this node to $\levelset(e)$.
 \State increment $n(S',i)$
 \State \textbf{if} $c_{S'}/n(S',i) < 2^i$ \textbf{then} add $S'$ to $\unstablelist(i)$ if not already in this list.
 \EndFor
\EndFor
\end{algorithmic}
\end{algorithm}

\begin{algorithm}
\caption{Implementation for {\sf Stabilize()}}
\label{alg:stabilizeimplement}
\begin{algorithmic}[1]
\While{there is a level $i$ such that $\unstablelist(i)$ is non-empty}
\State $S \leftarrow$ \texttt{dequeue}$(\unstablelist(i))$
\State $X \leftarrow \densitylevel_S[i]$ \Comment{these are the elements
  $S$ will cover}
\State $i^\star \gets$ highest level such that $2^{i^\star} <
c_{S}/n(S,i)$
\State Initialize $\templist \gets \emptyset$
\For{$e \in X$}
\State {\sf RemoveElement$(e,i)$}
\State suppose $\covernode(e)$ in list $\cov(S')$\footnote{Recall that every list contains a pointer to the beginning of the list. So using $\covernode(e)$, we can find which set is covering it.}
\Comment{i.e., $S'$ was covering $e$ at level $i$.}
\State add $S'$ to $\templist$ if not already in this list.
\State remove $\covernode(e)$ from list $\cov(S')$.
\State add new node $\covernode(e)$ to list $\cov(S)$.
\EndFor
\State {\sf InsertSet$(S, \cov(S), i^\star)$}
\For{every set $S' \in \templist$} \Comment{move $S'$ up or delete if empty}
\If{$|\cov(S')| = 0$}
\State remove $S'$ from the solution $\calS_t$.
\Else
\If{$c_{S'}/|\cov(S')| > 2^{i+10}$ } \Comment{move $S'$ to higher level}
\For{$e \in \cov(S')$}
\State \textsf{RemoveElement$(e,i)$}
\EndFor
\State {\sf InsertSet$(S', \cov(S'), i')$}
where $i'$ is the highest level such that $2^{i'} < c_{S'}/n(S',i).$
\EndIf
\EndIf
\EndFor
\EndWhile
\end{algorithmic}
\end{algorithm}

\begin{algorithm}
\caption{Implementation for Dynamic$(e_\bt, \pm)$}
\label{alg:dynamicimplement}
\begin{algorithmic}[1]
\If{the operation $\sigma_\bt$ is $(e_\bt, +)$}
\State  $S \gets$ cheapest set containing $e_t$
\State $\cov(S) \leftarrow \{e_t\}$. \Comment{add new copy of $S$ with one element}
\State update $\covernode(e_t)$ accordingly
\State $i \gets $ highest level such that $2^i \leq c_S = \dt(S)$
\State $\levelof(S) \leftarrow i$
\State {\sf InsertSet$(S, \cov(S), i)$}
\ElsIf{the operation $\sigma_\bt$ is $(e_\bt, -)$}
\State $S \gets$ set covering $e_t$, say at level $i$
\State {\sf RemoveElement$(e_t,i)$}
\State remove $e_t$ from $S$ and update $\cov(S)$
\If{$|\cov(S)| == 0$} \Comment{set $S$ is empty}
\State remove $S$ from solution $\calS_t$
\Else
\If{$c_{S}/|\cov(S)| > 2^{i+10} $ } \Comment{move $S$ up}
\For{$e \in \cov(S)$}
\State \textsf{RemoveElement$(e,i)$}
\EndFor
\State {\sf InsertSet$(S, \cov(S), i')$},
where $i'$ is the highest level such that $2^{i'} < c_{S}/n(S,i).$
\EndIf
\EndIf
\EndIf
\State {\sf Stabilize}()
\end{algorithmic}
\end{algorithm}

\subsection{Better Cost Analysis}
We now state and prove~\Cref{lem:cost-better}, which gets an improved competitive ratio of $O(\log \Delta_t)$: recall that $\Delta_t$ is the maximum set size at time $t$, i.e., $\max_{S \in \calF} |S \cap \At|$.
\begin{lemma}
  \label{lem:cost-better}
  The solution $\St$ has cost at most $O(\log \Delta_t) \, \Opt_\bt$.
\end{lemma}
\begin{proof}
  Our proof now proceeds via a dual fitting argument. Recall the dual of the set cover LP:
\begin{gather}
  \textstyle \max\{ \sum_{e \in A_t} \ytil_e \mid \sum_{e \in A_t \cap
    S} \ytil_e \leq c_S ~\forall S \in \calF, ~~~\ytil_e \geq 0 \} .\label{eq:6-new}
\end{gather}

We now exhibit a dual $\ytil_e$ such that (a) the cost of $\St$ can be bounded by $O(1) \sum_{e \in \At} \ytil_e$, and (b) $\ytil_e/(\log \Delta_t+2)$ is feasible for the dual LP \eqref{eq:6-new}. This will establish the proof of the lemma via standard LP duality.

Let us define $\ytil_e$ to be $2^{i}$ if $e$ is covered in level $i$. To show (a), consider any set $S \in \St$ which is at level $i$. Then, by invariant~(i), the density of set $S$ is at most $2^{i +10}$. Therefore, the number of elements $S$ currently covers (according to $\covt(\cdot)$) is at least $c_S/2^{i+10}$. Since each such element assigned to $S$ has dual value $\ytil_e = 2^i$, we get that $c_S \leq 2^{10} \cdot \sum_{e \in \covt(S)} \ytil_e$. Summing over all $S \in \St$, and noting that $\covt(\cdot)$ defines a partition of the set of elements $\At$, establishes (a).

To show (b), for any set $S \in \calF$, let us define $i^S_h := \max\{i: 2^i \leq c_S\}$, and $i^S_l := i^S_h - \ceil{\log \Delta_t}$. We first bound the sum of duals for elements in set $S$ that are covered at level $i^S_l$ or below: $\sum_{e \in \Et{\leq i^S_l} \cap S} \ytil_e \leq c_S$. Indeed, for any such element $e \in \Et{\leq i^S_l}$, we have $\ytil_e \leq c_S/\Delta_t$ and the sum is over at most $\Delta_t \geq |A_t \cap S|$ active elements in $S$. 

Next, we observe that for any level $i$, we have $|\Et{i} \cap S| \leq c_S/2^i$. Suppose not, and there exists $i$ such that $|\Et{i} \cap S| > c_S/2^i$. But then, this set $S$ violates invariant~(ii) at level $i$,  contradicting the stability of $\St$. This immediately gives us that $\Et{i} \cap S = \emptyset$ for $i > i^S_h$, since $2^i > c_S$. Furthermore, for all $i^S_l \leq i \leq i^S_h$, we have $\sum_{e \in \Et{i} \cap S} \ytil_e \leq c_S$ since $\ytil_e = 2^i$ for all such $e \in \Et{i} \cap S$. Therefore, summing over all $i$, we get that $\sum_{e \in S} \ytil_e \leq (\log \Delta_t + 2) c_S$. This establishes (b) and also completes the proof of the lemma.
\end{proof}

Using the above Lemma instead of~\Cref{lem:cost}, we get the following theorem.

\begin{theorem}
  There is a fully-dynamic $O(\log \Delta_t)$-factor approximation algorithm
  for set cover with total update time of $O(\sum_{t=1}^{T} (f
  \log ( \nt))$ over $T$ element arrivals and departures.
\end{theorem}


\section{Dynamic Greedy Algorithm (Recourse): Full Details}
\label{sec:logn1-full}

In this section, we prove~\Cref{thm:main-recourse}(i). Our algorithmic framework is identical to the one in~\Cref{sec:lognlogn-full}, with the crucial change being that we don't treat all elements identically. In fact, for each element $e \in \At$, we will also define its \emph{volume} $\vt(e,i)$ as a function of $e$ and the level $i$ it is located in. Similarly, for a set of elements $X \subseteq \At$ and some level $i$, we extend the definition of volume to $\vt(X,i) = \sum_{e \in X} \vt(e,i)$. Now, for any set $S \in \St$  in the current solution which covers the elements $\covt(S)$,  we define its  {\em level-$i$-current density} to be
$\dt(S,i) := \nicefrac{c_S}{\vt(\covt(S),i)}$. This now corresponds to the ratio of cost to the \emph{level-$i$ volume} of elements it covers.  Finally, to complete the description, we now define the volume of an element $e$ at some level $i$. Indeed, consider any element $e$, and let $S_e$ denote the minimum cost set which covers $e$. Then let $b(e)$ be the highest level $i$ such that $2^{i} \leq c_{S_e}$. Then, the volume $v(e,i)$ of element $e$ at a level $i$ is defined to be $2^{i-b(e)}$. Note that the
volume of an element is $1$ at the level $b(e)$ and decreases geometrically as it moves to lower indexed levels. For the remainder of the sub-section, $b(e)$ is said to be the \emph{base level} for element $e$.

Our algorithm would always try to find \emph{stable} solutions where no local improvement in the current density of elements is possible. We formalize this with the following two invariants which our algorithm satisfies at the beginning all time-steps:
\begin{itemize}
\item[(i)] A set $S \in \Lt{i}$ has current density $2^{i} \leq \dt(S,i) \leq  2^{i+10}$.
\item[(ii)] For each level $i \in \Z$, there exists no set $S \in \calF$
  such that the elements $S \cap \Et{i}$ can be brought down to level $i-1$ or below by adding a new copy of set $S$ to $\St$. Formally, there exists no $S \in \calF$ such that $\nicefrac{c_S}{\vt(S \cap \Et{i},i)} < 2^{i}$.
\end{itemize}

\begin{algorithm}
\caption{RecourseAlgo$(e_\bt, \pm)$}
\label{alg:dynamic2}
\begin{algorithmic}[1]
\If{the operation $\sigma_\bt$ is $(e_\bt, +)$}
\State  let $S$ be the cheapest set containing $e_t$
\State let $b(e)$ be the highest level $i$ such that $2^{i} \leq c_S$
\State Add a copy of $S$ to $\Lt{b(e)}$ and set $\covt(S) = \{e_\bt\}$
\ElsIf{the operation $\sigma_\bt$ is $(e_\bt, -)$}
\State  denote $\varphi_t(e)$ by $S$, and set $\covt(S) = \covt(S) \setminus \{e\}$; suppose $S$
belongs to level $\Lt{i}$
\If {$\covt(S)$ becomes empty}
\State remove $S$ from $\St$
\ElsIf{the current density $\dt(S,i)$ exceeds $2^{i+10}$}
\State move $S$ to the the highest level $\ell$ such that $2^{\ell} \leq \dt(S,\ell)$
\EndIf
\EndIf
\State call Procedure~{\sf Stabilize($t$)}
\end{algorithmic}
\end{algorithm}

\begin{algorithm}
  \caption{{\sf Stabilize($t$)}}
  \label{alg:fix-dp2}
  \begin{algorithmic}[1]
    \While{there exists set $S \in \calF$ and level $i$ such that $\nicefrac{c_S}{\vt(S \cap \Et{i},i)} < 2^{i}$}
    \State let $i^\star \leq i$ be the highest index such that $2^{i^\star} \leq \nicefrac{c_S}{\vt(S \cap \Et{i}, i^\star)} \leq 2^{i^\star+10}$
    \State add a copy of set $S$ to our solution and set $\covt(S)$ to $S \cap \Et{i}$
    \State assign $S$ to the level $i^\star$
    \While{there exists set $X \in \Lt{i}$ such that $\covt(X) \cap \covt(S) \neq \emptyset$}
    \State set $\covt(X) \leftarrow \covt(X) \setminus
    \covt(S)$, and update $\dt(X,i)$ accordingly
    \If{$\covt(X) = \emptyset$}
    \State remove $X$ from the solution
   \ElsIf{the current density of $X$ exceeds $2^{i+10}$}
\State move $X$ to the the highest level $\ell$ such that $2^{\ell} \leq \dt(X,\ell)$
    \EndIf
    \EndWhile
    \EndWhile
\end{algorithmic}
\end{algorithm}

The analysis of this algorithm involves showing the following properties:
\begin{itemize}
	\item {\bf Correctness.}
		For any unstable solution, the sequence of fix operations
		is finite. Furthermore, any subset can always be placed in some
		level.
	\item {\bf Competitive Ratio.}
		Any stable solution has total weight $O(\log \nt)$ times $\Opt_t$.
	\item {\bf Recourse.}
		The number of sets added by this algorithm to the solution,
		averaged over the element arrivals and departures, is $O(1)$.
\end{itemize}

The proof of termination of {\sf Stabilize} is identical to~\Cref{cl:terminate} and we omit it for avoiding redundancy. Next, we prove the validity of the algorithm by showing that every subset
can be placed at some level. In particular, this shows that the term $i^*$ is well defined in the algorithm {\sf Stabilize}.
\begin{lemma}
\label{lma:valid-real}
	Every subset covering any set of elements can be placed at some density
	level.
\end{lemma}
\begin{proof}
	Consider a set $S$ covering a subset $X$ of elements, and suppose it does not satisfy the density condition in invariant~(i) for any level. Since the $\vt(X,i)$ is an increasing function of $i$, there must be two adjacent levels $i$ and $i+1$ such that
	the set $S$ has too low a density for level $i+1$ and too high a
	density for level $i$. In other words, the former condition implies that $2^{i+1} > \nicefrac{c_S}{\vt(X,i+1)}$ and $2^{i+10}  < \nicefrac{c_S}{\vt(X,i)}$. But note that by the way we have defined $\vt$, we have $\vt(X, i) = 2 \vt(X, i+1)$,	which in turn implies that
	\begin{equation*}
		2^{i+10}  < \nicefrac{c_S}{\vt(X,i)} = 2 \nicefrac{c_S}{\vt(X,i+1)} < 2 \cdot 2^{i+1}  = 2^{i+2},
	\end{equation*}		
	which gives us the desired contradiction.
\end{proof}

\subsection{Bounding Cost}

Next we bound the cost of our solution. Let $\Opt_\bt$ denotes
the cost of the optimal solution at time $t$. Let $\dt$ denote
$\Opt_\bt/\nt$, and let $i_\bt$ be the index such that $2^{i_\bt -1 } < \dt \leq 2^{i_\bt}$.

We first begin with a simple but useful claim about the highest level an
element can belong in. (Same as Claim~\ref{cl:baselevel}.)
\begin{claim} \label{cl:baselevel-recourse} Consider any solution $\St$
  which satisfies the invariants~(i) and~(ii). Then, for all elements $e
  \in \At$, $e$ will be covered in a level at most $b(e)$.  So the
  volume of $e$ is at most $1$.
\end{claim}

\begin{proof}
Suppose not, and suppose there exists an element $e$ currently covered in level $\ell \geq b(e) +1$ in a solution $\St$ which satisfies invariants~(i) and~(ii). Then, let $S$ be the cheapest set containing $e$. Therefore, by definition of $b(e)$, we have that $2^{b(e)} \leq c_S < 2^{b(e)+1}$, and moreover, that $\vt(e,b(e)) = 1$. But now, we claim that $S$ along with $e$ would violate invariant~(ii) at level $\ell$. Indeed, we have that $c_S/\vt(e,\ell) \leq c_S/2 < 2^{b(e)} < 2^{\ell}$, which establishes the desired contradiction.
\end{proof}

The next lemma asserts that all density levels lower than $i_\bt$ can be more or less ignored
in calculating the cost of the solution.
\begin{lemma}
\label{lma:small}
	The total cost of all subsets in levels below $i_\bt$ is $O(\Opt_\bt)$.
\end{lemma}
\begin{proof}
	Every set $S$ at a level below $i_\bt$ has density at most
	$2^{i_\bt + 10} \leq 2 \cdot (\Opt_\bt/\nt) \cdot 2^{10}$. Since the volume of every element at any
	levels is at most $1$ (as it always appears in a level at most its base level from Claim~\ref{cl:baselevel}), and there are at most $\nt$ elements in the current active set $\At$,
	the total cost of sets in all levels including and below level $i_\bt$ is at most $\Opt_\bt \cdot 2^{11}$.
\end{proof}

We next bound the cost of sets in any single level.
\begin{lemma}
\label{lma:single-level}
	The total cost of all subsets in any level $i \geq i_\bt$ at any time step is $O(\Opt_\bt)$.
\end{lemma}
\begin{proof}
	Suppose not, and there exists some level $i > i_{\bt}$ such that the sum of costs of sets in density level $i$ is strictly greater
	than $\Opt_\bt \cdot 2^{10}$. Then, since the density of every set in level $i$
	is at most $2^{i+10}$, it follows that the total volume
	of elements in level $i$ is strictly greater than $\Opt_{\bt}/2^{i}$. Since these elements
	are covered by an optimal solution of total cost $\Opt_\bt$, it follows that
	there \emph{exists} some set $S \in \calF$ which would have current density in this level strictly less than $2^{i}$, which then contradicts invariant~(ii).
\end{proof}

Finally, we show that the levels above $i_{\bt} + \log \nt + 1$ are empty.
\begin{lemma}
\label{lma:top-level}
	There are no sets in $\St$ at levels above $i_{\bt} + \log \nt + 1$.
\end{lemma}
\begin{proof}
	This almost immediately follows from Claim~\ref{cl:baselevel}. Indeed, note that $\Opt_{\bt} \geq 2^{b(e)}$ for all $e \in \At$ since $2^{b(e)}$ defines a lower bound on the cost of the cheapest set covering $e$. Therefore, the highest level with any element is at most $\max_{e \in \At} b(e)$ which is at most $\log \Opt_{\bt} \leq \log  \dt + \log \nt \leq i_{\bt} + \log \nt$.
\end{proof}

\subsection{Bounding Recourse}
Finally, we show the recourse bound. This will be proved via a {\em token}
scheme, where the token invariant is that every element, when it first enters a level, has tokens equal
to its volume at the corresponding level; over time, it may gain more tokens as it stays at the level which it uses should it move to a higher level where our invariant will force it to have a higher token requirement. Clearly, when an element arrives, the invariant is satisfied as it enters its base level and therefore, requires one token which can be provided and charged to the element arrival.
This is the only external injection of tokens into the system. Hence, if
we are able to show a) that the token invariant can be maintained and b) whenever a
new set is added to the solution, we can \emph{expend} a constant $\lambda$ fraction of tokens from
the system, then we can claim that the total number of sets ever added (and therefore ever deleted) is
at most $1/\lambda$ times the number of element arrivals.

We now describe the token-based argument we use to bound the recourse, starting with the crucial invariant we maintain and then the token distribution scheme.

\medskip \noindent {\bf Token Invariant:} Consider any element $e \in \At$  which is covered at level $i$ in the current solution $\St$. Then, it has tokens at least as much as its level-$i$ volume $\vt(e,i)$.

\medskip \noindent {\bf Token Distribution Scheme:}
We inject tokens when elements arrive, and redistribute them when elements move sets or depart. Crucially, our redistribution would ensure that every time a new set is added to $\St$, a constant $\lambda$ units of tokens are expended from the system. More formally:

\begin{enumerate}
\item[(a)] {\bf Element arrival $\sigma_t = (e_\bt, +)$:} We give $e_\bt$ \emph{two units of tokens}, and $e_\bt$ immediately \emph{expends} one unit of token for the formation of the singleton set covering it.
\item[(b)] {\bf Element departure $\sigma_t = (e_\bt, -)$:} Suppose $e$ was covered by some set $S$ at level $i$. Then $e_\bt$ equally distributes its tokens (at least $\vt(e_\bt,i)$) to the remaining elements covered by $S$.
\item[(c)] {\bf {\sf Stabilize} operation:} Suppose we add a set $S$ to the solution $\St$ at some level $i^\star$ during an iteration of the {\bf While} loop in~{\sf Stabilize}. Consider any element $e$ now covered by $S$ but earlier covered by some set $U_j$ at level $i$. Suppose $e$ has $\tau_e \geq \vt(e,i)$ tokens (since it currently satisfies the token invariant). To satisfy $e$'s new token requirement, $e$  retains $\vt(e,i^\star)$ tokens. Then, $e$ \emph{expends $\nicefrac12 (\tau_e - \vt(e,i))$ tokens} from the system toward the creation of this set, and equally distributes the remaining $\nicefrac12 (\tau_e - \vt(e,i))$ tokens among the remaining elements still covered by $U_j$, i.e., the set of elements $\cov^\bt(U_j) \setminus
\cov^\bt(S)$.
\end{enumerate}

We now show that the above scheme will maintain the token invariant at all time-steps, and also guarantee a) that the number of tokens brought into the system is bounded by $O(t)$ after $t$ element operations, and b) that at least $\lambda = 2^{-12}$ tokens are expended any time a new set is added to $\St$. To this end, we begin with the following easy claims.

\begin{claim} \label{cl:rec-elt-arr}
The total number of tokens introduced into the system is at most $2t$.
\end{claim}

\begin{proof}
The only injection of tokens into the system is on element arrivals, when we introduce $2$ tokens into the system per arrival.
\end{proof}

\begin{claim} \label{cl:rec-set-form-full}
Whenever a new set is created, $\lambda \geq 2^{-11}$ units of tokens are expended from the system.
\end{claim}

\begin{proof}
If the new set is added when an element arrives, then we expend one token from the system by definition in the token distribution scheme.  So consider the case when a new set $S$ is formed at some time-step $t$ in some level $i$. We first show that the total level-$i$ volume of the elements $\covt(S)$ now covered by $S$ is at least $2^{-10}$. Indeed,
	consider an element $e\in \covt(S)$ with base level $b(e)$. Recall that the base level is defined according to the minimum
	cost set containing $e$, and so we have  $c_S \geq 2^{b(e)}$. But since the set is formed at level $i$, its current density satisfies
		$\dt(S, i) \leq 2^{i+10}$. It follows that the total level-$i$ volume $\vt(\covt(S),i)$ of elements covered by this set $S$ is at least
		$c_S / \dt(S, i) \geq 2^{b(e)-i-10}$. So if $i \leq b(e)$, we get the desired bound on the volume of the new set formed. On the other hand, if $i > b(e)$, then it already has $\vt(e,i) > 1$ and again we have $\vt(\covt(S),i) \geq 1$ in this case.

Next, note that when the new set is formed, each element $e$ newly covered by $S$ has dropped its level by at least $1$, since this is the criterion for forming new sets in {\sf Stabilize}. This implies that each such element had at least $\vt(e,i^\star+1) = 2 \vt(e,i^\star)$ tokens before the new set was formed. Then, it retains $\vt(e,i^\star)$ tokens to satisfy its new token requirement, and expends \emph{at least} $\nicefrac12 (2 \vt(e,i^\star) - \vt(e,i^\star) ) = \nicefrac12 \vt(e,i^\star)$ tokens for the formation of the new set. Since each newly covered element does the same, we get that the total tokens expended is at least $\nicefrac12 \sum_{e \in \covt(S)} \vt(e,i^\star)  \geq 2^{-11}$ from the above volume lower bound.
\end{proof}

\begin{claim} \label{cl:rec-token-invt}
The {\em Token Invariant} is always satisfied.
\end{claim}
\begin{proof}
We first show that the invariant is satisfied after a new element arrives, i.e., the operation $\sigma^\bt$ is $(e, +)$.
By definition, element $e$ gets covered by at level $b(e)$, and so, its token invariant requires it to have $\vt(e,b(e) = 1$ tokens which we give it when it arrives. Similarly, if the operation is $(e,-)$, we simply delete $e$, and there is one fewer token invariant to satisfy.  Now we show that the operation~{\sf Stabilize} maintains the invariant. To this end, suppose we find a set $S$ during
an iteration of the {\bf While} loop of procedure~{\sf Stabilize}. Firstly note that the elements which are covered by the new set have requisite number of tokens since they each drop their level by at least one and we ensure in the token scheme property (c) that they retain sufficiently many tokens to satisfy their token invariant at the new level.

	Now we turn our attention to the more challenging scenario when sets move up in level having lost many elements and their current density exceeds the threshold for their current level. So consider the setting when a set $S$ moves up from level $i$ to level $i+\ell$, and
    consider the first time $t_\ini$ when a set $S$ was added to level-$i$, and let $\vt(\cov_{t_\ini}(S),i)$ denote the initial level-$i$ volume of the elements it covers.
    Also consider the first time $t_\fin$ when it violates the invariant~(i) for level $i$, and so it moves to some level, say $i + \ell$. Let its level-$i$ volume of the remaining elements at this time be $\vt(\cov_{t_\fin} (S), i)$. We now make some observations regarding these volumes through the corresponding current densities.

    Firstly, because this set moves up from level $i$ its level-$i$ current density $\nicefrac{c_S}{\vt(\cov_{t_\fin} (S), i)}$ must be greater than $2^{i+10}$. That is,
     \begin{equation} \label{eq:curr1}
     {\vt(\cov_{t_\fin} (S), i)} < \nicefrac{c_S}{2^{i+10}} \, .
     \end{equation}

     Likewise, since it does not move up higher than $i+\ell$, its level-$i+\ell+1$-density $\nicefrac{c_S}{\vt(\cov_{t_\fin} (S), i+\ell+1)}$ must be strictly less than $2^{i+\ell+1}$ (recall that when a set is relocated, it
	is placed in the highest level that can accommodate it). Therefore we have,

\begin{equation}\label{eq:curr2}
     {\vt(\cov_{t_\fin} (S), i+\ell+1)} > \nicefrac{c_S}{2^{i+\ell+1}} \, .
     \end{equation}

Moreover, ${\vt(\cov_{t_\fin} (S), i+\ell+1)} = 2^{\ell+1} \cdot {\vt(\cov_{t_\fin} (S), i)}$ by definition of the volume function. This, along with~\Cref{eq:curr1,eq:curr2} implies that $2^{i+\ell+1}\cdot 2^{\ell+1} > 2^{i+10}$, i.e., $\ell > 5$.

Next, we note that when the set first entered
	level $i$ at time $t_\ini$, its level-$(i+1)$ density $\nicefrac{c_S}{\vt(\cov_{t_\ini}(S),i+1)}$ must have been less than $2^{i+1}$
	(otherwise we would have placed it at level $i+1$ or above); therefore, we have

     \begin{equation} \label{eq:curr3}
     {\vt(\cov_{t_\ini} (S), i+1)} > \nicefrac{c_S}{2^{i+1}} \, ,
     \end{equation}

and so it's level-$i$ volume satisfies
     \begin{equation} \label{eq:curr4}
     {\vt(\cov_{t_\ini} (S), i)} > \nicefrac{c_S}{2^{i+2}} \, ,
     \end{equation}

Similarly, when the set (and the remaining elements it covers) moves from level $i$ to level $i+\ell$ at time $t_\fin$,
its level-$(i+\ell)$ density is at least $2^{i+\ell}$, i.e.,
     \begin{equation} \label{eq:curr5}
     {\vt(\cov_{t_\fin} (S), i+\ell)} \leq \nicefrac{c_S}{2^{i+\ell}} \, ,
     \end{equation}

and hence its level-$i$ volume satisfies
     \begin{equation} \label{eq:curr6}
     {\vt(\cov_{t_\fin} (S), i)} \leq \nicefrac{c_S}{2^{i+2\ell}} \, .
     \end{equation}

So from~\cref{eq:curr4,eq:curr6}, we have that

	\begin{equation*}
		\nicefrac{\vt(\cov_{t_\ini} (S), i)}{\vt(\cov_{t_\fin} (S), i)} = 2^{2\ell-2} \geq 16 \cdot 2^{\ell}.
	\end{equation*}

	To complete the proof, note that all the elements in $\cov_{t_\ini}(S) \setminus \cov_{t_\fin}(S)$ together left $1/4$ of their $\vt(\cov_{t_\ini}(S) \setminus \cov_{t_\fin}(S),i)$ tokens with the remaining elements $\cov_{t_\fin}(S)$. So the tokens which the remaining elements gained is at least
\begin{eqnarray}
& & \nicefrac{1}{4} \left( \vt(\cov_{t_\ini}(S) \setminus \cov_{t_\fin}(S),i) \right)\\
&=& \nicefrac{1}{4} \big( \vt(\cov_{t_\ini}(S),i) - \vt(\cov_{t_\fin}(S),i) \big) \\
&\geq& \nicefrac{1}{4} \big( 16 \cdot 2^\ell \vt(\cov_{t_\fin}(S),i) - \vt(\cov_{t_\fin}(S),i) \big) \\
&\geq& 4 \cdot 2^\ell \cdot \vt(\cov_{t_\fin}(S),i)  \\
&\geq& \vt(\cov_{t_\fin}(S),i+\ell)
\end{eqnarray}
Hence, when the set (along with the remaining elements) moves up to level $i+\ell$,
	these elements have as many tokens as their volume in level $i+\ell$, ensuring that the token invariant holds.

\end{proof}

Thus, we have shown that the token invariant holds at all times. Therefore, by~\Cref{cl:rec-elt-arr,cl:rec-set-form,cl:rec-token-invt}, we get the following lemma.

\begin{lemma} \label{lem:rec-recourse}
The total number of sets added till time $T$ is at most $\nicefrac{2T}{\lambda}$.
\end{lemma}

Putting~\Cref{lem:cost,lem:rec-recourse} together, we get the following theorem:
\begin{theorem} \label{thm:wtd-rec}
For any sequence of $T$ insertions or deletions, there is an efficient algorithm which maintains a collection of set cover solutions $\St$ such that each $\St$ is $O(\log \nt)$-competitive for the active set $\At$, and the total recourse (i.e., number of sets added or deleted) is $O(T)$.
\end{theorem}

\subsection{Better Cost Analysis}
Much like in the update time model, we can again get an improved competitive ratio of $O(\log \Delta_t)$ analysis for the same algorithm.
\begin{lemma}
  \label{lem:cost-better-rec}
  The solution $\St$ has cost at most $O(\log \Delta_t) \, \Opt_\bt$.
\end{lemma}
\begin{proof}
  Our proof is a suitable adaptation of that of~\Cref{lem:cost-better}, but we provide complete details to make it self-contained. 
  
  Recall the dual of the set cover LP:
\begin{gather}
  \textstyle \max\{ \sum_{e \in A_t} \ytil_e \mid \sum_{e \in A_t \cap
    S} \ytil_e \leq c_S ~\forall S \in \calF, ~~~\ytil_e \geq 0 \} .\label{eq:6-new}
\end{gather}

We now exhibit a dual $\ytil_e$ such that (a) the cost $\St$ can be bounded by $O(1) \sum_{e \in \At} \ytil_e$, and (b) $\ytil_e/(\log \Delta_t+2)$ is feasible for the dual LP \eqref{eq:6-new}. This will establish the proof of the lemma.

To this end, we define $\ytil_e$ to be $2^{i} \vt(e,i)$ if $e$ is covered in level $i$. To show (a), consider any set $S \in \St$ which is in level $i$. Then, we have by invariant~(i), that the density of set $S$ is at most $2^{i +10}$. Therefore, the \emph{volume of elements} $S$ currently covers (according to $\covt(\cdot)$) is at least $c_S/2^{i+10}$. Since each such element assigned to $S$ has dual value $\ytil_e = 2^i \vt(e,i)$, we get that $c_S \leq 2^{10} \cdot \sum_{e \in \covt(S)} \ytil_e$. Summing over sets $S \in \St$, and noting that $\covt(\cdot)$ defines a partition of the set of elements $\At$, establishes (a).

To show (b), for any set $S \in \calF$, we define $i^S_h := \max \{i: 2^i\ \leq c_S\}$ and $i^S_l := i^S_h - \ceil{\log \Delta_t}$. We first bound the sum of duals of elements in $S$ that are covered in levels $i^S_l$ or below: $\sum_{e \in \Et{\leq i^S_l} \cap S} \ytil_e \leq c_S$. Indeed, any element $e \in \Et{\leq i^S_l}$ has $\ytil_e \leq c_S/\Delta_t$ since the volume of any element in a stable solution is at most $1$ by~\Cref{cl:baselevel-recourse}, and the sum is over $|A_t\cap S|\leq \Delta_t$ active elements in $S$. 

Next, we observe that for any level $i$, we have $\vt(\Et{i} \cap S,i) \leq c_S/2^i$. Suppose not, and there exists $i$ such that $\vt(\Et{i} \cap S,i) > c_S/2^i$. But then, this set $S$ violates invariant~(ii) at level $i$,  contradicting the stability of $\St$. This immediately gives us that $\Et{i} \cap S = \emptyset$ for $i > i^S_h$. To see this, note that if a level $i > i^S_h$ has an element $e$ in $S$, i.e., $e \in \Et{i} \cap S$, then the volume of $e$ will be strictly greater than $1$ since its base level $b(e) \leq i^S_h$. This contradicts our volume bound of $c_S/2^i \leq c_S/2^{i^S_h+1} < 1$ at all levels $i > i^S_h$. Moreover, this also gives us that for all $i^S_l \leq i \leq i^S_h$, we have $\sum_{e \in \Et{i} \cap S} \ytil_e \leq c_S$ since $\ytil_e = 2^i \vt(e,i)$ for all such $e \in \Et{i} \cap S$. Therefore, summing over all $i$, we get that $\sum_{e \in S} \ytil_e \leq (\log \Delta_t + 2) c_S$. This establishes (b) and also completes the proof of the lemma.
\end{proof}

Using the above Lemma instead of~\Cref{lma:small,lma:single-level}, we get the following theorem.

\begin{theorem} \label{thm:wtd-rec-better}
For any sequence of $T$ element arrivals and departures, there is an efficient algorithm that maintains a set cover solution $\St$ that is $O(\log \Delta_t)$-competitive for the active set $\At$ at each time $t$. Furthermore, the total recourse (i.e., number of sets added or deleted to this solution) over the entire sequence of $T$ element arrivals and departures is $O(T)$.
\end{theorem}



\newcommand{\leader}{{\ell}}
\newcommand{\FIX}{{\sf Stabilize}\xspace}
\newcommand{\outs}{{\tt out}\xspace}
\renewcommand{\ins}{{\tt in}\xspace}
\newcommand{\probe}{{\tt Probe}}
\newcommand{\node}{{\tt node}}
\newcommand{\nodelist}{{\tt nodelist}}
\newcommand{\lists}{{\tt lists}}
\newcommand{\updatedual}{{\sf UpdateDual}\xspace}
\newcommand{\oldval}{{\tt old}}

\newcommand{\spc}{\hspace*{0.2 in}}




\section{Dynamic Primal-Dual Algorithm (Update-Time): Full Details} \label{sec:pd-dyn}

In this section, we furnish complete details of our dynamic primal-dual algorithm in the update time model and the proof of~\Cref{thm:main-update}(ii).  Recall our notation: given a set system $(U,\calF)$, and an element $e$, let $\calF_e$ denote
the sets containing $e$, and let $f_e := |\calF_e|$.  Let $f_t = \max_{e
  \in A_t} f_e$ denote the maximum number of sets containing any active
element. As described in~\Cref{sec:short-pd-update}, we maintain our sets in levels, and they automatically induce levels for elements.

\begin{itemize}
\item \emph{Set and Element ``Levels''.}
   At all times, each set $S \in \calF$ resides at an integer level,
   denoted by $\levelt(S)$. For an element $e \in \At$, define
   $\levelt(e) := \max_{S \, : e \in S} \levelt(S)$ to be the largest
   level of any set covering it.
 \item \emph{Set and Element ``Dual Values''.}  Given levels for sets
   and elements, the \emph{dual value} of an element $e$ is defined to
   be $\yt(e) := 2^{-\levelt(e)}$. The dual value of
   a set $S \in \calF$ is the sum of dual values of its elements,
   $\yt(S) := \sum_{e \in S \cap \At} \yt(e)$.
\end{itemize}

Recall the dual of the set cover LP:
\begin{gather}
  \textstyle \max\{ \sum_{e \in A_t} \ytil_e \mid \sum_{e \in A_t \cap
    S} \ytil_e \leq c_S ~\forall S \in \calF, ~~~\ytil_e \geq 0 \} .\label{eq:6}
\end{gather}
Now our solution at time $t$ is simply all sets whose dual constraints
are \emph{approximately tight}, i.e., $\St = \{S \, : \, \yt(S) \geq
\nicefrac{c_S}{\pdapx}\}$ for $\pdapx := 32
f$. In addition, we will try to ensure that the duals  $\yt(e)$ we maintain will be approximately feasible for~\eqref{eq:6}, i.e., for every set $\yt(S) \leq \beta c_S$. Then, note that bounding the cost becomes a simple application of weak duality. The question then --- as with previous work as well, e.g.~\cite{Bhatta-icalp} --- is about how we maintain such a dual solution dynamically when elements arrive and depart.

We achieve this by defining a \emph{base level}
for every set to indicate non-tight dual constraints, and maintaining that all sets
\emph{above their base levels} always have approximately tight dual
constraints. Formally, the \emph{base level} for set $S$ is defined to
be $b(S) := - \ceil{ \log ( \pdapx {c_S}) } - 1$, and our solution
$\St$ consists of \emph{all sets} which
are located strictly above their respective base levels, i.e., $\St
= \{S \in \calF \, : \, \levelt(S) > b(S)\}$. To initialize, each
set $S$ is placed at level $b(S)$. We will maintain the invariant that every set lies either at its base level or above.

The reason we define base level this way is the following: suppose a new element $e$ arrives and it is uncovered, i.e., all its covering sets are at their base levels. Then, by the way we have defined level of an element, it would have a dual value of $\yt(e) = \nicefrac{1}{2^{b(S_e)}} > 2 \pdapx c_{S_e}$ where $S_e$ is the set with highest $b(S)$ value (or equivalently lowest cost) among all sets covering $e$. Now this means that the set $S_e$ would have $\yt(S_e) > 2 \pdapx c_{S_e}$ and so our algorithm would move it up in order to satisfy approximate dual feasibility, thereby including it in the solution.

 We ensure the approximate tightness (and approximate feasibility) of dual constraints using the \emph{stability property} below, which is again the key definition of our algorithm and similar to~\cite{Bhatta-icalp}.

\begin{itemize}
\item \emph{Stable Sets.} A set is \emph{stable} if it satisfies the following conditions:
 if $\levelt(S) > b(S)$ we have $y(S) \in
  [\nicefrac{c_S}{\pdapx}, \pdapx c_S]$, and if $\levelt(S) =
  b(S)$ we have $y(S) < \pdapx c_S$.
\end{itemize}

As mentioned in~\Cref{sec:short-pd-update}, the algorithm follows the principle of least effort toward ensuring stability of all sets: if at some point $\yt(S)$ is too large, $S$ moves up the least number of levels so that the resulting $\yt(S)$ falls within the admissible range, and similarly if $\yt(S)$ is too small, it moves down until $\yt(S) \geq \nicefrac{c_S}{\pdapx}$.  We re-state the high-level description of the algorithm for a self-contained presentation in this section.

\begin{leftbar}
  \noindent {\bf Arrival:} When $e$ arrives,
  define $\yt(e) := 1/2^{\max_{S: e \in S} \levelt(S)}$, update all
  $\yt(S)$, and run {\sf Stabilize}.

  \medskip \noindent {\bf Departure:} Delete $e$ from $\At$, update
  $\yt(S)$ for all sets, and run $\Stabi$.

  \medskip \noindent {\bf Stabilize:} While there exists some $S$ at level
  $\levelt(S)$ such that $\yt(S) > \pdapx c_S$: find the \emph{lowest
    level} $\ell' > \levelt(S)$ such that placing $S$ at level $\ell'$
  results in $\yt(S) \leq \pdapx c_S$.
  Analogously, if $\yt(S) < \nicefrac{c_S}{\pdapx}$, find the \emph{highest
  level} $\ell' < \levelt(S)$ such that placing $S$ at level $\ell'$
  results in $\yt(S) \geq \nicefrac{c_S}{\pdapx}$. If such an $\ell' \geq b(S)
  -1$, we place $S$ at level $b(S)$ and drop $S$ from $\St$.
\end{leftbar}

\medskip \noindent {\bf Data Structures.} The algorithm will maintain a queue $Q$ of sets for which one of the invariant conditions  is violated.
For each set $S$, the algorithm will maintain several doubly-linked lists, collectively denoted by
$\lists(S)$ : (i) $\outs(S)$: these are the elements in $S$ whose level is equal to $\levelt(S)$, (ii) for every $l > \levelt(S)$, a list $\ins_l(S)$ of elements in $S$ whose level is exactly $l$. Note that for every element $e$ and each set $S$ containing $e$, $e$ is present in exactly one of the lists among $\outs(S)$ and the different $\ins_{\ell}(S)$ for $\ell > \levelt(S)$. We use $\node(e,S)$ to refer to the physical node/cell containing $e$ in the appropriate list it belongs to. Finally, for every element $e$, we have a doubly linked list, $\nodelist(e)$, which contains $\node(e,S)$ for every set $S$ containing $e$.

When~\Cref{alg:fix}~(\FIX) is invoked, it considers a set $S$ in $Q$, and will either increase or decrease its level till it satisfies the stability property. We now describe how these steps work in detail.
There are two cases (details given in Algorithm~\FIX):

\begin{itemize}
\item Case (i) $\yt(S) > \beta c_S$ : In this case, we move the set $S$
  up. As it moves up, we need to keep track of change in $\yt(S)$. One
  naive way to do this would be to re-compute $\yt(S)$ each time it
  moves up. This could lead to high update time. Instead, we keep two
  running sums: $\yt^o(S)$, which is the contribution of elements in
  $\outs(S)$ toward $\yt(S)$, and $\yt^i(S)$, which is the remaining
  contribution to $\yt(S)$.  Now updating $\yt^o(S)$ and $\yt^i(S)$, as
  we move from a level $\ell-1$ to $\ell$, only requires time dependent
  on $\ins_{\ell}(S)$. We keep moving $S$ up till $S$ satisfies the
  stability condition. Note that this will happen before $S$ reaches the
  level~$\ceil{\log (n/c_S)}$. Indeed, if $S$ reaches this level, then
  $\yt(e) \leq \frac{c_S}{n}$, and so, $\yt(S) < c_S$. Since $\yt(S)$
  can decrease by at most a factor of $2$ when we go up one level,
  $\yt(S)$ was at most $2c_S$ just before moving to this level -- but
  then we would not have moved $S$
  up.
  Finally, when the upward move stops at a level $\ell^*$, we update the
  following: (i) set $\yt(e)$ values of all elements in $\outs(S)$ to be
  $1/2^{\ell^*}$; (ii) if $e$ was in the $\outs(S')$ list of some set
  $S'$, then remove it and add $e$ to the $\ins_{\ell^*}(S')$ list, or
  if $e$ was in the $\ins_{\ell'}(S')$ list for some $S'$ residing at a
  level below $l^*$, remove $e$ from this list and add it to the list
  $\ins_{\ell^*}(S')$; (iii) update the corresponding $\yt({S'})$ values
  for all $S' \in \calF_e$; and (iv) this may cause violation of
  stability property for some $S'$, in which case we add it to $Q$.
  Overall, this step requires some care in updating the data structures
  and we use $\nodelist(e)$ to guide us through the correct lists which
  $e$ previously belonged to, and change them. These steps are
  formalized in~\Cref{alg:fix}~{\FIX}. In
  Lines~\ref{code:up1}-\ref{code:up2}, we move the set $S$ up till it
  satisfies the stability property. In
  Lines~\ref{code:up3}-\ref{code:up4}, we go through all elements $e$ in
  $\outs(S)$ and update the $y_{S'}$ values of any set $S'$ containing
  any such element. This may also need placing $e$ in the correct list
  for $S'$, and adding $S'$ to $Q$
  (\Cref{alg:updatedual}~(\updatedual)). Note that
  \Cref{alg:updatedual}~(\updatedual) takes three parameters -- an
  element $e$, its old $\yt(e)$ value (before we moved $S$), and the new
  $\yt(e)$ value (after we have moved $S$).

\item
Case (ii) $\yt(S) < \frac{c_S}{\beta}$: In this case, we move $S$
down. We again follow a similar process as in case~(i). In case there
was a set $S'$ containing $e$ at level $\ell$, we need to remove it from
$\outs(S)$ and add it to $\ins_\ell(S)$.  We stop moving $S$ down it
satisfies the stability condition
(Lines~\ref{code:dn1}-\ref{code:dn2}). Note that it will trivially
satisfy stability property if it reaches the base level $b(S)$. Indeed,
if it reaches the base level, then $\yt(S)$ was at most
$\frac{c_S}{\pdapx}$ just before it moved to this level. Since $\yt(S)$
can increase by at most a factor $2$, it will satisfy the stability
property required at the base level. As in case~(i) above, updating
$\yt(e)$ value means that we need to call~\updatedual. For some
technical details, we update all the data structures each time $S$ moves
a single level, whereas in case~(i), we do these updates only after $S$
settles finally in a level $\ell^*$ after coming out of the loop at
Line~\ref{code:up1}.
\end{itemize}

Finally, we briefly mention how updates are handled in \Cref{alg:dynamicf}~({\sf Dynamic}) When a new element $e_t$ arrives at time $t$, we simply set its dual value $y_t$ according to levels of sets containing it. Further, for each set $S$ containing $e_t$, we update $y(S)$ and place $e_t$ in the appropriate list in $\lists(S)$. Finally, if $S$ does not satisfy stability property, we add it to $Q$. If this update operation was deletion of an element $e_t$, we simply remove it and update the $\yt(S)$ values for all sets containing it (and add it to $Q$ if needed).

\begin{algorithm}
\caption{{\sf Dynamic}$(e_\bt, \pm)$}
\label{alg:dynamicf}
\begin{algorithmic}[1]
\If{the operation $\sigma_\bt$ is $(e_\bt, +)$}
\State  Set $\levelt(e) = \max_{S:e \in S} \levelt(S)$, and define $\yt(e)$ to be $\nicefrac{1}{2^{\levelt(e)}}$.
\State  For all sets $S$ containing $e$
\State \spc Update $\yt(S) \leftarrow \yt(S) + \yt(e)$.
\State \spc Add $e$ to the appropriate list in $\lists(S)$, and build $\nodelist(e)$.
\State\spc  If $\yt(S) > \beta  c_S$, and $S$ is already not in $Q$, add it to $Q$.
\ElsIf{the operation $\sigma_\bt$ is $(e_\bt, -)$}
\State For all sets $S$ containing $e$ (traverse using $\nodelist(e)$)
\State  \spc Update $\yt(S) \leftarrow \yt(S) - \yt(e)$.
\State \spc Delete  $\nodelist(e)$ and all $\node(e,S)$.
\State \spc If $\yt(S) < \frac{c_S}{\beta}$, and $\levelt(S) > b_S$ and $S$ is not in $Q$, add $S$ to $Q$.
\EndIf
\State If $Q$ is non-empty, call \FIX.
\end{algorithmic}
\end{algorithm}

\begin{algorithm}
\caption{\FIX}
\label{alg:fix}
\begin{algorithmic}[1]
\While{$Q$ is non-empty}
\State $S \leftarrow$ first element of $Q$
\If{ $\yt(S) > \pdapx c_S$}   \Comment{$\yt(S)$ is too high, move $S$ up} \label{code:up1}
\State $\yt^o(S) \gets \sum_{e \in \outs(S)} \yt(e)$, $\yt^i(S) \gets \yt(S) - \yt^o(S)$ \label{ln:high1}
\Repeat
\State $\levelt(S) \leftarrow \levelt(S)+1$ \Comment{Move $S$ up one step} \label{ln:high1a}
\State $\yt^o(S) \leftarrow \yt^o(S)/2 +
2^{-\levelt(S)} \cdot | \ins_{\levelt(S)}(S)| $
\State $\yt^i(S) \leftarrow \yt^i(S) -
2^{-\levelt(S)} \cdot | \ins_{\levelt(S)}(S)| $
\State $\yt(S) \leftarrow \yt^o(S) + \yt^i(S) $
\State $\outs(S) \leftarrow \outs(S) \cup \ins_{\levelt(S)}(S). $ \label{ln:high1b}
\Until{$\yt(S) \leq \pdapx c_S $} \Comment{Found the right level} \label{code:up2}
\ForAll{$e \in \outs(S)$} \label{code:up3}
\State $\yt^\oldval(e) \leftarrow \yt(e)$ \Comment{Store the old value of
  $e$'s dual}
\State $\yt(e) \gets 2^{-\levelt(S)}$ \Comment{This is the new dual}
\State call $\updatedual(e, \yt^\oldval(e), \yt(e))$.
\EndFor \label{ln:high2} \label{code:up4}
\ElsIf {$\yt(S)< \nicefrac{c_S}{\pdapx}$ and $\levelt(S) > b(S)$}
\Repeat \label{code:dn1}
\Comment{$\yt(S)$ is too low, move down}
\State $\levelt(S) \leftarrow \levelt(S)- 1$ \label{ln:low1}
\ForAll{$e \in \outs(S)$}
\If{there is no other set containing $e$ at level $\levelt(S) + 1$ or higher}
\State $\yt^\oldval(e) \leftarrow \yt(e)$.
\State $\yt(e) \gets 2^{-\levelt(S)}$.
\State call $\updatedual(e, \yt^\oldval(e), \yt(e))$.
\Else
\State Remove $e$ from $\outs(S)$ and add it to $\ins_{\levelt(S)+1}(S)$.
\EndIf
\EndFor \label{ln:low2}
\Until{$\yt(S)< \nicefrac{c_S}{\pdapx}$ and $\levelt(S) > b(S)$} \label{code:dn2}
\EndIf
\EndWhile
\end{algorithmic}
\end{algorithm}

\begin{algorithm}
\caption{\updatedual$(e, \yt^\oldval(e), \yt(e))$}
\label{alg:updatedual}
\begin{algorithmic}[1]
\State For all nodes $(e,S)$ in $\nodelist(e)$ do
\State \spc  If $e$ is not in the correct list in $\lists(S)$ (according to $\yt(e)$ value)
\State \spc \spc \spc Delete $\node(e,S)$ and add a node corresponding to $e$ in the correct list in $\lists(S)$.
\State \spc  Update $\yt(S) \leftarrow \yt(S) - \yt^\oldval(e) + \yt(e)$.
\State \spc  If $S$ does not satisfy one of the invariant conditions and is not in $Q$, add $S$ to $Q$.
\end{algorithmic}
\end{algorithm}


\subsection{Analysis}
We now analyze the update algorithm.
Consider a time $t$ and  assume by induction
that the stability property  holds for every set $S$  before this operation. We first analyze the competitive ratio of our solution and the amortized update time till this time step. Then we will show that the invariants hold after the update operation at time $t$.

We first show that feasibility of our solution, and then bound the total cost.


\begin{claim}
Let $e \in \At$. Then there exists $S \in \calF_e$ with $\levelt(S) > b(S)$.
\end{claim}
\begin{proof}
Suppose not. So every set containing $e$ is at its base level (notice that a set is never placed below its base level in the algorithm). Then $\yt(e)$ will be  $\frac{1}{2^{b(S_e)}} \geq \beta c_{S_e}$, where $S_e$ is the cheapest set containing $e$,  implying that $\yt(S_e) \geq \yt(e) \geq \pdapx c_{S_e} $, hence $S$ violates the stability property.
\end{proof}

\begin{lemma} \label{lem:pd-cost-app}
The set cover algorithm is $O(f^3)$-competitive.
\end{lemma}
\begin{proof}
The proof follows from LP duality and the fact that all sets are stable. More formally, define quantities $z_e$ for every element $e \in A_t$ as $\nicefrac{\yt(e)}{\pdapx}$. We claim that $z_e$ variables are dual feasible. Indeed, for any set $S$,
$\sum_{e \in S} z_e = \nicefrac{\yt(S)}{\pdapx} \leq c_S$ (using stability of $S$).

Let $\calS_t$ denote our set cover solution at time $t$. We know from stability
that for any set $S \in \calS_t$, $\yt(S) \geq \nicefrac{c_S}{\pdapx}$.
Now $\sum_e z_e$ is a lower bound on the cost of an optimal solution.
Since each element
is in at most $f$ sets, we see that
$$ \sum_{e \in A_t} z_e = \sum_{e \in A_t} \frac{\yt(e)}{\pdapx} \geq
\sum_{S \in \calS_t} \frac{\yt(S)}{\pdapx \cdot f} \geq \sum_{S \in \calS_t} \frac{c_S}{\pdapx^2 \cdot f}. $$
Therefore, the dual objective value corresponding to $z_e$
 is $\Omega(1/f^3)$ times the cost of our solution.
\end{proof}

We now analyze the update time of the algorithm.

\begin{lemma}
\label{lem:outs1}
Suppose we move a set $S$  down from a level $\ell$ to $\ell-1$ during algorithm~\FIX. Let $\outs^{\old}(S)$ denote the list $\outs(S)$ when $S$ was at level $\ell$ before this move. Then $|\outs^{\old}(S)| \leq 2^{\ell} \left( \nicefrac{c_S}{\pdapx} \right)$. Further, the total update time incurred during this move (\cref{ln:low1}-\cref{ln:low2} of~\FIX) is $O(f |\outs^{\old}(S)|+1)$.
\end{lemma}
\begin{proof}
We moved $S$ down because $\yt(S) < \nicefrac{c_S}{\pdapx}$. Each element $e \in \outs^{\old}(S)$ has
$\yt(e)$ value equal to $2^{-\ell}$ since $\levelt(e) = \levelt(S)$ for all such elements. The first claim then follows. The second statement follows from the fact that we spend $O(f)$ time for each element in $\outs^{\old}(S)$ due to \updatedual. The extra ``+1'' term in the statement is to account for the case when $\outs^{\old}(S)$ is empty.
\end{proof}

\noindent
Now we consider the case when a set $S$ moves up. Instead of updating the data structures after incremental steps of $1$, we perform them once $S$ settles (i.e., after it has come out of the {\bf repeat}-{\bf until} loop in
Lines~\ref{code:up1}-\ref{code:up2} in \FIX ). We now bound the total update time of this step.

\begin{lemma}
\label{lem:outs2}
Suppose we move a set $S$ up from a a level $\ell$ to $\ell^*$ during
the {\bf repeat}-{\bf until} loop in
Lines~\ref{code:up1}-\ref{code:up2} in \FIX.
Let $\outs^{\new}(S)$ denote the list $\outs(S)$ when $S$ settled at level $\ell^*$ after this move. Then $|\outs^{\new}(S)| \leq 2^{\ell^*} \pdapx c_S$. Further, the total update time incurred during these steps (\cref{code:up1}-\cref{ln:high2} of~\FIX) can be implemented in  $O(f |\outs^{\new}(S)|+(\ell^*-\ell))$ time.
\end{lemma}

\begin{proof}
This iteration ends at level $\ell^*$ because
$\yt(S) \leq \pdapx c_S$. Since each element in $\outs^{\new}(S)$ has $\yt(e)$ value $1/2^{\ell}$, the first result follows.
Each iteration of the inner loop (\cref{ln:high1a}-\cref{ln:high1b}) when we move $S$ up from a level $l -1$ to $l$ takes time proportional to $\ins_{l}(S)+1$. Therefore, the overall time during the iterations of the inner while loop can be bounded by $ |\outs^{\new}(S)|+(\ell^*-\ell)$ when $S$ settles at level $\ell^*$, because $\outs^{\new}(S)$ eventually contains each of the intervening $\ins_l(S)$ sets. Similarly, updating
$\yt(e)$ and $y_{S'}$ values and running the \updatedual for each $e$ take time proportional to $f |\outs^{\new}(S)|$.
\end{proof}

To analyze these update costs, we use the following token scheme.

{\bf Token Distribution Scheme: } Every set $S$ maintains a certain non-negative number
of tokens at any point of time. Whenever we move a set $S$ up or down it will \emph{expend} some tokens to account for update time during these moves, and also {\em transfer} some tokens to some other sets. The only way tokens get injected into the system is when a element arrives or departs.
We fix a set $S$ for rest of the discussion.
We now give details of the token distribution scheme:
\begin{itemize}
\item[(i)] When a new element arrives or departs, it gives $20f$ tokens to each of the sets containing it.
\item[(ii)] Suppose the set $S$ moves down from a level $\ell$ to $\ell-1$. Consider the set
$\outs^{\old}(S)$ when $S$ was at level $\ell$. Then $S$ spends $1+f \cdot |\outs^{\old}(S)|$ tokens, and transfers $20f$ tokens to every set $S'$ such that $S' \in \calF_e$ for all $e \in \outs^{\old}(S)$.
\item[(iii)] Suppose the set $S$ moves up from level $\ell$ to $\ell^*$ during~\cref{code:up1}-\cref{ln:high2} in the algorithm~\FIX. Consider the set $\outs^{\new}(S)$ when $S$ reaches level $\ell^*$.
Then it spends $(\ell^*-\ell) + f \cdot |\outs^{\new}(S)|$ tokens, and transfers $1$ token to every set $S'$ such that $S' \in \calF_e$ for all $e \in \outs^{\new}(S)$.
\end{itemize}

We make some simple observations first.
\begin{claim} \label{cl:token-count}
The total number of tokens injected into the system is $O(f^2 T)$ after $T$ arrivals/departures.
\end{claim}
\begin{proof}
This follows from the definition of our token distribution scheme since we inject $O(f^2)$ tokens per arrival/departure.
\end{proof}

\begin{claim} \label{cl:work-to-tokens}
The total update time can be bounded by $O(1)$ times the total number of tokens expended.
\end{claim}
\begin{proof}
This follows as an immediate corollary of~\Cref{lem:outs1,lem:outs2} and the definition of our token distribution scheme.
\end{proof}

We now show that at any time, the number of tokens expended is at most the number of tokens remaining.  To do this, it will be convenient to divide the change in levels of a set $S$ into {\em epochs}. For a set $S$, consider the sequence of levels of $S$ as the algorithm progresses. This sequence can be broken into maximal sub-sequences of up and down moves. A maximal sub-sequence of up moves will be called an {\em up-epoch}. Define a {\em down-epoch} similarly, so every epoch is either an up-epoch or a down-epoch. Note that the moves (of $S$) during an epoch could have been made during different calls to algorithm~\FIX at different points in time.

\begin{lemma}
  \label{lem:epochdown-app}
  Consider a down-epoch for a set $S$ starting at level $\ell$. The number
  of tokens expended or transferred by $S$ during this epoch is at most
  $\max(1,2^{\ell+1} f c_S)$. Moreover, the total number of tokens that $S$ gained
  when it reached level $\ell$ at the beginning of this epoch is at
  least $\max(1,2^{\ell+1} f c_S)$.
\end{lemma}


\begin{proof}
We first assume that $\outs^{\old}(S)$ is non-empty.
Lemma~\ref{lem:outs1} shows that the total number of tokens needed by $S$ when it moves down from a level $l$ to $l-1$  during this epoch is at most $ 1+
f  |\outs^{\old}(S)| + 20f |\{S': S' \in \calF_e, e \in \outs^{\old}(S)\}| \leq
2f  |\outs^{\old}(S)| + 20f |\{S': S' \in \calF_e, e \in \outs^{\old}(S)\}| \leq
 22 f^2 2^{l} ( \nicefrac{c_S}{\pdapx} )$. Therefore, the total token requirement during this epoch is at most
$22 f^2 2^{\ell+1} ( \nicefrac{c_S}{\pdapx} ) \leq 2 f c_S \cdot 2^\ell$, using $\pdapx = 32 f$.
Let us now count how many tokens $S$ had at the beginning of this epoch. First observe that
$\ell > b(S)$, otherwise $S$ will not move down. So $S$ must have moved up from level $\ell-1$ to $\ell$ during the preceding up-epoch. The set $S$ must have moved up from level
$\ell-1$ because $\yt(S)$ was greater than $\pdapx c_S$ at that time. Since moving up cannot reduce
$\yt(S)$ by more than a factor of half, $\yt(S)$ must have been at least $( \nicefrac{\pdapx}{2} ) c_S$ when $S$ entered level $\ell$. Since then, $\yt(S)$ must have dropped to below $\nicefrac{c_S}{\pdapx}$ (otherwise it
will not move down). Thus $\yt(S)$ has decreased by more than $(\pdapx/2 - 1/\pdapx) c_S \geq 15 f c_S$ since $\pdapx = 32f$. Now $\yt(S)$ can decrease by
one of two events: (i) some element in $S$ gets removed, or (ii) the $\yt(e)$ value of
some element in $S$ decreases. In either case, the contribution to the decrease in $\yt(S)$ is at most $1/2^{\ell}$. Therefore, at least $15 \cdot 2^{\ell} f c_S$ such events must have happened. Each such event would give at least $1$ token to $S$ -- in case of element deletion, $S$ gets 1 token by rule~(i) of the token scheme. If
an element $e$ in $S$ moves up, it must be the case that some set $S'$ containing $e$ moves up -- during this process it would give $1$ token to $S$ (by rule~(iii)). Thus, $S$ gets a total of at least $15 \cdot 2^{\ell} f c_S$ while it is in level $\ell$. This proves the lemma.

If $\outs^{\old}(S)$ were empty, observe that $y_S$ will not change when $S$ moves down, and so, we will reach the base level for $S$ immediately (i.e., we need not
go through the {\bf repeat-until} loop in Lines~\ref{code:dn1}-\ref{code:dn2} in Algorithm~\FIX and can move to $b_S$ in one step). So, we need to
spend only 1 token (in terms of running time).
 Clearly, at least  event of kind~(i) or~(ii) as above must have occurred since $S$ moved to level $\ell$ first. Therefore, the set $S$ would have received at
least 1 token since then.
\end{proof}

Now we consider up-epochs.

\begin{lemma}
\label{lem:epochup-app}
Consider an up-epoch of a set $S$ ending at $\ell^*$.
Then  total number of tokens spent or transferred by $S$ in this epoch is at most $2^{\ell^*+8} f^2 c_S$. Further, the total number of tokens $S$ gained while it stays at level $\ell$ at the beginning of this epoch is at least $2^{\ell^*+8} f^2 c_S$.
\end{lemma}

\begin{proof}
  Consider an up-epoch where $S$ moves from $\ell_0 \rightarrow \ldots
  \rightarrow \ell_k = \ell^*$ via a sequence of up-moves.
  Let us see how many tokens are needed for the upward
  move $\ell_{i-1} \rightarrow \ell_i$.
    Using Lemma~\ref{lem:outs2} and our token scheme,
   $(\ell_{i} - \ell_{i-1}) + f |\outs^{\new}(S)|$ tokens are spent to account for running time, and $f |\outs^{\new}(S)|$ tokens are transferred. Therefore, the total token requirement during this upward move is at most $\ell_{i} - \ell_{i-1} + 2f |\outs^{\new}(S)| \leq (\ell_i - \ell_{i-1}) + 2f 2^{\ell_i} \pdapx c_S$. Summing over all $i$ gives us a bound of $(\ell^* - \ell_0) + \sum_{l \leq \ell} \left( 2f \pdapx c_S \cdot 2^{l} \right)\leq  (\ell^* - \ell_0) + 2^{\ell^*+7} f^2 c_S. $ Now we claim that $(\ell^* - \ell_0) \leq 2^{\ell^*+7} f^2 c_S$ and so, the sum can be bounded by
   $2^{\ell^*+8} f^2 c_S$. To see this, observe that
    $\ell_0 \geq b_S$, and so, $2^{\ell_0} \geq \frac{1}{4\beta c_S}$. Therefore,
$2^{\ell^*+7} f^2 c_S =  2^{\ell_0+7}
\cdot 2^{\ell^*-\ell_0} \cdot f^2 c_S \geq  2^7 \cdot \frac{1}{4\beta c_S} \cdot
\cdot 2^{\ell^*-\ell_0} f^2 c_S \geq (\ell^*-\ell_0)$,
because  $\beta = 32f$. This proves the first part of the lemma.

For the second part, consider the time when this epoch started. We claim that when $S$ had moved to level $\ell_0$ at the end of the previous epoch, then $\yt(S)$ was at most $\frac{2c_S}{\pdapx}$. Indeed, if this epoch started at the
beginning of the algorithm, then $\yt(S)$ was 0; otherwise it was preceded by a down-epoch. When we moved from level $\ell_0+1$ to $\ell_0$, $\yt(S)$ was less than $\frac{c_S}{\pdapx}$. Since
$\yt(S)$ can at most double when we move down one level, the claim follows. Let us then denote $\yt(S)$ at the beginning of this epoch as $y_0(S)$.

Now, let $S_0$ be the elements of $S$ which were active at the time $S$ moved to level $\ell_0$. Now, hypothetically, suppose, at that time instant, we had placed $S$ at level $\ell^*-1$ \emph{without changing the levels of other sets}. Let $\widetilde{\yt}_0(e)$ be the \emph{hypothetical dual values for elements in $S_1$ corresponding to these levels of sets}, and let $\widetilde{\yt}_0(S) = \sum_{e \in S_0} \widetilde{\yt}_0(e)$.
Clearly because $\ell^*-1 \geq \ell_0$ and the dual values decrease as a set moves up levels, we have $\widetilde{\yt}_0(S) \leq y_0(S) \leq \frac{2 c_S}{\pdapx}$.

Next, consider the time during this up-epoch when we move $S$ from $\ell^*-1$ to $\ell^*$, and let $S^*$ be the elements of $S$ active at this time. Similarly, let $\yt^*(e)$ be the dual values at this time for all elements in $S^*$, and let $\yt^*(S)$ denote
$\sum_{e \in S^*} \yt^*(e)$. Since the algorithm moved $S$ from $\ell^*-1$ to $\ell^*$, we have $\yt^*(S) > \pdapx c_S$.

To complete the proof, we consider the change from $\widetilde{\yt}_0(S)$ to $\yt^*(S)$ (of more than $(\pdapx - \nicefrac{2}{\pdapx}) c_S$): indeed, this can happen precisely because of two reasons:~(i) there are elements in $S^*$ which are not present in $S_0$ -- each such element contributes
at most $2^{-(\ell^*-1)}$ to $\yt^*(S)$, and (ii) there are elements in $S^* \cap S_0$ which have moved down since the beginning of this epoch. Again, any such element contributes at most $2^{-(\ell^*-1)}$ to $\yt^*(S)$. Thus, each of these kind of events will contribute at most
$2^{-(\ell^*-1)}$ to $\yt^*(S) - \widetilde{\yt}_0(S)$. It follows that there must be at least $2^{\ell^*-1} (\pdapx - 1) c_S \geq 15 f c_S  2^{\ell^*}$ events of either kind, using $\pdapx = 32f$.
Then, our token distribution scheme now says that $S$ would have collected at least
$20 f$ tokens from each such event, implying that $S$ gained a total of $300 f^2 c_S 2^{\ell^*}$ tokens  in this epoch, more than the number required, thus proving the lemma.
\end{proof}

Thus we have showed that after the end of any epoch, every set has more tokens with it than when the epoch began, even after expending tokens for updates. Therefore, the iterations in~\FIX will terminate because there are finite number of tokens. When~\FIX terminates, all sets will satisfy the stability property.
Thus,  from~\Cref{lem:pd-cost-app,lem:epochup-app,lem:epochdown,cl:work-to-tokens,cl:token-count}, we get the following theorem.

\begin{theorem}
There is an efficient $O(f^3)$-competitive algorithm for dynamic set cover with an amortized update time of $O(f^2)$.
\end{theorem}

\section{Dynamic Primal-Dual Algorithm (Recourse): Full Details}
\label{sec:f1}

In this section, we consider the recourse model and give an algorithm
with stronger guarantees than the one in the previous section. Our
algorithm is inspired by the following offline algorithm for set cover:

\smallskip
The algorithm maintains a tentative solution $\calS$, which is initially
empty. Now, the algorithm arranges the elements in an arbitrary order for
inspection. For each inspected element $e$, one of these two cases arise:
\begin{itemize}
	\item
	{\em $e$ is already covered by a set $S \in \St$:}~in this case, 
    the solution $\calS$ remains unchanged. 
    \item 
    {\em no set in $\calS$ covers $e$:}~in this case, the algorithm adds a
    randomly chosen set $S\in \calF_e$ to add to the solution $\calS$, where 
    set $S \in \calF_e$ is chosen with probability:
		$$p^e_S := \frac{1/c_S}{\sum_{S' \in \calF_e} 1/c_{S'}}.$$ 
    In this case, $e$ is said to be a {\em probed} element. 
\end{itemize}

\smallskip
Note the solution $\calS$ can be defined through a bijection $\varphi$ from 
the set of probed elements.
It is known that this offline algorithm is $O(f)$-competitive 
in expectation~\cite{Pitt}.

We now describe our adaptation of this algorithm to the {\em fully dynamic}
model. Recall that $A_t$ denotes the set of
active elements at time $t$.  At all times $t$, we maintain a partition
of $A_t$ into two sets $P_t$ (called the \emph{probed set}) and 
$Q_t$ (called the \emph{unprobed set}). Elements in the probed set are 
exactly those on which we have performed the random experiment 
outlined above. All other elements are in the unprobed set.
The algorithm maintains a bijection $\varphi$ from the probed set 
$P_t$ to $\calS_t$, the set cover solution at time $t$.
In other words, for every probed element $e\in P_t$, 
there is a unique set $\varphi(e)$ in $\calS_t$ and vice-versa. 
Elements can move from the unprobed set to the probed set over time,
i.e., an element in $Q_t$ can be in $P_{t'}$ at a later time $t' > t$. 
But, once an element enters the probed set, it cannot move back to 
the unprobed set in the future. In other words, 
$P_t\cap A_{t'}\subseteq P_{t'}$ for all $t' \geq t$.

We now describe the procedures which will be used in our algorithm. At certain times, our
algorithm may choose to \emph{probe} an unprobed element $e$. This will happen when there is
no set in the current solution covering $e$.

\smallskip\noindent {\bf Probing an element.} When an unprobed element
$e \in Q_{t-1}$ is probed by the algorithm at time $t$, a set $S$ 
containing $e$ in randomly chosen with probability:
$$p^e_S := \frac{1/c_S}{\sum_{S'\in \calF_e} 1/c_{S'}}.$$ 
This chosen set $S$ is
added to the current solution of the algorithm: 
$\calS_t = \calS_{t-1}\cup \{S\}$. 
Element $e$ moves from the unprobed set to the probed set:
$P_t = P_{t-1} \cup \{e\}$ and $Q_t = Q_{t-1} \setminus \{e\}$. As long
as $e$ remains in the active set of elements $\calA_t$, i.e., does not 
depart from the instance, the set $S$
also remains in the solution $\St$. We say that $e$ is {\em responsible} for
$S$ and denote $\varphi(e) := S$.

\noindent
Having defined the process of probing an element, we describe when an
element is probed. These probes are triggered by element insertions and
deletions as described below. 

\smallskip
\noindent
{\bf Element Arrivals.}
Suppose element $e$ arrives at time $t$.  There are two cases:
\begin{itemize}
\item 
If $e$ is already covered in
the current solution $\calS_{t-1}$, it is added to the unprobed set,
i.e., $Q_t = Q_{t-1}\cup \{e\}$. The solution $\St$ remains unchanged,
i.e., $\St = \calS_{t-1}$. 
\item
If $e$ is not covered in the current
solution, then it is probed, which adds a set $\varphi(e)$ to $\calS_t$
as described above. We then set $P_t = P_{t-1} \cup \{e\}$.
\end{itemize}

\smallskip
\noindent
{\bf Element Departures.}
Suppose element $e$ departs from the instance at time $t$. Again,
there are two cases:
\begin{itemize}
\item
If $e$ is currently an unprobed element, then we set 
$Q_t = Q_{t-1}\setminus \{e\}$ and the 
solution remains unchanged: $\St = \calS_{t-1}$. 
\item
If $e$ is currently a probed element, 
then we set $P_t = P_{t-1}\setminus \{e\}$. In addition,
the set $\varphi(e)$ is also removed from the solution
$\calS_{t-1}$. This might lead to some elements in $Q_t$ becoming
uncovered. (Note that each remaining element is $P_t$ is still 
covered, by the sets they are respectively responsible for in
the current solution.) We pick the element that arrived the 
earliest\footnote{This can be any arbitrary but fixed order 
on the elements.} among the uncovered elements and move it
to the probed set. On probing this element, a new set covering it 
is added to the solution. This set might cover some of the uncovered
elements, which are then removed from the uncovered set. The
process continues iteratively, probing the uncovered element 
that arrived the earliest in each step. Once there is no uncovered
element left, we denote the new solution by $\S_t$.
\end{itemize}

\medskip
\noindent
{\bf Analysis.}
First, we observe that the algorithm maintains a feasible solution by
definition. In particular, it maintains the invariants that:
\begin{itemize}
\item For every element $e$ in the probed set $P_t$, there is a unique set
  $\varphi(e)$ in the solution $\St$. Moreover, this mapping is a bijection,
  implying $\calS_t =  \{\varphi(e): e \in P_t\}$. 
  (Note that an element can have
  multiple sets in $\calS_t$ covering it, but exactly one of these is
  defined as $\varphi(e)$. Conversely, a set in $\St$ covers multiple 
  elements, but only one of them maps to it by function $\varphi$.)
\item Every element $e$ in the unprobed set is covered in the solution
  $\calS$.
\end{itemize}

We first show that the recourse is $O(1)$, and next that the algorithm
is $f$-competitive.

\smallskip
\noindent
{\bf Recourse Bound.}
As discussed in Section~\ref{sec:notation}, it suffices to only count
set deletions. Indeed, each set added was either deleted (and hence can
be charged to the deletion), or still exists in $\St$ (but this number
is at most the number of active elements $n_t$, by the bijectiveness of
the mapping $\varphi$).  To complete the proof, note that no set is
deleted from $\St$ on an element arrival, and at most one set,
$\varphi(e)$, is deleted when $e$ departs.

\smallskip
\noindent
{\bf Competitive Ratio.}
To show the competitive ratio, it is useful to denote the set of elements
in $S$ at time $t$ by $S_t$. Note that $S_t$ evolves over time.
The main property that we prove is
the following.

\begin{lemma}
\label{lma:set-count}
	At time $t$, for any set $S_t$,
	$$\ex\left[\sum_{e\in S_t \cap P_t} c_{\varphi(e)}\right] \leq f \cdot c_{S_t}.$$
\end{lemma}
\begin{proof}
  Since the adversary is oblivious of the random choices made by the
  algorithm, we can consider a fixed input sequence. Further, we
  condition on the random choices of elements not in $S_t$ (irrespective
  of whether they are probed or not).  We will show the desired bound
  under this conditioning, and so the result will follow once we remove
  the conditioning.

  The execution of our algorithm can now be seen as a tree. Given the
  fixed input sequence, and coin tosses of elements not in $S_t$, we can
  determine the first element probed by the algorithm. Similarly, given
  the random choices of the first $i$ elements in $S_t$ which get
  probed, we can determine which element in $S_t$ will be probed
  next. Therefore, let us define a sequence of random variables $X_1,
  X_2, \ldots$ and $Y_1, Y_2, \ldots$, where $X_i$ is the $i^{th}$
  element of $S_t$ that gets probed, and $Y_i$ is the set selected
  by this probed element. If we probe less than $i$ elements,
  then $X_i$ and $Y_i$ are defined as $\bot$. Now observe that given
  $X_1, Y_1, \ldots, X_{i-1}, Y_{i-1}$, we can determine the identity of
  $X_i$. Also, we need to define these random variables for $i \leq
  |S_t|$ only. For notational convenience, let $m$ denote $|S_t|$.
  Finally, observe that
  \begin{gather*}
    \ex\left[\sum_{e\in S_t \cap P_t} c_{\varphi(e)}\right] = \sum_i \ex\left[c_{Y_i}\right],
  \end{gather*}
  where $c_{Y_i}$ is 0 if $Y_i = \bot$. Now we state the induction
  hypothesis.
  \begin{center}
    IH($i$): $\ex\left[c_{Y_i}+ \ldots + c_{Y_m}|X_1, Y_1, \ldots, X_{i-1},
    Y_{i-1}\right] \leq f \cdot c_{S_t}. $
  \end{center}
  Let us check this for $i=m$. Suppose we are given $X_1, Y_1, \ldots,
  X_{m-1}, Y_{m-1}$. Then we know the identity of $X_m$. If $X_m =
  \bot$, we do not incur any cost, and so the statement holds
  trivially. Else, suppose $X_m = e \in S_t$.  Then the expected cost of
  $Y_m$ is
  \begin{gather*}
    \sum_{S':e \in S'} c_{S'} \cdot \frac{1/c_{S'}}{\sum_{S'': e \in
        S''} 1/c_{S''}} = \frac{f}{\sum_{S'': e \in S''} 1/c_{S''}} \leq
    f \cdot c_{S_t}.
  \end{gather*}
  Now suppose the statement holds for $i+1$, and we want to prove it for
  $i$. Suppose we are given the random variables $X_1, Y_1, \ldots,
  X_{i-1}, Y_{i-1}$. This determines $X_i$ also.  
  Therefore, we can write the desired sum as
  \begin{gather*}
    \ex\left[c_{Y_i}\mid X_1, Y_1, \ldots, X_{i-1}, Y_{i-1}, X_i\right] +
    \ex\left[c_{Y_{i+1}} + \ldots + c_{Y_m} \mid X_1, Y_1, \ldots, X_{i-1},
    Y_{i-1}, X_i\right].
  \end{gather*}
  Assume $X_i \neq \bot$, otherwise both terms are 0 are we would be done.
  Let $e$ denote $X_i$, and $\Delta_e$ denote $\sum_{S': e \in S'} 1/c_{S'}$.
  The first term is at most $f/\Delta_e$, the argument being similar to 
  the base case, $i=m$. Let $q_e$ denote the probability that element
  $e$ chooses set $S$. So
  $q_e = \frac{1}{c_{S_t} \Delta_e}$ and we can write the second term as
  \begin{gather*}
    \sum_{S': e \in S', S' \neq S_t} \ex\left[c_{Y_{i+1}} + \ldots +
    c_{Y_m} \mid X_1, Y_1, \ldots, X_i, Y_i=S'\right] \cdot \pr\left[Y_i =
    S' \mid X_1, Y_1, \ldots, X_{i-1}, Y_{i-1}, X_i\right].
  \end{gather*}
  Using the induction hypothesis and independence, this can be bounded by
  \begin{gather*}
    \sum_{S': e \in S', S' \neq S_t} f\cdot c_{S_t} \cdot \pr[Y_i = S'|
    X_i] = f\cdot c_{S_t} \cdot (1 - q_e).
  \end{gather*}
  So, the overall sum is at most
  \begin{gather*}
    \frac{f}{\Delta_e} + \left( 1 - \frac{1}{c_{S_t} \Delta_e} \right) f
    \cdot c_{S_t} = f \cdot c_{S_t}.
  \end{gather*}
  This proves the lemma.
\end{proof}

To complete the analysis, we use a dual fitting argument. Define the dual
of an element $e$ as $\nicefrac{c_{\varphi(e)}}{f}$ (and 0 if the element is 
unprobed). The above lemma implies that this dual is feasible in 
expectation. Note that the dual objective is a $\nicefrac1f$ fraction of the 
cost of algorithm's solution. Hence, by standard LP duality, the 
competitive ratio of the algorithm is $f$ in expectation.

\section{A Scaling-Based Algorithm}
In this section, we turn our focus to non-amortized bounded-recourse algorithms. We give algorithms which run in exponential time, and the competitive ratio matches that of the dynamic greedy algorithms from~\Cref{sec:greedy}. However, we are able to get $O(1)$ non-amortized $O(1)$ recourse. Further, we get a similar result for the fractional set cover problem as well. The notion of recourse in the fractional setting can be defined in a natural manner as follows -- the number of changes while moving from one fractional solution to another, i.e., recourse, is defined as the $L_1$-distance between the corresponding  fractional solution vectors. Our techniques rely on reducing a general set cover instance to several instances where  sets have {\em nice}  cost structure in a particular instance. As a corollary, we get improved results for the unweighted set cover problem -- the competitive ratio improves to $O(1)$.  
 Formally, the main results of this section will be:
\begin{theorem}
  \label{thm:scaling-results}
  There are deterministic online algorithms which perform $O(1)$
  recourse operations per step, and achieve the following guarantees:
  \begin{OneLiners}
  \item[(i)] $O(\log n)$-competitiveness for fully-dynamic (integral)
    set cover in exponential time,
  \item[(ii)] $O(\log n)$-competitiveness for fully-dynamic (fractional)
    set cover in poly-time.
  \item[(iii)] For the fully-dynamic unweighted set cover problem, we get similar results as~(i) and~(ii) above along with improvement in competitive ratio to $O(1)$. 
  \end{OneLiners}
\end{theorem}

To get these results, we show a reduction from general instances to
certain ``nice'' instances where the cheapest set covering every element
of the universe has the same cost. This reduction loses a logarithmic
factor in the competitiveness. Call a set-cover instance $\I = (U,
\calF)$ \emph{well-scaled} if for each element $e$, the cost of the
cheapest set containing $e$ is the same; i.e., for each $e, e'\in U$,
$\min_{S \in \calF: e \in S} c(S) = \min_{S' \in \calF: e' \in S'}
c(S')$. In particular, an instance of unweighted set cover is  a well-scaled instance. We now show how any set cover instance can be reduced to a small number of such well-scaled instances. 

\subsection{A Reduction to Well-Scaled Instances}
\label{sec:reduction-scaling}

In this section, we show that an on-line algorithm for well-scaled instances can be used for general instances. We incur a logarithmic loss in competitive ratio, but the recourse bound remains unchanged. 
\begin{lemma}
  \label{lem:scaling}
  Suppose we have an $\alpha(t)$-competitive online algorithm
  $r(t)$-recourse algorithm $\calA$ for well-scaled instances of length
  $t$, where $\alpha()$ and $r()$ are  monotone increasing. 
  Then there is an $O(\alpha(t) \log t)$-competitive algorithm
  $r(t)$-recourse online algorithm for all dynamic set cover instances of length~$t$.
\end{lemma}


\begin{proof}
  First, we can assume that all set costs $c(S)$ are powers of $2$,
  losing only a factor of $2$ in the competitiveness. We show how to
  partition the input sequence $\seq$ into several well-scaled
  input sequences $\seq^{(0)}, \seq^{(1)}, \ldots$, one
  for each power of $2$, in an online fashion: indeed, when we see an
  element $e$ such that the cheapest set covering $e$ has cost $2^k$, we
  send $e$ to the sequence $\seq^{(k)}$. By construction, each
  of the sequences $\seq^{(k)}$ is well-scaled. We run the
  algorithm $\calA$ on each of these independently.

  At any time $t$, let $\ell$ be the integer such that $2^\ell :=
  \max_{e \in \At} \min_{S \in \calF: e \in S} c(S)$. Clearly, $2^\ell$ is
  a trivial lower bound on $\Opt_t$. Notice that by definition, every element in $\At$ can be individually covered by a set of cost at most $2^\ell$.  We will construct sub-instances $\seq^{(k)}$, where $c \leq \ell$. We now show that the optimal cost for sub-instances
  $\seq^{(k)}$, where $k \leq \ell-\log t$ is very small. Indeed, the optimal cost of
  sub-instance $\seq^{(\ell-\log t - c)}$, where $c \geq 0$ is an integer,  is at most $t \cdot 2^{\ell-\log t - c} \leq 2^{\ell-c} \leq  2^{-c} \cdot \Opt_t$.
  Hence the total cost incurred by the algorithms run on
  instances $\seq^{(\ell - \log_2 t)}, \seq^{(\ell - \log_2
    t - 1)}, \ldots$ is at most $\alpha(t) \Opt_t \cdot \sum_{c \geq 1}
  2^{-c} \leq \alpha(t) \Opt_t$. 
  Moreover, the optimal cost of each of the
  other sub-instances $\seq_{\ell}, \ldots, \seq_{\ell -
    \log_2 t+1}$ is at most $\Opt_t$; and so our algorithm incurs cost at most
  $\alpha(t) \Opt_t$ for each of these sub-instances. Combining these, we get the total cost over
  all these scales is at most $(\log_2 t + 2) \cdot \alpha(t) \Opt_t$.

  Now we look at the recourse cost. Let $t_k$ denote the length of the sub-instance $\seq^{(k)}$.
  Then, the total recourse cost is at most $\sum_k t_k \cdot r(t_k) \leq r(t) \cdot t$. 
  Note that if the recourse of
  algorithm $\calA$ is non-amortized, so is the recourse of the combined
  algorithm.
\end{proof}

\subsection{An Algorithm for Well-Scaled Instances}
\label{sec:algo-unit-cost}

We now show how to get an online $O(1)$-competitive algorithm with constant
amortized recourse for any well-scaled input instances.
This algorithm will solve set cover instances exactly and
hence not run in polynomial time. We will then make this algorithm have 
non-amortized worst-case recourse budget per operation by losing only a 
constant factor in the competitive ratio.

By scaling, we can assume that for each element the cheapest set
covering it has unit cost. Let us recap some notation from the earlier sections: each element operation $\sigma_t$ is either an insertion $(e_t, +)$ or a deletion $(e_t, -)$, and $A_t$ denotes the set of active elements at time $t$. The algorithm maintains a solution $\calS_t$ at time $t$ (after processing the request $\sigma_t$). The online algorithm runs in phases. Initially, when an element $e$ arrives, we will start the first phase, and pick the cheapest set containing $e$. We shall use $\tau(i)$
to denote the time $t$ at which phase $i$ begins.  So $\tau(1)$ is 1. The algorithm will maintain the following invariant for all phases $i$, $i \geq 1$: the solution $\calS_{\tau(i)}$ at the beginning of phase $i$ is an optimal solution for the instance $A_{\tau(i)}$, and so has cost $\Opt(A_{\tau(i)}). $

The update procedure for maintaining the solution is described in Algorithm~\ref{alg:scaling}. 

\begin{algorithm}
\caption{Dynamic$(e_\bt, \pm)$}
\label{alg:scaling}
\begin{algorithmic}
\If{the operation $\sigma_\bt$ is $(e_\bt, +)$}
\State Pick the cheapest (i.e., unit cost) set covering $e_t$.
\ElsIf{the operation $\sigma_\bt$ is $(e_\bt, -)$}
\State Do nothing 
\EndIf
\State If  the number of inserts or the number of deletes in the
  current phase $i$ is at least $\frac{1}{2} \Opt(A_{\tau(i)})$ 
 \State \ \ \ \ \ \ \emph{Recompute} the solution, i.e., we drop all the sets 
 $ \calS_{t-1}$, and start the new phase
  $i+1$. 
  \State \ \ \ \ \ \ \ (Hence, $\tau(i+1) := t$ and $\calS_t$ is an optimal solution for 
  $A_t$)
\end{algorithmic}
\end{algorithm}







Note that the invariant is maintained by definition of the algorithm. We
now show the competitiveness and the recourse guarantee of the
algorithm.

\begin{lemma}
  \label{lem:scale-comp}
  At each time $t$, our solution is $3$-competitive, and total number of
  sets added until time $t$ is at most $4t$.
\end{lemma}

\begin{proof}
  Suppose time $t$ falls within phase $i$, i.e., $t \in [\tau_i, \tau_{i+1})$.  Let $L := t - \tau(i)$ be the length of the phase until
  now, with $L_{ins}$ inserts and $L_{del}$ deletes, where $L_{ins} +
  L_{del} = L$. Moreover, by the criterion for ending the phase we know
  that $L_{ins}, L_{del} \leq \frac12 \Opt(A_{\tau(i)})$.

  The  solution at time $t$ consists of the optimal solution on
  $A_{\tau(i)}$, plus unit-cost sets for each of the $L_{ins}$
  insertions, and hence has cost $\Opt(A_{\tau(i)}) + L_{ins} \leq
  \frac32 \Opt(A_{\tau(i)})$. But also observe the following:
  \begin{gather}
    \Opt(A_t) \geq \Opt(A_{\tau(i)}) - L_{del}. \label{eq:1} \\    
    \Opt(A_t) \geq \frac12 \Opt(A_{\tau(i)}). \label{eq:2} \\    
    \Opt(A_{\tau(i+1)}) \geq \frac12 \Opt(A_{\tau(i)}). \label{eq:3} 
  \end{gather}
Indeed, to see the first inequality above, notice that one feasible solution for covering $A_{\tau(i)}$ is to take the optimal solution for $A_t$ and to add one unit-cost set for each of the deleted elements. The second inequality follows by plugging in $L_{del} \leq \frac12 \Opt(A_{\tau(i)})$, and the third follows by setting $t = \tau(i+1)$. 
Using~\eqref{eq:2}, we immediately get that our solution costs at most $3 \cdot \Opt(A_t)$.

  For the recourse bound, we inductively assume the recourse bound is
  true until the end of phase $i-1$. If request $t$ does not result in a
  phase ending, we only add a single new set, and so the recourse bound continues
  to hold. Suppose phase $i$ ends
  when request $t$ is received, so $\tau(i+1) = t$. Let $L = t -
  \tau(i)$ be the length of the phase. We add one new set for the
  $L_{ins}$ requests. Moreover, since
  \begin{gather}
    \Opt(A_{\tau(i+1)}) \leq \Opt(A_{\tau(i)}) + L_{ins} \label{eq:4}
  \end{gather}
  (by an argument identical to~(\ref{eq:1})), and since
  each set has at least unit cost, the number of sets in the new optimal
  solution is at most $\Opt(A_{\tau(i+1)}) \leq \Opt(A_{\tau(i)}) +
  L_{ins}$. Moreover, since the phase ended, at least one of $L_{ins}$
  or $L_{del}$ reached $\frac12 \Opt(A_{\tau(i)})$. So the total
  number of sets added in this phase is at most \[ L_{ins} +
  (\Opt(A_{\tau(i)}) + L_{ins}) \leq 2\Opt(A_{\tau(i)}) \leq
  4\max(L_{ins}, L_{del}) \leq 4L. \]
  This completes the induction for phase $i$, and hence the proof.
\end{proof}
Note that any set that is deleted has to be previously added, so the
bound on the amortized number of set additions and deletions per step is
$8$.

\subsubsection{An algorithm for fractional set cover}
\label{sec:algo-unit-cost-fract}

Using ideas identical to Section~\ref{sec:algo-unit-cost}, we can also get a \emph{polynomial-time} $O(1)$-competitive $O(1)$-recourse algorithm for the fractional set cover problem on well-scaled instances by just using the optimal fractional solution $\lp(A_{\tau(i)})$ at the beginning of each phase, and by ending
phase $i$ when the number of inserts or deletes becomes $\frac12
\lp(A_{\tau(i)})$. Here the recourse is measured in terms of the sum of 
$\ell_1$-distances between the $\{x_S\}$ vectors at consecutive
times. Similarly, we can use the ideas from Section~\ref{sec:reduction-scaling}
to get a a poly-time fractional $O(\log t)$-competitive $O(1)$-recourse
algorithm for all set cover instances. In fact, we can de-amortize the
recourse as indicated in the following section. 

\subsection{De-Amortizing the Algorithm}
\label{sec:deamort-unitcost}

We now show how to de-amortize the algorithms above, so that we perform
only a constant number of set additions or deletions per step, while
hurting the competitiveness only by a constant. The approach is based on
two simple observations about Algorithm~\ref{alg:scaling} 
(recall that $\calS_t$ denotes the solution maintained by our
algorithm at time $t$):

\begin{claim}
\label{cl:deamort}
The length of phase $i$, $\tau(i+1)-\tau(i)$ is at least $\frac12 \cdot |\Opt(A_{\tau_i})|$ and 
at least $\frac{1}{6} \cdot |\calS_{\tau_i-1}|$. 
\end{claim}

\begin{proof}
The first observation follows from the fact that $\tau(i+1) - \tau(i) \geq \frac12 \cdot \Opt(A_{\tau_i})$, 
and the fact that $\Opt(A_{\tau_i}) \geq |\Opt(A_{\tau_i})|$ (because each set has cost at least 1). 
Now we show the second statement. Note that the cost of our solution at the end 
of phase $i-1$, i.e., $\calS_{\tau(i)-1}$, is at most $\frac32 \Opt(A_{\tau(i-1)})$. Indeed, when phase $i-1$ started, we had a solution of cost $\Opt(A_{\tau(i-1)})$, 
and then this phase inserted at most $\frac12 \cdot \Opt(A_{\tau(i-1)})$ elements.
 The length of phase $i$ is at least
$\frac12 \Opt(A_{\tau(i)}) \geq \frac14 \Opt(A_{\tau(i-1)})$
   using~(\ref{eq:1}) with LHS being $\Opt(A_{\tau(i)})$.
\end{proof}

In Algorithm~\ref{alg:scaling}, we were purging all the sets in $\calS_{\tau(i)-1}$
and adding new sets from $\Opt(A_{\tau(i)})$. The claim above shows that if we instead purge 12 sets from $\calS_{\tau(i)-1}$ and add 4 news sets from $\Opt(A_{\tau(i)})$ during each time step of phase $i$, then we will be done half-way during this phase, i.e., things will not spill over to the next stage. Thus, we will get a non-amortized constant recourse algorithm. In terms of competitive ratio, we will just need to account for the fact that we carry over  sets from 
$\calS_{\tau(i)-1}$ during phase $i$. We now give details of this idea. 

As indicated above, the new algorithm is as follows: use Algorithm~\ref{alg:scaling} (in background) to figure out phases. Let $\calS'(t)$ be the solution maintained by the algorithm at time $t$ (we shall use $\calS(t)$ to denote the solution maintained by Algorithm~\ref{alg:scaling}). At the start of phase $i$, we merely mark all the sets in $\calS'(\tau(i))$ as ``dirty'' -- note that we do not remove these sets yet. At each time $t$ during phase $i$, if a new element arrives, choose the cheapest set containing it (of cost 1). 
Besides this, if $\calS'(t)$ does not contain all the sets in $\Opt(A_{\tau(i)})$, we bring in 4 new sets from $\Opt(A_{\tau(i)})$ into $\calS'(t)$. In case, $\calS'(t)$  already contains all of 
$\Opt(A_{\tau(i)})$, we purge 12 dirty sets from $\calS'(t)$ (if there are any sets marked dirty). 
Note that we are allowing our algorithm to keep multiple copies of a set -- a set could be marked dirty, and another copy of it could be in $\Opt(A_{\tau(i)})$ (and so not marked dirty in our solution). Similarly, we are allowing duplicates when we pick a set on each element arrival. 

Now we analyze our algorithm. We maintain the following invariant for all $i$: at the end of phase $i-1$, i.e., at time $\tau(i)-1$, the sets $\calS(\tau(i)-1)$ and $\calS'(\tau(i)-1)$ are identical. This follows immediately from Claim~\ref{cl:deamort}.

The cost of sets maintained at any time $t$ during phase $i$ is at most
$\frac32 \Opt(A_{\tau(i-1)}) + \Opt(A_{\tau(i)}) + \frac12
\Opt(A_{\tau(i)})$---the first expression upper bounds the cost of the
dirty sets which remained at the end of phase $i-1$ (which is same as 
$\calS(\tau(i)-1)$) , the second bounds the cost of the optimal solution to be brought in at the beginning of phase $i$, and the third bounds the cost of sets added in this phase. But note that $\Opt(A_{\tau(i-1)}) \leq 2 \, \Opt(A_{\tau(i)})$ from~\eqref{eq:3}. Hence the overall sum is $O(1) \Opt(A_{\tau(i)})$ which is in turn $o(\Opt(A_t))$ from~\eqref{eq:2}. Hence at any time $t$ the
de-amortized algorithm is constant-competitive, and adds and deletes at
most a constant number of sets at each time step. This proves 
Theorem~\ref{thm:scaling-results}(i). Part~(ii) follows in a similar manner. 
Part~(iii) of the theorem follows from the fact that unweighted instances are special cases of 
well-scaled instances. 





\section{The Combiner Algorithm} \label{sec:combiner}

There are two classical approximation algorithms for set cover: one
which gives a $\ln n$-approximation, and the other which gives a
$f$-approximation where $f$ is the maximum frequency of any element,
i.e., the maximum number of sets which can cover a single
element. Therefore by computing both the solutions and outputting the
one with least cost, we obtain an approximation ratio of $\min(\ln n,
f)$. It is then natural to ask if we can obtain such a guarantee in the
online-recourse or fully dynamic models for set cover.

In this section, we give a {\em combiner algorithm} which takes in
fully-dynamic (or online-recourse) algorithms with these kinds of
different guarantees and combines them into a single fully-dynamic (or
online-recourse) algorithm.

\begin{theorem} \label{thm:best-of-two} Let $n = \max_{t=1}^{T} \nt$
  denote the maximum universe size that will be seen in a particular
  dynamic instance. Suppose we are given different algorithms for
  fully-dynamic set cover: (a)~an algorithm $G$ with competitive ratio
  $O(\log \nt)$ at any time $t$ and amortized work (or recourse) $W_G$ per
  insert/delete, and (b)~a family of algorithms $PD_f$ which works on
  instances where element frequencies are bounded by $f$, and has
  competitive ratio $O(f^c)$ for some constant $c \geq 1$ and amortized
  work (or recourse) $W_{PD}$ per insert/delete, we can combine them into an efficient
  fully-dynamic algorithm $C$ with competitive ratio $O(\min( \log \nt,
  \ft^c))$ and amortized work (or recourse) $O(W_G + W_{PD})$.
\end{theorem}

Our idea is the following: We partition the elements into different
groups, with group ${\cal F}^i_t$ denoting the collection of elements in
$\At$ of frequency between $[2^i, 2^{i+1})$ for $i \geq 0$. Let $\lt$ be
such that $2^{c\,\lt-1} < \log \nt \leq 2^{c\,\lt}$. Then, we cover the
elements in ${\cal F}^0_t, {\cal F}^1_t, \ldots, {\cal F}^\lt_t$ by
individual runs of algorithm $PD_f$ (where we use $PD_{f}$ with $f =
2^{i+1}$ for the elements in $\calF^i_t$), and the rest of the elements
by a single run of algorithm $G$.

\begin{lemma} \label{lem:cost-best} The cost of any solution respecting
  the above invariant is $O(\min( \log \nt, \ft)) \Opt_\bt$ at any time
  $t$.
\end{lemma}

\begin{proof}
  Firstly, note that if $(\ft)^{c} < \log_2 \nt$, then by the algorithm
  above, we will not cover any element using algorithm $G$. So if $\ell
  := \lceil \log \ft \rceil$, then the total cost of our solution is at
  most $\sum_{i=0}^{\ell} (2^{i+1})^c \cdot \Opt_\bt \leq O(\ft^c) \cdot
  \Opt_\bt = O(\min (\log \nt, \ft^c))\, \Opt_\bt$. On the other hand,
  if $\ft \geq \log \nt$, then run algorithm $PD$ only on elements in
  the groups ${\cal F}^0_t, {\cal F}^1_t, \ldots, {\cal F}^\lt_t$, so
  the total cost of these solutions is at most $\sum_{i=0}^{\lt}
  (2^{i+1})^c \,\Opt_\bt = O(\log \nt)\, \Opt_\bt$, and algorithm $G$
  maintains a solution of cost at most $O(\log \nt)\, \Opt_\bt$ on the
  remaining elements. So again the total cost is at most $O(\min (\log
  \nt, \ft)) \Opt_\bt$, thereby completing the proof.
\end{proof}

\begin{lemma} \label{lem:work-best}
There is an efficient way to maintain the above invariant with amortized work $O(W_G + W_{PD})$ per element operation.
\end{lemma}

\begin{proof}
Firstly, when an element arrives, we insert into exactly one group, and thereby feed it into one of algorithms $G$ or $PD$. Here we incur work of $\max (W_G, W_{PD}) \leq W_G + W_{PD}$. Similarly, when an element departs, we delete it from exactly one group, and again incur work of $\max (W_G, W_{PD}) \leq W_G + W_{PD}$. We now show how we can also maintain the invariant when the values $\lt$ and $\nt$ change, which would sometimes force us to recompute the solution to satisfy the invariant. Indeed, if $\lt$ increases from say $k$ to $k+1$, we will have to remove the elements in group ${\cal F}^{k+1}_t$ from the run of algorithm $G$ and feed them to the run of algorithm $PD$ corresponding to group ${\cal F}^{k+1}_t$. But this work can be amortized to all the element arrivals which caused $\nt$ to essentially square itself, so as to to increase $\log \nt$ by a factor of two, which in turn makes $\lt$ increase by $1$. Similarly, if $\lt$ decreases from say $k+1$ to $k$, we will have to remove the elements in group ${\cal F}^{k+1}_t$ from the corresponding run of algorithm $PD$ and feed them to the run of algorithm $G$. But this work can be amortized to all the element departures which caused $\log \nt$ decrease by a factor of two, which in turn made $\lt$ decrease by $1$.
\end{proof}

\Cref{thm:best-of-two} then follows from~\Cref{lem:cost-best,lem:work-best}.

\subsection{Knowledge of $f$, and Dependence on $f_t$}
\label{sec:unknown-f}

In Section~\ref{sec:short-pd-update} and Appendix~\ref{sec:pd-dyn}, we
assumed the algorithm was given a value $f$ such that the frequencies
$f_e \leq f$ for all elements $e$, and then we showed an
$O(f^c)$-competitiveness guarantee for the algorithm, for some constant
$c \geq 1$. The idea used for the combiner also gives, in a black-box
fashion, an algorithm that does not require this knowledge up-front;
indeed, the algorithm is $O(f_t^c)$-competitive, where $f_t := \max_{e
  \in A_t} f_e$ is the maximum frequency of any element active at
time~$t$.

The idea is simple: for each integer $i \in \{0, 2, \ldots, \lceil
\log_2 m \rceil$ we run a copy $\calA_i$ of the given
$O(f^c)$-competitive algorithm $\calA$. For each update $(e, \pm)$, we
feed it to $\calA_i$ if $f_e \in (2^{i-1}, 2^i]$; let $\pmb{\sigma}_i$
be the subsequence of the input $\pmb{\sigma}$ given to
$\calA_i$. Clearly this partitioning can be done with constant extra
update time (and no recourse) per element if element frequencies are
given, else we can calculate these frequencies in $\sum_e f_e$ time.
The set cover solution we output
is the union of the solutions maintained by all these copies. To show
competitiveness, look at time $t$, and let $\ell := \lceil \log_2 f_t
\rceil $. There are no elements to be covered in the copies of $\calA_i$
for $i > \ell$, so the cost of the solution is at most
\begin{gather*}
  \sum_{i = 0}^{\ell} (2^i)^c \cdot \Opt_t(\pmb{\sigma}_i) \leq \sum_{i =
    0}^{\ell} (2^i)^c \cdot \Opt_t \leq O(f_t)^c \cdot \Opt_t.
\end{gather*}


\section{Applications}

\subsection{Dynamic $k$-Coverage}
\label{sec:coverage}

In the $k$-coverage problem, given a set system $(U, \calF)$ and an
integer $k$, the goal is to pick $k$ sets from $\calF$ that maximize the
size of their union. Note that all sets have unit cost in this model.
The following fact is well-known: for the $k$-coverage problem, the
greedy algorithm is a $(1-1/e)$-approximation, and this is the best
possible unless $P=NP$.

Here, the greedy algorithm picks a set which covers the maximum number
of yet-uncovered elements. Now, suppose $U_i$ is the set of
uncovered elements after $i$ sets have been picked (so $U_0 = U$), and
we pick as the $i+1^{st}$ set some set $S_i \in \calF$ that satisfies
$|S_i \cap U_i| \geq \alpha \max_{T \in \calF} |T \cap U_i|$ for some
constant fraction $\alpha > 0$. It turns out that this
``approximately greedy'' algorithm also has a constant approximation
ratio. We formally show this first.

Let a sequence of sets $S_1, S_2, \ldots, S_k$ be \emph{$\alpha$-approximately-greedy} 
if the following property holds:
for all $1 \leq i \leq k$, the residual coverage of the $i^{th}$ set $S_i$ is at least $\alpha$ times the residual coverage of the best possible set, given that $S_1, S_2, \ldots, S_{i-1}$ are already chosen. That is, $|S_i \setminus (S_1 \cup S_2 \cup \ldots S_{i-1})| \geq \alpha \cdot |S' \setminus (S_1 \cup S_2 \cup \ldots S_{i-1})|$ for all $S' \in \calF$. 

Note that if $\alpha = 1$, we get the greedy property. 
We first show the approximation factor of an approximately 
greedy solution for the $k$-coverage problem.

\begin{lemma}
\label{lma:approx-greedy}
Consider an $\alpha$-approximately-greedy sequence of sets 
$S_1, S_2, \ldots, S_k$. Then these sets give a 
$(1-e^{-\alpha})$-approximate solution to the $k$-coverage
problem.
\end{lemma}
\begin{proof}
Let $N_i$ denote the total coverage of the first $i$ sets in the sequence, 
i.e., $N_i := |S_1 \cup S_2 \cup \ldots S_{i}|$.  From the $\alpha$-approximate greedy property, and the fact that the optimal solution covers $\Opt$ elements with $k$ sets, we get that for each $i$, $$N_i - N_{i-1} \geq \alpha\cdot \frac{(\Opt - N_{i-1})}{k} \,.$$
Adding $\Opt$ on both sides and re-arranging terms, we get:
\begin{eqnarray}
\nonumber \Opt - N_{i} &\leq& \Opt - N_{i-1} - \alpha\cdot \frac{(\Opt - N_{i-1})}{k}  \\
\nonumber &=& \left(1 - \frac{\alpha}{k}\right) (\Opt - N_{i-1}) 
\end{eqnarray}
Using this inequality iteratively for $i=1,2,\ldots,k$, we get:
$$\Opt - N_k \leq \left(1 - \frac{\alpha}{k}\right)^k \cdot \Opt \leq e^{-\alpha}  \cdot \Opt,$$ 
which implies that $N_k \geq \Opt ( 1 - \exp(-\alpha))$, completing the proof.
\end{proof}

Now, we use this notion of approximate greediness to design an 
algorithm for $k$-coverage in the fully-dynamic setting.
Our dynamic greedy framework with $\vt(e) = 1$ for all elements naturally
suggests the following algorithm: \emph{pick the $k$ sets with smallest
  densities}.  Recall that the density is now $\rho_t(S) :=
1/|\cov(S)|$, since all sets have unit cost.

\begin{theorem}
  The algorithm above is a constant-competitive fully-dynamic algorithm
  for $k$-coverage. It has recourse $O(\log n)$, and can be implemented
  with update time $O(f \log n)$.
\end{theorem}

\begin{proof}
We first make the algorithm more concrete to specify tie breaking issues. Our algorithm maintains a solution to the corresponding set cover instance where the cost of every set is 1.
For each level $i$,
we have a (doubly linked) list $\calS_{t,i}$ of sets in $\calS_t$ which are in level $i$.
Whenever a set gets added or removed at some level $i$ in our solution, we either add or
remove a new location in $\calS_{t,i}$  {\em without} changing the relative ordering of the remaining sets in this list. This can be easily done because each set at level $i$ in $\calS_t$ maintain a pointer to its location in $\calS_{t,i}$. Now, when we need to specify the solution for the $k$-coverage instance, we scan the lists $\calS_{t,i}$ (starting from the smallest $i$) from left to right till we get  $k$ sets. It is easy to see that whenever we change our solution by $t$ sets, $\Omega(t)$ new sets would have changed their levels. Therefore, the recourse and update time bounds for this algorithm follow from the corresponding bounds for the set cover instance.

It remains to show the approximate greediness of the solution, so that we can 
use Lemma~\ref{lma:approx-greedy} and prove the competitive ratio.
Let $T_1, \ldots, T_k$ be the $k$ sets in the solution $\calS_t$,
and let their respective levels be $\ell_1, \ell_2, \ldots, \ell_k$. 
Define $S_i := \covt(T_i)$; since $\cup_{i=1}^k S_i \subseteq \cup_{i=1}^k T_i$,
any approximation factor that we show for $S_i$ also holds for $T_i$. 
By the density range of level
$\ell_i$, we know that $|S_i| \geq 2^{-\ell_i-10}$. 
(Note that all sets are of unit cost, so densities 
are reciprocals of coverage.) Moreover, since
every element is covered by a unique set, $S_i$'s are
disjoint. It follows that 
$|S_i| = |S_i \setminus (S_1 \cup S_2 \cup \ldots S_{i-1})|$.
For any set $S' \in \calF$ that is not among $S_1, S_2, \ldots, S_{i-1}$, 
let $S''$ denote $S' \setminus (S_1 \cup \ldots \cup S_{i-1})$. Observe that all elements
in $S''$ are at levels $\ell_i$ or above. So, by the stability property of the dynamic set cover algorithm, for any $j\geq 0$, the number of elements in $S''$ that are at level $\ell_i+j$ 
is at most $2^{-\ell_i-j}$, for all $j \geq 0$. Summing over all $j \geq 0$, we get that $|S''| \leq 2\cdot 2^{-\ell_i}$. Thus, $\alpha = 2^{-11}$ suffices for 
$\alpha$-approximate greediness.

The theorem now follows from Lemma~\ref{lma:approx-greedy}.
\end{proof}


\subsection{Dynamic Non-Metric Facility Location}
\label{sec:fl}

We now show how our dynamic greedy framework can be applied to the \emph{dynamic non-metric facility location} problem. In the offline problem, we are given a set of facilities $\calF$, and a collection of clients $\calC$. For each facility $i \in \calF$ there is a facility opening cost $f_i$, and for every $i \in \calF, j \in \calC$, there is a connection cost $c_{i,j} \geq 0$ of 
connecting client $j$ to facility $i$. The goal is to open a set of facilities $X \subseteq \calF$ and assign every client $j$ to some open facility $\varphi(j) \in X$, to minimize the total cost $ \sum_{i \in X} f_i + \sum_{j} c_{\varphi(j),j}$. If all $c_{ij}$ are either $0$ or $\infty$, the problem reduces to set cover. As in set cover, this problem also admits an 
$O(\log n)$-approximation offline, where $n$ is the number of clients. 

In the online setting of this problem, clients arrive online and must be irrevocably 
connected to facilities. Similarly, facilities once opened cannot be closed. For this problem,~\cite{ABABN} gave an $O(\log n \log m)$-competitive algorithm, when there are $m$ facilities and $n$ clients. We now show that with $O(1)$ recourse (i.e., we can open/close 
$O(1)$ facilities per client), we can improve this bound to get a competitive ratio of $O(\log n)$.

Our techniques also illustrate one benefit of~\Cref{thm:main-recourse}(a), that the competitive ratio does not depend on the number of sets $m$. Indeed, the facility location problem can be modeled as a set cover problem with \emph{exponential} number ofsets as follows: each client $j$ corresponds to an element, and for each subset of clients $J' \subseteq \calC$ and each facility $i$, we have a set $S(J',i)$ with cost $\sum_{j \in J'} c_{i,j'} + f_i$. This set contains exactly the elements in $J'$. By construction, a feasible set cover solution for this instance corresponds to a feasible facility location solution with an identical objective value. Therefore, we can apply our framework (specifically~\Cref{thm:main-recourse}(a)) to get $O(\log n)$ competitive algorithms with $O(1)$-amortized recourse. 

\newpage
{\small
\bibliography{bib}
\bibliographystyle{amsalpha}
}






\end{document}